\documentclass[journal]{IEEEtran}
\usepackage{amsmath,amsfonts}
\usepackage{algorithmic}
\usepackage{algorithm}
\usepackage{array}
\usepackage[caption=false,font=normalsize,labelfont=sf,textfont=sf]{subfig}
\usepackage{textcomp}
\usepackage{stfloats}
\usepackage{url}
\usepackage{verbatim}
\usepackage{graphicx}
\usepackage{cite}
\usepackage{balance}
\usepackage{enumitem}

\newcommand{\In}{\mathrm{in}}
\newcommand{\Out}{\mathrm{out}}
\newcommand{\ato}{\overset{\mathrm{a.s.}}{\to}}
\newcommand{\aeq}{\overset{\mathrm{a.s.}}{=}}
\newcommand{\plkto}{\overset{\mathrm{PL}(k)}{\to}}
\newcommand{\plto}{\overset{\mathrm{PL}(2)}{\to}}
\newcommand{\pleto}{\overset{\mathrm{PL}(2+\epsilon)}{\to}}

\newtheorem{definition}{Definition}
\newtheorem{assumption}{Assumption}
\newtheorem{theorem}{Theorem}
\newtheorem{lemma}{Lemma}
\newtheorem{proposition}{Proposition}

\begin{document}

\title{Decentralized Generalized Approximate Message-Passing for Tree-Structured Networks}

\author{Keigo~Takeuchi,~\IEEEmembership{Member,~IEEE}
\thanks{
This work was supported in part by the Grant-in-Aid 
for Scientific Research~(B) (Japan Society for the Promotion of Science (JSPS)
KAKENHI), Japan, under Grant 21H01326/23K20932. An earlier version of this 
paper was presented in part at the 2024 IEEE International Conference on 
Acoustics, Speech and Signal Processing [DOI: 10.1109/ICASSP48485.2024.10446135].}
\thanks{The author is with the Department of Electrical and Electronic Information Engineering, Toyohashi University of Technology, Toyohashi 441-8580, Japan (e-mail: takeuchi@ee.tut.ac.jp).}
}

\markboth{IEEE transactions on information theory}%
{Takeuchi: Decentralized Generalized Approximate Message-Passing for Tree-Structured Networks}

\IEEEpubid{0000--0000/00\$00.00~\copyright~2021 IEEE}

\maketitle

\begin{abstract}
Decentralized generalized approximate message-passing (GAMP) is proposed for compressed sensing from distributed generalized linear measurements in a tree-structured network. Consensus propagation is used to realize average consensus required in GAMP via local communications between adjacent nodes. Decentralized GAMP is applicable to all tree-structured networks that do not necessarily have central nodes connected to all other nodes. State evolution is used to analyze the asymptotic dynamics of decentralized GAMP for zero-mean independent and identically distributed Gaussian sensing matrices. The state evolution recursion for decentralized GAMP is proved to have the same fixed points as that for centralized GAMP when homogeneous measurements with an identical dimension in all nodes are considered. Furthermore, existing long-memory proof strategy is used to prove that the state evolution recursion for decentralized GAMP with the Bayes-optimal denoisers converges to a fixed point. These results imply that the state evolution recursion for decentralized GAMP with the Bayes-optimal denoisers converges to the Bayes-optimal fixed point for the homogeneous measurements when the fixed point is unique. Numerical results for decentralized GAMP are presented in the cases of linear measurements and clipping. As examples of tree-structured networks, a one-dimensional chain and a tree with no central nodes are considered.
\end{abstract}


\section{Introduction}
\subsection{Background}
\IEEEPARstart{A}{pproximate} message-passing (AMP)~\cite{Donoho09} is a 
powerful iterative algorithm for signal recovery from linear 
measurements~\cite{Donoho06,Candes061}. In particular, AMP using the 
Bayes-optimal denoiser---called Bayes-optimal AMP---is regarded as an 
asymptotically exact approximation of loopy belief 
propagation~\cite{Kabashima03}. Applications of AMP contain 
compressive imaging~\cite{Som12,Tan15}, radar~\cite{Anitori13}, 
sparse superposition codes~\cite{Rush17,Barbier17}, and  
low-rank matrix estimation~\cite{Lesieur17,Montanari21}.

State evolution~\cite{Bayati11,Bayati15,Takeuchi19}, motivated by 
\cite{Bolthausen14}, allows us to analyze 
the asymptotic dynamics of AMP rigorously when the sensing matrix has 
independent and identically distributed (i.i.d.) zero-mean sub-Gaussian 
elements. The asymptotic dynamics of AMP is characterized with a 
discrete-time dynamical system---called state evolution recursion. 
When the state evolution recursion has a unique fixed point,   
Bayes-optimal AMP was proved in \cite{Bayati11,Bayati15} to 
achieve the Bayes-optimal performance~\cite{Reeves19,Barbier20} 
asymptotically.  

Generalized AMP (GAMP)~\cite{Rangan11} is a generalization of AMP to 
the case of generalized linear measurements, which allow us to treat general 
noise beyond the additive noise in the linear measurements. GAMP expands 
applications 
of AMP to one-bit compressed sensing~\cite{Kamilov121,Kamilov122}, 
phase retrieval~\cite{Schniter14,Ma19}, and peak-to-average power ratio (PAPR) 
reduction~\cite{Bao16,Chen16}. Like AMP, the asymptotic 
dynamics of GAMP was analyzed via state evolution~\cite{Javanmard13}. 
When the state evolution recursion of GAMP has a unique fixed point, 
Bayes-optimal GAMP was proved in \cite{Barbier19} to achieve the 
theoretically optimal performance in terms of 
the minimum mean-square error (MMSE).  

Distributed algorithms are more desirable than centralized algorithms that 
run on a single node exploiting the full information about the sensing matrix 
and all measurements. To help readers understand the motivation of 
distributed algorithms, this paper presents a practical example in which 
distributed algorithms are needed, while further investigation for this 
example is out of the scope of this paper. 

Consider cell-free massive multiple-input multiple-output (MIMO) 
uplink~\cite{Marzetta10,Ngo17}, in which users distributed over a wide area 
are served by multiple access points with massive antenna arrays. The access 
points are connected to a central processing unit via a fronthaul network. 
When all channel matrices and received signals are available at the central 
processing unit, cell-free massive MIMO can be regarded as virtual massive 
MIMO, in which the users are served by the central processing unit with 
a virtually single massive antenna array. In this centralized system, however,  
a heavy load is concentrated on the central processing unit. 
From a practical point of view, it is desired to balance the processing load 
and communication traffic via distributed algorithms.  

\IEEEpubidadjcol

Distributed algorithms are separated into two types of algorithms. 
A first type contains distributed algorithms that run on a 
central node and multiple remote nodes, which correspond to a central 
processing unit and access points in the cell-free massive MIMO example, 
respectively. Each remote node only uses local 
measurements to compute a local estimate, which is aggregated in the central 
node. The central node combines the local estimates to obtain a global 
estimate, which is fed back to the remote nodes. 
The iteration between the central and remote nodes is repeated until the 
algorithm reaches a final result. Iterative thresholding algorithms for 
compressed sensing, such as iterative shrinkage thresholding algorithm 
(ISTA)~\cite{Daubechies04}, fast iterative shrinkage thresholding algorithm 
(FISTA)~\cite{Beck09}, and iterative hard thresholding 
(IHT)~\cite{Blumensath09}, can be implemented as this type of distributed 
algorithms. See \cite{Patterson13} for distributed IHT.

In the other type of distributed algorithms---called decentralized 
algorithms in this paper, algorithms run on multiple nodes in an ad hoc 
network with no central nodes. In the cell-free massive MIMO example, 
a central processing unit does not perform signal estimation anymore. It 
configures dynamically an appropriate network for multiple access points 
to perform signal estimation in a distributed manner. 
 
In decentralized algorithms, distributed protocols for average 
consensus~\cite{Saber04,Xiao04} are utilized to reach the same 
result as the corresponding centralized algorithm only by sharing processing 
results with adjacent nodes locally. As this type of distributed algorithms for 
compressed sensing, distributed least absolute shrinkage and selection operator 
(LASSO)~\cite{Mateos10}, distributed spectrum sensing~\cite{Bazerque10}, 
distributed basis pursuit~\cite{Mota12}, and distributed alternating 
direction method of multipliers (ADMM)~\cite{Mota13,Shi14} were proposed. 

AMP was extended to distributed 
AMP~\cite{Han14,Han16,Ma17,Guo22,Bai22,Hayakawa18} 
exploiting a central node. More precisely, distributed AMP in 
\cite{Han14,Han16,Ma17} utilizes feedback from the central node to refine 
messages in each remote node, like distributed IHT~\cite{Patterson13}, while 
distributed AMP in \cite{Guo22,Bai22} exploits no feedback from the central 
node. Hayakawa {\em et al}.~\cite{Hayakawa18} proposed decentralized AMP 
(D-AMP) for tree-structured networks with no central nodes via consensus 
propagation~\cite{Moallemi06}. Consensus propagation can achieve perfect 
summation consensus for any tree-structured network: All nodes in the network 
can compute the exact value for the summation $\sum_{l}a[l]$ via consensus 
propagation when node~$l$ has a value $a[l]$. 
However, D-AMP was shown in numerical 
simulations~\cite{Hayakawa18} to have poor performance compared to that of 
centralized AMP~\cite{Donoho09}, unless approximately perfect consensus 
is achieved before denoising in each AMP iteration. This convergence property 
of D-AMP is different from that of conventional decentralized 
algorithms~\cite{Mateos10,Bazerque10,Mota12,Mota13,Shi14} that realize 
average consensus and signal estimation simultaneously. 

\subsection{Contributions}
This paper proposes decentralized GAMP (D-GAMP) for compressed sensing in 
tree-structured networks with no central nodes. Each node only utilizes local 
measurements to compute the GAMP iteration. Messages in each node are 
shared with adjacent nodes at every fixed time interval via consensus 
propagation~\cite{Moallemi06}. D-GAMP repeats the local GAMP 
iteration and consensus propagation until the algorithm converges. 

D-GAMP realizes average consensus and signal estimation simultaneously via 
consensus propagation between adjacent nodes, like conventional decentralized 
algorithms~\cite{Mateos10,Bazerque10,Mota12,Mota13,Shi14}. 
As a result, D-GAMP can reduce the total number of iterations for 
consensus propagation. 

D-GAMP allows different nodes to use a 
different number of inner GAMP iterations. This flexibility is useful to 
reduce latency when different nodes have different processing capability. 
Waiting for processing in the other nodes can be a cause of latency. To 
circumvent this waiting issue, D-GAMP shares messages between adjacent nodes 
at every fixed time interval, rather than after a common number of GAMP 
iterations in all nodes. 

The convergence property of Bayes-optimal D-GAMP is analyzed with a 
long-memory proof strategy~\cite{Takeuchi221,Takeuchi222,Liu221}, 
which is a general strategy for proving the convergence of state evolution 
recursion. Rigorous state evolution requires evaluation of the covariance 
matrix between estimation errors in all previous iterations. In the proof 
strategy, the covariance matrix is utilized to prove the convergence of 
state evolution recursion with respect to the mean-square errors (MSEs), 
i.e.\ its diagonal elements. When the Bayes-optimal denoisers are used 
in terms of MMSE, the covariance matrix has a special structure that 
implies the convergence for the sequence of its diagonal elements. 
See \cite{Takeuchi202,Takeuchi21,Fan22,Venkataramanan22,Skuratovs22,Liu222} 
for the original purpose of long-memory message-passing, 
i.e.\ complexity reduction.

The main contributions of this paper are as follows: 
A first contribution (Theorems~\ref{theorem_SE}, \ref{theorem_variance}, 
and \ref{theorem_SE_tech}) is rigorous state evolution of D-GAMP. 
This paper proposes and analyzes a general error model that contains 
the error model of D-GAMP and is applicable to general ad hoc networks 
without tree structure. While D-GAMP assumes tree-structured networks 
in consensus propagation, as considered in \cite{Hayakawa18}, 
the state evolution result can be utilized to design another sophisticated 
protocols for average consensus in general ad hoc networks. 

A second contribution (Theorems~\ref{theorem_fixed_point} and 
\ref{theorem_convergence}) is the convergence analysis of D-GAMP 
for tree-structured networks. This paper proves that state evolution 
recursion for D-GAMP has the same fixed point as that of the corresponding 
centralized GAMP~\cite{Rangan11} when all nodes have homogeneous measurements 
with an identical dimension. On the basis of the long-memory proof 
strategy~\cite{Takeuchi222}, the state evolution recursion 
for D-GAMP is proved to converge toward a fixed point when the Bayes-optimal 
inner and outer denoisers are used in terms of MMSE. These results imply that 
the state evolution for Bayes-optimal D-GAMP converges to the Bayes-optimal 
fixed point~\cite{Barbier19} for the homogeneous measurements 
when the Bayes-optimal fixed point is unique. 

The last contribution is numerical results for D-GAMP. As examples of 
tree-structured networks, a one-dimensional chain and a tree with no central 
nodes are considered. For the linear measurements, D-GAMP is numerically shown 
to reduce the total number of inner iterations for consensus propagation 
compared to conventional D-AMP~\cite{Hayakawa18}. 
Furthermore, D-GAMP is shown to converge toward 
the performance of the corresponding centralized GAMP~\cite{Rangan11} for 
finite-sized measurements with clipping when homogeneous measurements with 
an identical dimension in all nodes are considered. 

Part of these contributions were presented in \cite{Takeuchi24}. 

\subsection{Related Work}
From a technical point of view, state evolution in this paper is compared 
to existing state evolution. This paper generalizes state evolution 
for GAMP~\cite{Rangan11,Javanmard13} to that for D-GAMP. 
The proof strategy in this paper is essentially different from 
in \cite{Rangan11,Javanmard13}: 
Rangan~\cite{Rangan11} considered vector-valued AMP to analyze the asymptotic 
dynamics of GAMP. GAMP for rectangular sensing matrices was analyzed via 
state evolution of GAMP for symmetric sensing matrices in \cite{Javanmard13}. 
This paper establishes state evolution of D-GAMP for rectangular matrices 
directly by defining the general error model appropriately. In this sense, 
the definition of the general error model is a key contribution in the 
state evolution analysis. 

Another related work is graph-based AMP~\cite{Gerbelot23}, which is a 
message-passing algorithm for signal recovery from generalized linear 
measurements in nodes on any directed graph. State evolution for graph-based 
AMP is available to that of GAMP for multi-layer generalized linear 
measurements~\cite{Manoel17}, multi-layer generative priors~\cite{Aubin20}, 
graph-based priors~\cite{Ma23}, and for spatially coupled generalized 
linear measurements~\cite{Cobo23}.   

The general error model proposed in this paper allows us to analyze 
long-memory message-passing while graph-based AMP postulates conventional 
memoryless message-passing. Owing to this flexibility in the general error 
model, this paper can incorporate a consensus protocol with inner 
iterations---such as consensus propagation---in GAMP modules to 
guarantee the convergence of the state evolution recursion for Bayes-optimal 
D-GAMP. 

\subsection{Organization}
The remainder of this paper is organized as follows: After summarizing the 
notation used in this paper, Section~\ref{sec2} formulates signal 
reconstruction from generalized linear measurements distributed in an ad hoc 
network without central nodes. The network is modeled as a directed and 
connected graph in graph theory. In particular, this paper focuses on 
a tree-structured network, i.e.\ an undirected and connected graph without 
cycles.  

D-GAMP based on consensus propagation~\cite{Moallemi06} is proposed in 
Section~\ref{sec3}. It is regarded as a generalization of 
D-AMP~\cite{Hayakawa18} to the generalized linear measurements. The proposed 
D-GAMP is more flexible in terms of the iteration schedule than 
D-AMP~\cite{Hayakawa18}, as well as graph-based AMP~\cite{Gerbelot23}. 

Section~\ref{sec4} presents the main results of this paper while the proofs of 
theorems are summarized in Appendices. The asymptotic 
dynamics of D-GAMP is analyzed via state evolution. When a tree-structured 
network is assumed to justify use of consensus propagation, the long-memory 
proof strategy~\cite{Takeuchi222} is utilized to prove that state evolution 
recursion for Bayes-optimal D-GAMP converges toward a fixed point. 
In particular, the fixed point is equal to the Bayes-optimal fixed 
point~\cite{Barbier19} when the fixed point is unique.  

Section~\ref{sec5} presents numerical results for D-GAMP. A one-dimensional 
chain and a tree with $8$ nodes are considered as examples of tree-structured 
networks. Section~\ref{sec6} concludes this paper. 

\subsection{Notation}
Throughout this paper, the transpose and determinant of a matrix 
$\boldsymbol{M}$ are denoted by $\boldsymbol{M}^{\mathrm{T}}$ and 
$\det\boldsymbol{M}$, respectively. The notation $\boldsymbol{O}$ represents 
an all-zero matrix. The Kronecker delta is denoted by $\delta_{i,j}$. 
For $\{a_{i}\}_{i=1}^{n}$, the notation $\mathrm{diag}\{a_{1},\ldots, a_{n}\}$ 
represents the diagonal matrix having the $i$th diagonal element  $a_{i}$. 
The norm $\|\cdot\|$ means the Euclidean norm. The notation 
$\boldsymbol{o}(1)$ denotes a vector of which the Euclidean norm converges 
almost surely toward zero.  

For a vector $\boldsymbol{v}_{\mathcal{I}}$ with a set of indices $\mathcal{I}$, 
the $n$th element $[\boldsymbol{v}_{\mathcal{I}}]_{n}$ of 
$\boldsymbol{v}_{\mathcal{I}}$ is written as $v_{n,\mathcal{I}}$. Similarly, 
the $t$th column of a matrix $\boldsymbol{M}_{\mathcal{I}}$ is represented as
$\boldsymbol{m}_{\mathcal{I},t}$. 

The notation $\mathcal{N}(\boldsymbol{\mu},\boldsymbol{\Sigma})$ represents 
the Gaussian distribution with mean $\boldsymbol{\mu}$ and covariance 
$\boldsymbol{\Sigma}$. The almost sure convergence and equivalence are 
denoted by $\ato$ and $\aeq$, respectively. 

For a scalar function $f:\mathbb{R}\to\mathbb{R}$ and a vector 
$\boldsymbol{x}\in\mathbb{R}^{n}$, the notation $f(\boldsymbol{x})$ means 
the element-wise application of $f$ to $\boldsymbol{x}$, i.e.\ 
$[f(\boldsymbol{x})]_{i}=f(x_{i})$. The arithmetic mean of 
$\boldsymbol{x}\in\mathbb{R}^{n}$ is written as 
$\langle\boldsymbol{x}\rangle=n^{-1}\sum_{i=1}^{n}x_{i}$. For a multi-variate 
function $f:\mathbb{R}^{t}\to\mathbb{R}$, the notation $\partial_{i}$ represents 
the partial derivative of $f$ with respect to the $i$th variable. 

The notation $\boldsymbol{M}^{\dagger}=(\boldsymbol{M}^{\mathrm{T}}
\boldsymbol{M})^{-1}\boldsymbol{M}^{\mathrm{T}}$ represents the pseudo-inverse of 
a full-rank matrix $\boldsymbol{M}\in\mathbb{R}^{m\times n}$ satisfying 
$m\geq n$. The matrix $\boldsymbol{P}_{M}^{\parallel}=\boldsymbol{M}
(\boldsymbol{M}^{\mathrm{T}}\boldsymbol{M})^{-1}\boldsymbol{M}^{\mathrm{T}}$ is 
the projection onto the space spanned by the columns of $\boldsymbol{M}$ 
while $\boldsymbol{P}_{M}^{\perp}=\boldsymbol{I} - \boldsymbol{P}_{M}^{\parallel}$
is the projection onto the corresponding orthogonal complement. 

\section{Measurement Model} \label{sec2}
This paper considers the reconstruction of an unknown signal vector 
$\boldsymbol{x}\in\mathbb{R}^{N}$ from measurements in an ad hoc network 
with $L$ nodes. While D-GAMP postulates tree-structured networks in consensus 
propagation~\cite{Moallemi06}, state evolution analysis is performed for 
general ad hoc networks. Thus, this section presents the definition of 
a general ad hoc network. 

The ad hoc network is modeled as a directed and connected graph 
$\mathfrak{G}=(\mathcal{L}, \mathcal{E})$ 
with the set of nodes $\mathcal{L}=\{1,\ldots,L\}$, the set of 
edges $\mathcal{E}\subset\mathcal{L}\otimes\mathcal{L}$, and no self-loops.  
When the pair $(l_{1}, l_{2})\in\mathcal{E}$ exists, there is an edge connected 
from node~$l_{1}$ to node $l_{2}$. 
Since the graph has no self-loops,  $(l, l)\not\in\mathcal{E}$ holds 
for all $l\in\mathcal{L}$.  The incoming neighborhood 
$\mathcal{N}[l]=\{l'\in\mathcal{L}: (l', l)\in\mathcal{E}\}
\subset\mathcal{L}$ of node~$l$ represents the set of nodes that have 
incoming edges to node $l$ while the outgoing neighborhood 
$\tilde{\mathcal{N}}[l]=\{l'\in\mathcal{L}: (l,l')\in\mathcal{E}\}$ is 
the set of nodes that have outgoing edges from node~$l$. 
Since the graph has no self-loops, 
we have $l\notin\mathcal{N}[l]$ and $l\notin\tilde{\mathcal{N}}[l]$ 
for all $l\in\mathcal{L}$. A central node $l\in\mathcal{L}$ is defined as 
a node that is connected to all other nodes, i.e.\ 
$\mathcal{N}[l]=\tilde{\mathcal{N}}[l]=\mathcal{L}\backslash\{l\}$. 
Throughout this paper, the existence of central nodes is not assumed. 

Node~$l$ acquires an $M[l]$-dimensional measurement vector 
$\boldsymbol{y}[l]\in\mathbb{R}^{M[l]}$, 
modeled as the generalized linear measurements 
\begin{equation}
\label{measurement}
\boldsymbol{y}[l]
= g[l](\boldsymbol{z}[l], \boldsymbol{w}[l]), \quad
\boldsymbol{z}[l]=\boldsymbol{A}[l]\boldsymbol{x}.
\end{equation} 
In (\ref{measurement}), $\boldsymbol{w}[l]\in\mathbb{R}^{M[l]}$ and 
$\boldsymbol{A}[l]\in\mathbb{R}^{M[l]\times N}$ denote an unknown noise vector 
and a sensing matrix in node~$l$, respectively. The signal vector 
$\boldsymbol{x}$ is measured via the linear mapping $\boldsymbol{z}[l]$. 
The measurement vector $\boldsymbol{y}[l]$ is an element-wise application 
of a measurement function $g[l]:\mathbb{R}^{2}\to\mathbb{R}$ to the two vectors 
$\boldsymbol{z}[l]$ and $\boldsymbol{w}[l]$.  In particular, 
$g[l](z,w)=z+w$ corresponds to conventional linear measurements.  

The goal of this paper is to reconstruct the signal vector $\boldsymbol{x}$ 
under the following assumptions: 
\begin{itemize}
\item Node~$l$ only has the information about the local measurement vector 
$\boldsymbol{y}[l]$ and sensing matrix $\boldsymbol{A}[l]$, as well as 
the measurement model~(\ref{measurement}).

\item Node~$l$ can send messages to the outgoing neighborhood 
$\tilde{\mathcal{N}}[l]$ and receive them from the incoming neighborhood 
$\mathcal{N}[l]$. 

\item The communication link between nodes is error-free and latency-free.   
\end{itemize}

The first two assumptions are practical assumptions for decentralized 
algorithms in ad hoc networks. The last assumption may be reasonable for 
reliable wired networks or future wireless networks. 

\section{Distributed GAMP} \label{sec3}
\subsection{Overview}
The proposed D-GAMP algorithm consists of two parts: GAMP 
iteration~\cite{Rangan11} in each node and consensus 
propagation~\cite{Moallemi06} between nodes. GAMP in node~$l\in\mathcal{L}$ 
is composed of two modules, called outer and inner 
modules\footnote{
In GAMP~\cite{Rangan11}, the $M[l]$-dimensional measurement space was 
referred to an output space while the $N$-dimensional signal 
space was called an input space. In this paper, they are referred to as 
outer and inner spaces, respectively. The terminology ``inner module''  
does not mean that the inner module is physically located inside the outer 
module.}. The outer module 
utilizes the measurement vector $\boldsymbol{y}[l]$ to compute an 
estimator of $\boldsymbol{z}[l]$ in (\ref{measurement}) while the inner module 
uses prior information on the signal vector $\boldsymbol{x}$ to compute an 
estimator of $\boldsymbol{x}$. 

For consensus propagation all nodes share messages with their adjacent nodes. 
Each node utilizes the messages sent from its adjacent nodes to 
update the current message. Message transmission for consensus propagation 
is repeated $J$ times. While consensus propagation requires  
tree-structured networks, D-GAMP will be applicable to general ad hoc networks 
when consensus propagation is replaced with another sophisticated protocol 
for average consensus. However, research in this direction is beyond 
the scope of this paper. 

The outer module, consensus propagation, and the inner module are 
executed in this order. After that, node~$l$ repeats $T[l]-1$ iterations 
between the outer and inner modules in a fixed time interval, without 
performing consensus propagation. Different $T[l]$ is used for 
different $l$ since the nodes might not have identical processing capability. 
In other words, $T[l]$ corresponds to the number of iterations which 
node~$l$ can repeat in the fixed time interval. 
After the fixed time interval, the outer module, consensus propagation, 
and the inner module are executed again. 
Then, each node repeats $T[l]-1$ GAMP iterations to refine the estimation of 
the signal vector. Such iteration rounds are repeated 
until the consensus is achieved.

Let $T=\max_{l\in\mathcal{L}}T[l]$ denote the maximum number of GAMP iterations 
among all nodes. For notational convenience, we define a message $a_{t}[l]$ 
in iteration~$t$ after every $T[l]$ GAMP iterations for node~$l$ as that  
in the corresponding $T[l]$th iteration, i.e.\ $a_{t}[l]=a_{T[l]-1}[l]$ for 
$t\in\{T[l],\ldots,T-1\}$. This notation allows us to use  
the common number of GAMP iterations $T$ in all nodes. When the total 
number of GAMP iterations~$t$ is counted, consensus propagation is 
performed in iteration $t=iT$ for all non-negative integers $i$.  

To represent messages in the iteration where consensus propagation is 
performed, the following underline notation is used throughout this paper:  
\begin{definition}
For integers $t, T\in\mathbb{N}$, and variables $\{a_{\tau}\in\mathbb{R}\}$, 
the notation $\underline{a}_{t}$ is defined as 
$\underline{a}_{t}=a_{iT}$ with $i=\lfloor t/T \rfloor$. 
For vectors $\{\boldsymbol{v}_{\tau}\}$ the notation 
$\underline{\boldsymbol{v}}_{t}=\boldsymbol{v}_{iT}$ is defined in the same 
manner while the notation 
$\underline{\boldsymbol{M}}_{t}=[\underline{\boldsymbol{m}}_{0},\ldots,
\underline{\boldsymbol{m}}_{t-1}]$ is used 
for a matrix $\boldsymbol{M}_{t}=[\boldsymbol{m}_{0},\ldots,
\boldsymbol{m}_{t-1}]$. 
\end{definition}

\subsection{Outer Module}
For iteration~$t\in\{0,1,\ldots\}$, 
suppose that the outer module has an estimator 
$\tilde{\boldsymbol{z}}_{t}[l]\in\mathbb{R}^{M[l]}$ of $\boldsymbol{z}[l]$, 
an estimator $v_{t}[l]>0$ of $M^{-1}[l]\|\tilde{\boldsymbol{z}}_{t}[l] 
- \boldsymbol{z}[l]\|^{2}$, and 
an estimator $\hat{\boldsymbol{x}}_{t}[l]\in\mathbb{R}^{N}$ of 
$\boldsymbol{x}$ sent from the inner module, as well as the measurement vector $\boldsymbol{y}[l]$. As initial conditions, 
$\tilde{\boldsymbol{z}}_{0}[l]=\boldsymbol{0}$, $v_{0}[l]=(LM[l])^{-1}
\mathbb{E}[\|\boldsymbol{x}\|^{2}]$, and 
$\hat{\boldsymbol{x}}_{0}[l]=\boldsymbol{0}$ are used 
for all $l\in\mathcal{L}$. 

The outer module computes an estimator 
$\hat{\boldsymbol{z}}_{t}[l]\in\mathbb{R}^{M[l]}$ of $\boldsymbol{z}[l]$ 
and a message $\boldsymbol{x}_{t}[l]\in\mathbb{R}^{N}$ as follows: 
\begin{equation} \label{z_hat}
\hat{\boldsymbol{z}}_{t}[l] 
= f_{\Out}[l](\tilde{\boldsymbol{z}}_{t}[l], \boldsymbol{y}[l]; v_{t}[l]),
\end{equation}
\begin{equation} \label{x}
\boldsymbol{x}_{t}[l]
= \frac{1}{L}\hat{\boldsymbol{x}}_{t}[l]
- \frac{1}{\xi_{\Out,t}[l]}
\boldsymbol{A}^{\mathrm{T}}[l]\hat{\boldsymbol{z}}_{t}[l], 
\end{equation}
with 
\begin{equation} \label{xi_out}
\xi_{\Out,t}[l] = \langle \partial_{1}f_{\Out}[l]
(\tilde{\boldsymbol{z}}_{t}[l], \boldsymbol{y}[l]; v_{t}[l]) \rangle. 
\end{equation}
Here, the scalar function $f_{\Out}[l](\cdot, \cdot; v_{t}[l])
:\mathbb{R}^{2}\to\mathbb{R}$ is an outer denoiser. 
The notation $\partial_{1}$ denotes the partial derivative of $f_{\Out}[l]$ 
with respect to the first variable. 
The parameter $\xi_{\Out,t}[l]\in\mathbb{R}$ has been designed so as to 
realize asymptotic Gaussianity of estimation errors before inner denoising. 
The outer module sends the messages $\boldsymbol{x}_{t}[l]$,  
$\hat{\boldsymbol{z}}_{t}[l]$, and $\xi_{\Out,t}[l]$ to the inner module. 

The update rules in the outer module are similar to those in centralized 
GAMP~\cite{Rangan11}. Intuitively, the Onsager correction in (\ref{x}) 
eliminates intractable memory terms in each iteration. 
Since clean messages after the Onsager correction are shared with adjacent 
nodes, consensus propagation does not affect the update rules in the outer 
module. The correctness of this intuition is proved via state evolution. 

When the remainder of $t$ divided by 
$T$ is larger than or equal to the actual number of iterations $T[l]$, 
all messages are fixed to $\hat{\boldsymbol{z}}_{t}[l] 
=\hat{\boldsymbol{z}}_{iT+T[l]-1}[l]$, $\boldsymbol{x}_{t}[l]
=\boldsymbol{x}_{iT+T[l]-1}[l]$, and $\xi_{\Out,t}[l]=\xi_{\Out,iT+T[l]-1}[l]$ 
for $i=\lfloor t/T \rfloor$. Thus, 
we define the set of iteration indices 
$\mathcal{T}_{t}[l]$ as $\mathcal{T}_{0}[l]=\emptyset$ and  
\begin{equation} \label{T_set}
\mathcal{T}_{t}[l]
=\cup_{i=0}^{\lfloor t/T \rfloor}\{iT,\ldots,\min\{t,iT + T[l]\}-1\}  
\end{equation} 
for $t>0$, by eliminating from $\{0,\ldots,t-1\}$ the indices for which 
the messages are fixed. For iteration~$t$, the set $\mathcal{T}_{t}[l]$ 
contains all iterations in which the messages in node~$l$ are updated. 

\subsection{Consensus Propagation} 
The centralized GAMP~\cite{Rangan11} uses the messages 
$\tilde{\boldsymbol{x}}_{t} = \sum_{l\in\mathcal{L}}\boldsymbol{x}_{t}[l]$, 
$\eta_{t}=L$, and $\sigma_{t}^{2}=L^{-1}\sum_{l\in\mathcal{L}}\xi_{\Out,t}^{-1}[l]$ 
in the inner module. However, computation of these messages requires 
a central node that receives the messages 
$\{\boldsymbol{x}_{t}[l]: l\in\mathcal{L}\}$ and 
$\{\xi_{\Out,t}[l]: l\in\mathcal{L}\}$ from all nodes. For a tree-structured 
network with no central nodes, consensus propagation~\cite{Moallemi06} can 
be used to compute the messages $\tilde{\boldsymbol{x}}_{t}$, 
$\eta_{t}$, and $\sigma_{t}^{2}$ in a decentralized manner.   

Consensus propagation with $J$ inner iterations is 
performed in every $T$ iterations. 
More precisely, node~$l$ shares messages with adjacent nodes 
for consensus propagation if $t$ is divisible by $T$. 

Focus on inner iteration~$j\in\{1,\ldots,J\}$ and suppose that node~$l$ 
has messages $\{\underline{\boldsymbol{x}}_{t,j-1}[l'\rightarrow l]
\in\mathbb{R}^{N}: l'\in\mathcal{N}[l]\}$ and 
$\{\underline{\sigma}_{t,j-1}^{2}[l'\rightarrow l]\in\mathbb{R}: 
l'\in\mathcal{N}[l]\}$ 
sent from the adjacent nodes in the preceding inner iteration of consensus 
propagation and messages $\{\underline{\eta}_{t,j-1}[l'\rightarrow l]
\in\mathbb{R}: l'\in\mathcal{N}[l]\}$ computed in node~$l$, as well as 
the messages $\underline{\boldsymbol{x}}_{t}[l]$ 
and $\underline{\xi}_{\Out,t}[l]$ computed in the outer module. 
Node~$l$ computes the following messages 
$\underline{\boldsymbol{x}}_{t,j}[l\rightarrow l']$ and 
$\underline{\sigma}_{t,j}^{2}[l\rightarrow l']$,  
which are sent to node~$l'\in\tilde{\mathcal{N}}[l]$, as well as  
$\underline{\eta}_{t,j}[l\rightarrow l']$. 
\begin{equation} \label{x_ll'}
\underline{\boldsymbol{x}}_{t,j}[l\rightarrow l'] 
= \underline{\boldsymbol{x}}_{t}[l] 
+ \sum_{\tilde{l}'\in\mathcal{N}[l]\backslash\{l'\}}
\underline{\boldsymbol{x}}_{t,j-1}[\tilde{l}'\rightarrow l], 
\end{equation}
\begin{equation}
\underline{\eta}_{t,j}[l\rightarrow l']
= 1 + \sum_{\tilde{l}'\in\mathcal{N}[l]\backslash\{l'\}}
\underline{\eta}_{t,j-1}[\tilde{l}'\rightarrow l], 
\end{equation}
\begin{equation}
\underline{\sigma}_{t,j}^{2}[l\rightarrow l'] 
= \frac{1}{\underline{\xi}_{\Out,t}[l]}
+ \sum_{\tilde{l}'\in\mathcal{N}[l]\backslash\{l'\}}
\underline{\sigma}_{t,j-1}^{2}[\tilde{l}'\rightarrow l], 
\end{equation}
with $\underline{\boldsymbol{x}}_{t,0}[l'\rightarrow l]
=\underline{\boldsymbol{x}}_{t-T,J}[l'\rightarrow l]$, 
$\underline{\eta}_{t,0}[l'\rightarrow l]
=\underline{\eta}_{t-T,J}[l'\rightarrow l]$, and 
$\underline{\sigma}_{t,0}^{2}[l'\rightarrow l]
=\underline{\sigma}_{t-T,J}^{2}[l'\rightarrow l]$, which are the messages 
sent from the adjacent nodes in the preceding round of consensus propagation. 
As initial conditions, $\underline{\boldsymbol{x}}_{t-T,J}[l'\to l]
=\boldsymbol{0}$, $\underline{\eta}_{t-T,J}[l'\rightarrow l]=0$, 
and $\underline{\sigma}_{t-T,J}^{2}[l'\rightarrow l]=0$ 
are used for all $t<T$. 

After $J$ inner iterations for consensus propagation, node~$l$ 
receives the messages $\{\underline{\boldsymbol{x}}_{t,J}[l'\rightarrow l]
\in\mathbb{R}^{N}: l'\in\mathcal{N}[l]\}$ and 
$\{\underline{\sigma}_{t,J}^{2}[l'\rightarrow l]>0: 
l'\in\mathcal{N}[l]\}$ from the adjacent nodes. Then, 
the following messages $\overline{\underline{\boldsymbol{x}}_{t}[l]}$, 
$\overline{\underline{\eta}_{t}[l]}$, and 
$\overline{\underline{\sigma}_{t}^{2}[l]}$ are computed 
and sent to the inner module:
\begin{equation} 
\overline{\underline{\boldsymbol{x}}_{t}[l]} 
= \sum_{l'\in\mathcal{N}[l]}\underline{\boldsymbol{x}}_{t,J}[l'\rightarrow l],
\label{consensus}
\end{equation}
\begin{equation} 
\overline{\underline{\eta}_{t}[l]} = \sum_{l'\in\mathcal{N}[l]}
\underline{\eta}_{t,J}[l'\rightarrow l], 
\end{equation}
\begin{equation}
\overline{\underline{\sigma}_{t}^{2}[l]} 
= \sum_{l'\in\mathcal{N}[l]}\underline{\sigma}_{t,J}^{2}[l'\rightarrow l].
\end{equation}
The message $\overline{\underline{\boldsymbol{x}}_{t}[l]}$ is used in the 
inner module as an extrinsic estimate of $\boldsymbol{x}$ while 
$\overline{\underline{\sigma}_{t}^{2}[l]}$ is an estimator of the 
corresponding extrinsic MSE. After the aggregation of all messages, 
the messages $\overline{\underline{\boldsymbol{x}}_{t}[l]}$, 
$\overline{\underline{\eta}_{t}[l]}$, and 
$\overline{\underline{\sigma}_{t}^{2}[l]}$ converge to 
$\sum_{l'\neq l}\underline{\boldsymbol{x}}_{t}[l']$, $L-1$, and  
$\sum_{l'\neq l}\underline{\xi}_{\Out,t}^{-1}[l']$ 
for tree-structured networks, respectively. 
 
Consensus propagation is a belief-propagation-based 
algorithm~\cite{Moallemi06}, so that consensus propagation converges for all 
tree-structured networks. To understand consensus propagation intuitively, 
consider a one-dimensional chain network with $L$ nodes, in which the nodes 
are aligned in ascending order. Let $x[l]$ denote a value associated with 
node~$l$. An intuitive method to share the summation $S=\sum_{l=1}^{L}x[l]$ is 
as follows: Compute the forward message $m_{\mathrm{f}}[l]=\sum_{l'=1}^{l}x[l']$ 
and backward message $m_{\mathrm{b}}[l]=\sum_{l'=l}^{L}x[l']$ in node~$l$ as 
$m_{\mathrm{f}}[l] = x[l] + m_{\mathrm{f}}[l-1]$ and 
$m_{\mathrm{b}}[l] = x[l] + m_{\mathrm{b}}[l+1]$, respectively. The summation 
is computed as $S=m_{\mathrm{f}}[l-1] + x[l] + m_{\mathrm{b}}[l+1]$ 
in node~$l$. Consensus propagation is obtained by generalizing this simple 
consensus algorithm to the tree-structured case. 

\subsection{Inner Module}
In iteration~$t$, suppose that the inner module has the messages  
$\boldsymbol{x}_{t}[l]$, $\hat{\boldsymbol{z}}_{t}[l]$, and 
$\xi_{\Out,t}[l]$ sent from the outer module and the messages 
$\overline{\underline{\boldsymbol{x}}_{t}[l]}$, 
$\overline{\underline{\eta}_{t}[l]}$, and 
$\overline{\underline{\sigma}_{t}^{2}[l]}$ computed in consensus 
propagation. The inner module first computes the following messages: 
\begin{equation} 
\tilde{\boldsymbol{x}}_{t}[l] 
= \boldsymbol{x}_{t}[l] 
+ \overline{\underline{\boldsymbol{x}}_{t}[l]},
\label{x_tilde}
\end{equation}
\begin{equation} \label{eta}
\eta_{t}[l] = 1 + \overline{\underline{\eta}_{t}[l]}, 
\end{equation}
\begin{equation} \label{sigma_t}
\sigma_{t}^{2}[l] = \frac{1}{L}\left(
 \frac{1}{\xi_{\Out,t}[l]} + \overline{\underline{\sigma}_{t}^{2}[l]}
\right), 
\end{equation}
which should converge to $\tilde{\boldsymbol{x}}_{t}
=\sum_{l'\in\mathcal{L}}\boldsymbol{x}_{t}[l']$, $L$, and  
$L^{-1}\sum_{l'\in\mathcal{L}}\xi_{\Out,t}^{-1}[l']$ for tree-structured networks 
after the aggregation of all messages, respectively, 
when $t$ is divisible by $T$. Then, an estimator 
$\hat{\boldsymbol{x}}_{t+1}[l]\in\mathbb{R}^{N}$ of the signal vector 
$\boldsymbol{x}$ is computed as 
\begin{equation} \label{x_hat}
\hat{\boldsymbol{x}}_{t+1}[l] 
= f_{\In}[l](\tilde{\boldsymbol{x}}_{t}[l]; \eta_{t}[l], \sigma_{t}^{2}[l]),  
\end{equation}
where $f_{\In}[l](\cdot;\eta_{t}[l],\sigma_{t}^{2}[l]):\mathbb{R}\to\mathbb{R}$ 
denotes an inner denoiser. The parameter $\eta_{t}[l]$ denotes 
the number of messages aggregated in consensus propagation. 
The parameter $\sigma_{t}^{2}[l]$ corresponds to an estimator of the 
MSE for the message $\tilde{\boldsymbol{x}}_{t}[l]$. See Section~\ref{sec4}  
for its precise meaning revealed via state evolution. 

The estimator $\hat{\boldsymbol{x}}_{t+1}[l]$ depends on the node 
index~$l$ while the original signal vector $\boldsymbol{x}$ is independent 
of $l$. When there are no central nodes for aggregating 
$\{\tilde{\boldsymbol{x}}_{t}[l]: l\in\mathcal{L}\}$, the estimator 
$\hat{\boldsymbol{x}}_{t+1}[l]$ can be used when an estimator of 
$\boldsymbol{x}$ is needed in node~$l$. 

To refine the estimator of $\boldsymbol{x}$, the 
inner module computes the following messages:
\begin{equation} \label{z_tilde}
\tilde{\boldsymbol{z}}_{t+1}[l] 
= \boldsymbol{A}[l]\hat{\boldsymbol{x}}_{t+1}[l]
+ \frac{N\xi_{\In,t}[l]}{LM[l]\xi_{\Out,t}[l]}\hat{\boldsymbol{z}}_{t}[l], 
\end{equation}
\begin{equation} \label{v_t}
v_{t+1}[l] = \frac{N}{M[l]}\frac{\sigma_{t}^{2}[l]\xi_{\In,t}[l]}{\eta_{t}[l]}, 
\end{equation}
with 
\begin{equation} \label{xi_in}
\xi_{\In,t}[l] = \langle \partial_{1}f_{\In}[l](\tilde{\boldsymbol{x}}_{t}[l];
\eta_{t}[l], \sigma_{t}^{2}[l]) \rangle. 
\end{equation}
The message $\tilde{\boldsymbol{z}}_{t+1}[l]$ is an estimator of 
$\boldsymbol{z}[l]$. The message $\xi_{\In,t}[l]\in\mathbb{R}$ has been 
designed so as to realize asymptotic Gaussianity of estimation errors 
before outer denoising. The message $v_{t+1}[l]$ corresponds to an estimator 
of the MSE for $\tilde{\boldsymbol{z}}_{t+1}[l]$. 
See Section~\ref{sec4} for its precise meaning revealed via 
state evolution.

For $t-\lfloor t/T\rfloor T\geq T[l]$, all messages are fixed to 
$\hat{\boldsymbol{x}}_{t+1}[l]=\hat{\boldsymbol{x}}_{iT+T[l]}[l]$, 
$\tilde{\boldsymbol{z}}_{t+1}[l] =\tilde{\boldsymbol{z}}_{iT+T[l]}[l]$, 
and $v_{t+1}[l] = v_{iT+T[l]}[l]$, as fixed in the outer module.  
The inner module feeds the messages $\hat{\boldsymbol{x}}_{t+1}[l]$, 
$\tilde{\boldsymbol{z}}_{t+1}[l]$, and $v_{t+1}[l]$ back to 
the outer module to refine the estimator of $\boldsymbol{x}$. 

To understand D-GAMP, assume that the summation consensus 
$\overline{\underline{\boldsymbol{x}}_{t}[l]}= \sum_{l'\neq l}
\underline{\boldsymbol{x}}_{t}[l']$, $\hat{\underline{\boldsymbol{x}}}_{t}[l']=
\hat{\underline{\boldsymbol{x}}}_{t}[l]$, and $\underline{\xi}_{\Out,t}[l']
=\underline{\xi}_{\Out,t}[l]$ have been achieved for all $l, l'\in\mathcal{L}$ 
when $t$ is sufficiently large. Then, (\ref{x_tilde}) reduces to 
$\tilde{\boldsymbol{x}}_{t}[l] 
= \boldsymbol{x}_{t}[l] + \sum_{l'\neq l}\underline{\boldsymbol{x}}_{t}[l']$. 
Using the definition~(\ref{x}) of $\boldsymbol{x}_{t}[l]$ yields 
\begin{equation}  \label{x_tilde_consensus}
\tilde{\boldsymbol{x}}_{iT}[l] 
= \hat{\boldsymbol{x}}_{iT}[l] 
- \frac{1}{\xi_{\Out,iT}[l]}\sum_{l'\in\mathcal{L}}\boldsymbol{A}^{\mathrm{T}}[l']
\hat{\boldsymbol{z}}_{iT}[l']
\end{equation}
for $t=iT\in\mathbb{N}$. The update rules~(\ref{z_tilde}) and 
(\ref{x_tilde_consensus}) for $\tilde{\boldsymbol{z}}_{t}[l]$ and 
$\tilde{\boldsymbol{x}}_{iT}[l]$ are equivalent to those 
in centralized GAMP~\cite{Rangan11}. 
 
Conventional D-AMP~\cite{Hayakawa18} was designed under the implicit 
assumption of perfect consensus in each iteration. 
As a result, multiple inner iterations $J$ for consensus propagation were 
considered to realize the perfect summation consensus 
$\overline{\underline{\boldsymbol{x}}_{t}[l]}=\sum_{l'\in\mathcal{L}}
\underline{\boldsymbol{x}}_{t}[l']$ for all $l\in\mathcal{L}$ 
approximately. However, such a protocol requires heavy communications 
between adjacent nodes. 

D-GAMP with $T=1$ is equivalent to D-AMP~\cite{Hayakawa18} 
when the linear measurement model $g[l](z,w)=z+w$ is considered. 
Interestingly, state evolution in this paper reveals 
the correctness of the Onsager correction in D-GAMP. As a result, 
D-AMP~\cite{Hayakawa18} also has the correct Onsager correction while 
perfect consensus was implicitly assumed in designing D-AMP. 
Nonetheless, numerical simulations in \cite{Hayakawa18} showed poor 
performance of D-AMP with a few inner iterations of consensus propagation. 
This conflict between theoretical and numerical results may 
be because small $M[l]=6$ was simulated in \cite{Hayakawa18}. If much 
larger systems were simulated, good performance might be observed. 

\section{Main Results} \label{sec4}
\subsection{Definitions and Assumptions}
The dynamics of D-GAMP is analyzed via state evolution in the large system 
limit for fixed $L$, where the dimensions $\{M[l]\}$ and $N$ tend to infinity 
while the ratio $\delta[l]=M[l]/N$ is kept constant for all $l\in\mathcal{L}$. 
To present a rigorous result, we first define an empirical convergence in 
terms of separable, pseudo-Lipschitz, and proper 
functions~\cite{Bayati11,Takeuchi201}. 

\begin{definition}[Separability]
A vector-valued function $\boldsymbol{f}=(f_{1},\ldots,f_{N})^{\mathrm{T}}$ 
with $f_{n}:\mathbb{R}^{N}\to\mathbb{R}$ is 
said to be separable if the $n$th function $f_{n}(\boldsymbol{x})$ depends 
only on the $n$th element of $\boldsymbol{x}=(x_{1},\ldots,x_{N})^{\mathrm{T}}
\in\mathbb{R}^{N}$ for all $n$, i.e.\ $f_{n}(\boldsymbol{x})=f_{n}(x_{n})$. 
\end{definition}

\begin{definition}[Pseudo-Lipschitz]
A function $f:\mathbb{R}^{t}\to\mathbb{R}$ is said to be pseudo-Lipschitz of 
order~$k$ if there is some Lipschitz constant $C>0$ such that the following 
inequality holds: 
\begin{equation}
|f(\boldsymbol{x}) - f(\boldsymbol{y})|
\leq C\|\boldsymbol{x} - \boldsymbol{y}\|(1 + \|\boldsymbol{x}\|^{k-1} 
+ \|\boldsymbol{y}\|^{k-1})
\end{equation}
for all $\boldsymbol{x}, \boldsymbol{y}\in\mathbb{R}^{t}$. 
\end{definition}

By definition, any pseudo-Lipschitz function of order~$k=1$ is 
Lipschitz-continuous. A separable vector-valued function $\boldsymbol{f}$ is 
said to be pseudo-Lipschitz of order~$k$ if all element functions of 
$\boldsymbol{f}$ are pseudo-Lipschitz of order~$k$. In this paper, 
piecewise pseudo-Lipschitz functions are considered to include practical 
denoisers in the proposed framework of state evolution. 

\begin{definition}[Proper] \label{proper}
A separable, pseudo-Lipschitz, and vector-valued function 
$\boldsymbol{f}=(f_{1},\ldots,f_{N})^{\mathrm{T}}$ is said to be proper if 
the Lipschitz constant $C_{n}>0$ of the $n$th function 
$f_{n}:\mathbb{R}\to\mathbb{R}$ satisfies 
\begin{equation}
\lim_{N\to\infty}\frac{1}{N}\sum_{n=1}^{N}C_{n}^{k}<\infty \quad 
\hbox{for all $k\in\mathbb{N}$.} 
\end{equation}
\end{definition}

Definition~\ref{proper} is used to analyze separable, pseudo-Lipschitz, and 
vector-valued functions $\boldsymbol{f}$ in the same manner as in the 
common function case~\cite{Takeuchi201}: 
\ $f_{n}=f$ for a pseudo-Lipschitz function $f$. 

\begin{definition}
Random vectors $(\boldsymbol{v}_{1},\ldots,\boldsymbol{v}_{t})
\in\mathbb{R}^{N\times t}$ are said to converge jointly toward 
random variables $(V_{1},\ldots,V_{t})$ in the sense of $k$th-order 
pseudo-Lipschitz if the limit 
$\lim_{N\to\infty}N^{-1}\sum_{n=1}^{N}\mathbb{E}[f_{n}(V_{1},\ldots,V_{t})]$ exists 
and the following almost sure convergence holds: 
\begin{equation}
\lim_{N\to\infty}\frac{1}{N}\sum_{n=1}^{N}\left\{
 f_{n}(v_{n,1},\ldots,v_{n,t}) - \mathbb{E}[f_{n}(V_{1},\ldots,V_{t})]
\right\} \ato 0
\end{equation}
for all separable and piecewise proper pseudo-Lipschitz functions 
$\boldsymbol{f}=[f_{1},\ldots,f_{N}]^{\mathrm{T}}$ of order~$k$. This convergence 
in the sense of $k$th-order pseudo-Lipschitz is called the 
$\mathrm{PL}(k)$ convergence and denoted by 
$(\boldsymbol{v}_{1},\ldots,\boldsymbol{v}_{t})\plkto (V_{1},\ldots,V_{t})$. 
\end{definition}

The goal of state evolution is to prove asymptotic Gaussianity 
for the messages  $\tilde{\boldsymbol{z}}_{t}[l]$ and 
$\tilde{\boldsymbol{x}}_{t}[l]$ just before outer and inner denoising, 
respectively: The $\mathrm{PL}(2)$ convergence results  
$\tilde{\boldsymbol{z}}_{t}[l]\plto Z_{t}[l]$ and 
$\tilde{\boldsymbol{x}}_{t}[l]-L^{-1}\bar{\eta}_{t}[l]
\boldsymbol{x}\plto H_{t}[l]$ hold in the 
large system limit for some Gaussian random variables $Z_{t}[l]$, $H_{t}[l]$, 
and deterministic variable $\bar{\eta}_{t}[l]$ defined shortly. 
We summarize assumptions to justify the $\mathrm{PL}(2)$ convergence. 

\begin{assumption} \label{assumption_x}
For some $\epsilon>0$, the $\mathrm{PL}(2+\epsilon)$ convergence holds 
for the signal vector $\boldsymbol{x}$, 
i.e.\ $\boldsymbol{x}\pleto X$ for some random variable $X$.  
\end{assumption}

When $\boldsymbol{x}$ has i.i.d.\ elements with a bounded $(2+\epsilon)$th 
moment, the $\mathrm{PL}(2+\epsilon)$ convergence holds for $X$ 
that follows the distribution for the elements of $\boldsymbol{x}$. 

\begin{assumption} \label{assumption_w}
For some $\epsilon>0$, the $\mathrm{PL}(2+\epsilon)$ convergence holds 
for the noise vectors $\{\boldsymbol{w}[l]: l\in\mathcal{L}\}$ 
i.e.\ $\{\boldsymbol{w}[l]\}\pleto \{W[l]\}$ for some absolutely 
continuous random variables $\{W[l]: l\in\mathcal{L}\}$.  
\end{assumption}

Independent Gaussian noise 
$\boldsymbol{w}[l]\sim\mathcal{N}(\boldsymbol{0},
\sigma^{2}[l]\boldsymbol{I}_{M[l]})$ with some variance $\sigma^{2}[l]$ 
satisfies Assumption~\ref{assumption_w} for independent Gaussian 
random variables $W[l]\sim\mathcal{N}(0,\sigma^{2}[l])$.  

\begin{assumption} \label{assumption_A}
The sensing matrices $\{\boldsymbol{A}[l]: l\in\mathcal{L}\}$ are 
independent. Each matrix $\boldsymbol{A}[l]\in\mathbb{R}^{M[l]\times N}$ has 
independent zero-mean Gaussian elements with variance $(LM[l])^{-1}$. 
\end{assumption}

Assumption~\ref{assumption_A} is an important assumption to analyze the 
dynamics of D-GAMP via state evolution. The independent assumption cannot be 
relaxed for AMP~\cite{Takeuchi19}. More precisely, the empirical eigenvalue 
distribution of $\boldsymbol{A}^{\mathrm{T}}[l]\boldsymbol{A}[l]$ needs to 
converge in probability to that for zero-mean i.i.d.\ Gaussian sensing 
matrices.  

\begin{assumption} \label{assumption_Lipschitz}
The composition $f_{\Out}[l](\theta, g[l](z, w); v_{t}[l])$ of the measurement 
function $g[l]$ in (\ref{measurement}) and outer denoiser $f_{\Out}[l]$ in 
(\ref{z_hat}) is piecewise Lipschitz-continuous with respect to 
$(\theta, z, w)\in\mathbb{R}^{3}$. 
The inner denoiser $f_{\In}[l](u; \eta_{t}[l], \sigma_{t}^{2}[l])$ 
in (\ref{x_hat}) is piecewise Lipschitz-continuous with respect to 
$u\in\mathbb{R}$. 
\end{assumption}

The everywhere Lipschitz-continuity was assumed in conventional state evolution 
analysis~\cite{Bayati11,Takeuchi201}. Nonetheless, this paper postulates 
the piecewise Lipschitz-continuity to include practical outer denoisers 
in the proposed framework of state evolution. This slight generalization 
does not cause any gaps in state evolution analysis since any 
piecewise Lipschitz-continuous function has all properties required in 
state evolution, such as almost everywhere differentiability and the 
boundedness of derivatives---satisfied for all Lipschitz-continuous 
functions. Intuitively, there are no technically significant differences  
between the singularities at the origin of the two functions e.g.\ 
$f_{1}(x)=|x|$ and $f_{2}(x)=-x$ for all $x<0$ and $f_{2}(x)=x+1$ for 
all $x\geq 0$ unless $x=0$ occurs with a finite probability.  

\begin{assumption} \label{assumption_tree}
The graph $\mathfrak{G}=(\mathcal{L},\mathcal{E})$ is a tree, i.e.\ 
an undirected and connected graph with no cycles. 
\end{assumption}

Assumption~\ref{assumption_tree} is used in justifying consensus propagation 
while it is not required in proving the asymptotic Gaussianity. Note that 
the incoming neighborhood $\mathcal{N}[l]$ is equal to the outgoing 
neighborhood $\tilde{\mathcal{N}}[l]$ under Assumption~\ref{assumption_tree}. 

\subsection{State Evolution}
State evolution recursion for D-GAMP is given via four 
kinds of scalar zero-mean Gaussian random variables $\{Z[l]\}$, 
$\{Z_{t}[l]\}$, $\{H_{t}[l]\}$, and $\{\tilde{H}_{t}[l]\}$, associated 
with $\boldsymbol{z}[l]$, $\tilde{\boldsymbol{z}}_{t}[l]$, 
$\boldsymbol{x}_{t}[l]$, and $\tilde{\boldsymbol{x}}_{t}[l]$ 
in (\ref{measurement}), (\ref{z_tilde}), 
(\ref{x}), and (\ref{x_tilde}), respectively. The random variables  
$\{Z[l]:l\in\mathcal{L}\}$ are independent of $\{W[l]\}$ in 
Assumption~\ref{assumption_w} and independent zero-mean Gaussian random 
variables with variance 
\begin{equation} \label{Z}
\mathbb{E}[Z^{2}[l]] = \frac{1}{L\delta[l]}\mathbb{E}[X^{2}],
\end{equation} 
with $X$ defined in Assumption~\ref{assumption_x}. 
To define statistical properties of the other random variables, 
we first define two variables 
$\bar{\xi}_{\Out,t}[l]$ and $\bar{\zeta}_{t}[l]$ in the outer module as 
\begin{equation} \label{xi_out_bar}
\bar{\xi}_{\Out,t}[l] 
= \mathbb{E}\left[
 \partial_{1}f_{\Out}[l](Z_{t}[l], Y[l]; \bar{v}_{t}[l])
\right],
\end{equation}
\begin{equation} \label{zeta_bar}
\bar{\zeta}_{t}[l] = - \mathbb{E}\left[
 \left.
  \frac{\partial}{\partial z}f_{\Out}[l](Z_{t}[l],g[l](z,W[l]);\bar{v}_{t}[l])
 \right|_{z=Z[l]}
\right],
\end{equation}
with $Y[l] = g[l](Z[l], W[l])$, in which $Z_{t}[l]$ and $\bar{v}_{t}[l]$ are 
defined shortly. The variable $\bar{\xi}_{\Out,t}[l]$ is 
the asymptotic alternative of $\xi_{\Out,t}[l]$ in (\ref{xi_out}) while 
$\bar{\zeta}_{t}[l]$ is used in the inner module. 

We define random variables $\{H_{t}[l]\}$ and $\{\tilde{H}_{t}[l]\}$. 
The random variables $\{H_{t}[l]\}$ are independent of $X$ in 
Assumption~\ref{assumption_x} and zero-mean Gaussian random variables 
with covariance 
\begin{align} \label{H_covariance}
\mathbb{E}[H_{\tau}[l']H_{t}[l]] 
= \frac{\delta_{l,l'}}{L}\mathbb{E}\left[
 f_{\Out}[l](Z_{\tau}[l], Y[l]; \bar{v}_{\tau}[l])
\right.
\nonumber \\
\left.
 f_{\Out}[l](Z_{t}[l], Y[l]; \bar{v}_{t}[l])
\right]
\end{align}
for all $\tau\in\{0,\ldots,t\}$. 
Furthermore, $\tilde{H}_{t}[l]$ is defined 
recursively as follows:
\begin{equation} \label{H_tilde_tree}
\tilde{H}_{t}[l] = \frac{H_{t}[l]}{\bar{\xi}_{\Out,t}[l]}
+ \sum_{l'\in\mathcal{N}[l]}\underline{H}_{t,J}[l'\rightarrow l],
\end{equation}
with 
\begin{equation} \label{H_ll'_tree}
\underline{H}_{t,j}[l\rightarrow l'] 
= \frac{\underline{H}_{t}[l]}{\underline{\bar{\xi}}_{\Out,t}[l]}
+ \sum_{\tilde{l}'\in\mathcal{N}[l]\backslash\{l'\}}
\underline{H}_{t,j-1}[\tilde{l}'\rightarrow l]
\end{equation}
for $j\in\{1,\ldots,J\}$. 
As an initial condition, $\underline{H}_{t,0}[l'\rightarrow l]
=\underline{H}_{t-T,J}[l'\rightarrow l]$ is used, as well as  
$\underline{H}_{t-T,J}[l'\rightarrow l]=0$ for all $t<T$. 
The random variables $\{\tilde{H}_{t}[l]\}$ are independent of $X$ in 
Assumption~\ref{assumption_x} and zero-mean 
Gaussian random variables since they are a linear combination 
of $\{H_{t}[l]\}$. 

We next define four variables $\bar{\eta}_{t}[l]$,
$\bar{\sigma}_{t}^{2}[l]$, $\bar{\xi}_{\In,t}[l]$, and $\bar{v}_{t+1}[l]$ 
in the inner module. The variable $\bar{\eta}_{t}[l]$ corresponds to 
the effective amplitude of $X$, given by 
\begin{equation}
\bar{\eta}_{t}[l]
= \frac{\bar{\zeta}_{t}[l]}{\bar{\xi}_{\Out,t}[l]} + \sum_{l'\in\mathcal{N}[l]}
\bar{\underline{\eta}}_{t,J}[l'\rightarrow l], 
\label{eta_bar}
\end{equation}
with 
\begin{equation} \label{eta_ll'}
\bar{\underline{\eta}}_{t,j}[l\rightarrow l']
= \frac{\bar{\underline{\zeta}}_{t}[l]}{\bar{\underline{\xi}}_{\Out,t}[l]}
+ \sum_{\tilde{l}'\in\mathcal{N}[l]\backslash\{l'\}}
\bar{\underline{\eta}}_{t,j-1}[\tilde{l}'\rightarrow l].  
\end{equation}
As an initial condition, $\bar{\underline{\eta}}_{t,0}[l'\rightarrow l]
=\bar{\underline{\eta}}_{t-T,J}[l'\rightarrow l]$ is used, as well as 
$\bar{\underline{\eta}}_{t-T,J}[l'\rightarrow l]=0$ for all $t<T$. 
Similarly, $\bar{\sigma}_{t}^{2}[l]$ represents the asymptotic alternative of 
$\sigma_{t}^{2}[l]$ in (\ref{sigma_t}),   
\begin{equation} \label{sigma_t_bar}
\bar{\sigma}_{t}^{2}[l] = \frac{1}{L}\left(
 \frac{1}{\bar{\xi}_{\Out,t}[l]} + \sum_{l'\in\mathcal{N}[l]}
 \underline{\bar{\sigma}}_{t,J}^{2}[l'\rightarrow l]
\right), 
\end{equation}
with 
\begin{equation} \label{sigma_ll'_bar}
\underline{\bar{\sigma}}_{t,j}^{2}[l\rightarrow l'] 
= \frac{1}{\underline{\bar{\xi}}_{\Out,t}[l]}
+ \sum_{\tilde{l}'\in\mathcal{N}[l]\backslash\{l'\}}
\underline{\bar{\sigma}}_{t,j-1}^{2}[\tilde{l}'\rightarrow l]. 
\end{equation}
As an initial condition, $\underline{\bar{\sigma}}_{t,0}^{2}[l'\rightarrow l]
=\underline{\bar{\sigma}}_{t-T,J}^{2}[l'\rightarrow l]$ is used, as well as 
$\underline{\bar{\sigma}}_{t-T,J}^{2}[l'\rightarrow l]=0$ for all $t<T$. 
The variable $\bar{\xi}_{\In,t}[l]$ is the asymptotic 
alternative of $\xi_{\In,t}[l]$ in (\ref{xi_in}), given by 
\begin{equation} \label{xi_in_bar}
\bar{\xi}_{\In,t}[l] 
= \mathbb{E}\left[
 \partial_{1}f_{\In}[l]\left(
  \frac{\bar{\eta}_{t}[l]}{L}X + \tilde{H}_{t}[l]; 
  \bar{\eta}_{t}[l], \bar{\sigma}_{t}^{2}[l]
 \right)
\right]. 
\end{equation}
The variable $\bar{v}_{t+1}[l]$ is the asymptotic alternative of 
$v_{t+1}[l]$ in (\ref{v_t}), given by 
$\bar{v}_{0}[l]=(L\delta[l])^{-1}\mathbb{E}[X^{2}]$ and 
\begin{equation} \label{v_t_bar}
\bar{v}_{t+1}[l] = \frac{\bar{\sigma}_{t}^{2}[l]
\bar{\xi}_{\In,t}[l]}{\delta[l]\eta_{t}[l]}
\end{equation}
for $t\geq0$, where $\eta_{t}[l]$ is defined in (\ref{eta}). 

Finally, we define $Z_{0}[l]=0$ and the random variables $\{Z_{t+1}[l]\}$, 
which are independent of $\{W[l]\}$ and correlated with $\{Z[l]\}$. More 
precisely, $\{Z_{t+1}[l]\}$ are zero-mean Gaussian random variables with 
covariance 
\begin{align}
&\mathbb{E}[Z[l']Z_{t+1}[l]] 
\nonumber \\
&= \frac{\delta_{l,l'}}{L\delta[l]}\mathbb{E}\left[
 Xf_{\In}[l]\left(
  \frac{\bar{\eta}_{t}[l]}{L}X + \tilde{H}_{t}[l]; \eta_{t}[l], 
   \bar{\sigma}_{t}^{2}[l]
 \right)
\right], \label{Z_Zt}
\end{align}
\begin{align} 
&\mathbb{E}[Z_{\tau+1}[l']Z_{t+1}[l]] 
\nonumber \\
&= \frac{\delta_{l,l'}}{L\delta[l]}
\mathbb{E}\left[
 f_{\In}[l]\left(
  \frac{\bar{\eta}_{\tau}[l]}{L}X + \tilde{H}_{\tau}[l]; \eta_{\tau}[l], 
 \bar{\sigma}_{\tau}^{2}[l]
 \right)
\right. \nonumber \\
&\left.
 \cdot f_{\In}[l]\left(
  \frac{\bar{\eta}_{t}[l]}{L}X + \tilde{H}_{t}[l]; \eta_{t}[l], 
  \bar{\sigma}_{t}^{2}[l]
 \right)
\right]
\label{Z_covariance}
\end{align}
for all $\tau\in\{0,\ldots,t\}$. 

The definitions~(\ref{Z})--(\ref{Z_covariance}) provide state evolution 
recursion for D-GAMP. The significance of these definitions 
is presented in the following theorem:

\begin{theorem} \label{theorem_SE}
Suppose that Assumptions~\ref{assumption_x}, \ref{assumption_w}, 
\ref{assumption_A}, and \ref{assumption_Lipschitz} hold. 
Then, for all iterations $t=0,1,\ldots$ D-GAMP satisfies 
\begin{align}
&(\boldsymbol{z}[l], \{\tilde{\boldsymbol{z}}_{t}[l]\}_{l\in\mathcal{L}}, 
\{\boldsymbol{w}[l]\}_{l\in\mathcal{L}})
\nonumber \\
&\plto (Z[l], \{Z_{t}[l]\}_{l\in\mathcal{L}}, \{ W[l]\}_{l\in\mathcal{L}}), 
\end{align}
\begin{equation}
(\boldsymbol{x}, \{\tilde{\boldsymbol{x}}_{t}[l] 
- L^{-1}\bar{\eta}_{t}[l]\boldsymbol{x}\}_{l\in\mathcal{L}}) 
\plto (X, \{\tilde{H}_{t}[l]\}_{l\in\mathcal{L}})
\end{equation}
in the large system limit, where the zero-mean Gaussian random variables 
$Z_{t}$, $Z_{t}[l]$ and $\tilde{H}_{t}[l]$ are given via 
(\ref{Z})--(\ref{Z_covariance}) to represent state evolution recursion.  
\end{theorem}
\begin{IEEEproof}
See Appendix~\ref{appen_proof_theorem_SE}. 
\end{IEEEproof}

Theorem~\ref{theorem_SE} implies the asymptotic Gaussianity 
for the input messages to the outer and inner denoisers. 
In particular, the error covariance for D-GAMP converges almost surely to 
\begin{align}
&\frac{1}{N}(\hat{\boldsymbol{x}}_{\tau+1}[l] - \boldsymbol{x})^{\mathrm{T}}
(\hat{\boldsymbol{x}}_{t+1}[l] - \boldsymbol{x})
\nonumber \\
&\ato \mathbb{E}\left[
 \left\{
  f_{\In}[l]\left(
   \frac{\bar{\eta}_{\tau}[l]}{L}X + \tilde{H}_{\tau}[l];
   \eta_{\tau}[l], \bar{\sigma}_{\tau}^{2}[l]
  \right) - X
 \right\}
\right.
\nonumber \\
&\left.
 \cdot\left\{
  f_{\In}[l]\left(
   \frac{\bar{\eta}_{t}[l]}{L}X + \tilde{H}_{t}[l];
   \eta_{t}[l], \bar{\sigma}_{t}^{2}[l]
  \right) - X
 \right\}
\right]
\nonumber \\
&\equiv \mathrm{cov}_{\tau+1,t+1}[l], \label{error_covariance}
\end{align}
\begin{align}
&- \frac{1}{N}\boldsymbol{x}^{\mathrm{T}}
(\hat{\boldsymbol{x}}_{t+1}[l] - \boldsymbol{x})
\nonumber \\
&\ato \mathbb{E}\left[
 X^{2} - Xf_{\In}[l]\left(
  \frac{\bar{\eta}_{t}[l]}{L}X + \tilde{H}_{t}[l];
  \eta_{t}[l], \bar{\sigma}_{t}^{2}[l]
 \right) 
\right]
\nonumber \\
&\equiv \mathrm{cov}_{0,t+1}[l], \label{error_covariance0}
\end{align}
\begin{equation} \label{error_covariance00}
\frac{1}{N}\boldsymbol{x}^{\mathrm{T}}\boldsymbol{x}
\ato \mathbb{E}[X^{2}] \equiv \mathrm{cov}_{0,0}[l] 
\end{equation}
in the large system limit.

It is possible to derive a closed form with respect to the covariance 
$\mathbb{E}[\tilde{H}_{\tau}[l]\tilde{H}_{t}[l]]$ 
when the network is a tree. 
\begin{theorem} \label{theorem_variance}
Suppose that Assumption~\ref{assumption_tree} holds and let 
$M_{t}[l]=f_{\Out}[l](Z_{t}[l], Y[l]; \bar{v}_{t}[l])$. Then, the covariance 
$\bar{\Sigma}_{\tau,t}[l]
=\mathbb{E}[\tilde{H}_{\tau}[l]\tilde{H}_{t}[l]]$ is given by 
\begin{equation} \label{H_variance}
\bar{\Sigma}_{\tau,t}[l]
= \frac{\mathbb{E}[M_{\tau}[l]M_{t}[l]]}
{L\bar{\xi}_{\Out,\tau}[l]\bar{\xi}_{\Out,t}[l]}
+ \sum_{l'\in\mathcal{N}[l]}\underline{\bar{\Sigma}}_{\tau,t,J}[l'\rightarrow l]
\end{equation}
for all $\tau\in\{0,\ldots,t\}$, with 
\begin{align} 
\underline{\bar{\Sigma}}_{\tau,t,j}[l\rightarrow l']
&= \frac{\mathbb{E}[\underline{M}_{\tau}[l]\underline{M}_{t}[l]]}
{L\underline{\bar{\xi}}_{\Out,\tau}[l]\underline{\bar{\xi}}_{\Out,t}[l]}
\nonumber \\
&+ \sum_{\tilde{l}'\in\mathcal{N}[l]\backslash\{l'\}}
\underline{\bar{\Sigma}}_{\tau,t,j-1}[\tilde{l}'\rightarrow l]. 
\label{H_variance_ll'}
\end{align}
As an initial condition, 
$\underline{\bar{\Sigma}}_{\tau,t,0}[l'\rightarrow l]
=\underline{\bar{\Sigma}}_{\tau-T,t-T,J}[l'\rightarrow l]$
is used, as well as $\underline{\bar{\Sigma}}_{\tau-T,t-T,J}[l'\rightarrow l]]=0$ 
for all $t<T$. 
\end{theorem}
\begin{IEEEproof}
See Appendix~\ref{proof_theorem_variance}. 
\end{IEEEproof}

In evaluating the variance $\bar{\Sigma}_{t,t}[l]$ in (\ref{H_variance}) 
for $\tau=t$, the covariance $\mathbb{E}[Z_{\tau}[l]Z_{t}[l]]$ 
for $\tau\neq t$ is not needed. In other words, the variance variables 
$\mathbb{E}[Z^{2}[l]]$, $\mathbb{E}[Z[l]Z_{t}[l]]$, $\mathbb{E}[Z_{t}^{2}[l]]$, 
and $\bar{\Sigma}_{t,t}[l]$ satisfy closed-form state evolution recursion 
with respect to these variables. Nonetheless, we have evaluated the 
covariance between messages in all preceding iterations to follow the 
long-memory proof strategy~\cite{Takeuchi222}. 

We next investigate fixed points of the state evolution recursion 
for D-GAMP in tree-structured networks when homogeneous measurements are 
considered. 

\begin{theorem} \label{theorem_fixed_point}
Let $\delta[l]=\delta$ and $W[l]\sim W$ i.i.d.\ 
in Assumption~\ref{assumption_w} for all $l\in\mathcal{L}$, 
with some random variable $W$. Furthermore, 
consider $g[l]=g$, $f_{\Out}[l]=f_{\Out}$, and $f_{\In}[l]=f_{\In}$ with 
identical functions $g$, $f_{\Out}$ and $f_{\In}$ for all nodes 
$l\in\mathcal{L}$. If Assumption~\ref{assumption_tree} holds, 
then the state evolution recursion with respect to the variance variables 
for D-GAMP has the same fixed points as those for centralized 
GAMP~\cite{Rangan11}.  
\end{theorem}
\begin{IEEEproof}
See Appendix~\ref{proof_theorem_fixed_point}. 
\end{IEEEproof}

Theorem~\ref{theorem_fixed_point} implies that the consensus is achieved 
for tree-structured networks if D-GAMP converges. To realize the consensus 
for general ad hoc networks, one may replace consensus propagation 
with a distributed protocol in \cite{Xiao04}, as used in 
\cite{Mateos10,Bazerque10,Mota12,Mota13,Shi14}, 
\begin{equation} \label{x_tilde_naive}
\tilde{\boldsymbol{x}}_{t}[l] 
= \boldsymbol{x}_{t}[l] + \gamma\sum_{l'\in\mathcal{N}[l]}
(\boldsymbol{x}_{t}[l'] - \boldsymbol{x}_{t}[l])
\end{equation}
for $T=1$, with $\gamma>0$. However, this naive protocol cannot achieve 
the same performance as that for centralized GAMP. 
\begin{theorem} \label{theorem_naive}
Let $T=1$ and replace the update rule of $\tilde{\boldsymbol{x}}_{t}[l]$ in 
(\ref{x_tilde}) with the distributed protocol~(\ref{x_tilde_naive}). 
Suppose that Assumptions~\ref{assumption_x}, \ref{assumption_w}, 
\ref{assumption_A}, and \ref{assumption_Lipschitz} hold. Then, the fixed 
points of state evolution recursion for D-GAMP with the distributed 
protocol~(\ref{x_tilde_naive}) are different from those for centralized 
GAMP. 
\end{theorem}
\begin{IEEEproof}
See Appendix~\ref{proof_theorem_naive}. 
\end{IEEEproof}

The intuition of Theorem~\ref{theorem_naive} is as follows: To achieve the 
performance of centralized GAMP, the effective signal-to-noise ratio (SNR) 
$L^{-2}\bar{\eta}_{t}^{2}[l]/\bar{\Sigma}_{t,t}[l]$ in the inner denoiser has 
to converge toward the same 
fixed point as that for centralized GAMP. This convergence is realizable 
for consensus propagation since both signal power $L^{-2}\bar{\eta}_{t}^{2}[l]$ 
and noise power $\bar{\Sigma}_{t,t}[l]$ are updated via consensus propagation, 
as shown in (\ref{eta_bar}) and (\ref{H_variance}). However, the 
distributed protocol~(\ref{x_tilde_naive}) results in different protocols 
for the signal and noise power. As a result, the consensus for the signal 
and noise power is not achievable simultaneously. 

\subsection{Bayes-Optimal Denoisers} 
We consider the Bayes-optimal denoisers in terms of MMSE. D-GAMP using 
the Bayes-optimal inner and outer denoisers---called Bayes-optimal 
D-GAMP---has three advantages: A first advantage is the optimality in terms 
of the MSE performance. A second advantage is that the state evolution 
recursion is simplified. This simplification is due to the fact that the 
update rules in D-GAMP are matched to the state evolution recursion. 
The last advantage is the convergence guarantee for the state evolution 
recursion. The convergence is systematically proved via the long-memory 
strategy~\cite{Takeuchi222}. 

We first focus on the Bayes-optimal inner denoiser. 
The inner denoiser is designed so as to minimize the 
MSE~(\ref{error_covariance}) with $\tau=t$. 
We know that the MSE is minimized if the inner denoiser is 
the posterior mean estimator of $X$ given a scalar measurement $U_{t}[l]$,  
\begin{equation} \label{inner_measurement}
U_{t}[l] = \frac{\bar{\eta}_{t}[l]}{L}X + \tilde{H}_{t}[l], 
\end{equation}
where $\tilde{H}_{t}[l]\sim\mathcal{N}(0,\bar{\Sigma}_{t,t}[l])$ is 
independent of $X$. Thus, the Bayes-optimal inner denoiser is defined as 
the posterior mean of $X$ given $U_{t}[l]$, 
\begin{equation} \label{inner_denoiser} 
f_{\In}[l](u; \bar{\eta}_{t}[l], \bar{\Sigma}_{t,t}[l]) 
= \mathbb{E}[X | U_{t}[l] = u]. 
\end{equation}

We present an existing result~\cite[Lemma~2]{Takeuchi222} 
for the Bayes-optimal inner denoiser~(\ref{inner_denoiser}), 
which is a key lemma to evaluate the covariance~(\ref{Z_covariance}) 
in the long-memory proof strategy~\cite{Takeuchi222}. 

\begin{lemma}[\cite{Takeuchi222}] \label{lemma_inner_covariance} 
Consider the Bayes-optimal inner denoiser~(\ref{inner_denoiser}). 
For given $t\geq 0$, assume $\bar{\Sigma}_{\tau,t}[l]=\bar{\Sigma}_{t,t}[l]$ 
in (\ref{H_variance}) for all $\tau\in\{0,\ldots,t\}$ and 
$\bar{\Sigma}_{\tau',\tau'}[l]>\bar{\Sigma}_{\tau,\tau}[l]$ for all 
$\tau'<\tau\leq t$. If $\bar{\sigma}_{t}^{2}[l]$ 
in (\ref{sigma_t_bar}) is equal to $\bar{\Sigma}_{t,t}[l]$, then we have 
$\mathrm{cov}_{\tau+1,t+1}[l]=\mathrm{cov}_{t+1,t+1}[l]$ in (\ref{error_covariance})
for all $\tau\in\{0,\ldots,t\}$. 
\end{lemma}

The assumptions in Lemma~\ref{lemma_inner_covariance} can be understood 
as follows: They imply the cascaded representation of $U_{\tau}[l]$ and 
$U_{t}[l]$: 
\begin{equation}
U_{\tau}[l] = U_{t}[l] + \Delta\tilde{H}_{\tau,t}[l],  
\quad 
U_{t}[l] = \frac{\bar{\eta}[l]}{L}X + \tilde{H}_{t}[l], 
\end{equation}
where $\tilde{H}_{t}[l]\sim\mathcal{N}(0,\bar{\Sigma}_{t,t}[l])$ and 
$\Delta\tilde{H}_{\tau,t}[l]\sim\mathcal{N}(0, 
\bar{\Sigma}_{\tau,\tau}[l] - \bar{\Sigma}_{t,t}[l])$ 
are independent of all random variables. Since $U_{\tau}[l]$ is a noisy 
measurement of $U_{t}[l]$, the measurement $U_{\tau}[l]$ provides no additional 
information on $X$ when $U_{t}[l]$ is observed. Thus, we have 
$\mathbb{E}[X | U_{\tau}[l], U_{t}[l]] = \mathbb{E}[X | U_{t}[l]]
=f_{\In}[l](U_{t}[l]; \bar{\eta}_{t}[l], \bar{\Sigma}_{t,t}[l])$, 
which is used to prove Lemma~\ref{lemma_inner_covariance}.  

We next design the outer denoiser so as to maximize the signal-to-noise 
ratio (SNR) $L^{-2}\bar{\eta}_{t}^{2}[l]/\bar{\Sigma}_{t,t}[l]$ in 
the measurement model~(\ref{inner_measurement}) for the inner denoiser. 
Using the definitions of $\bar{\eta}_{t}[l]$ and $\bar{\Sigma}_{t,t}[l]$ 
in (\ref{eta_bar}) and (\ref{H_variance}), we find that the SNR 
$L^{-2}\bar{\eta}_{t}^{2}[l]/\bar{\Sigma}_{t,t}[l]$ after consensus propagation 
is maximized when the individual SNR 
$L\bar{\zeta}_{t}^{2}[l]/\mathbb{E}[M_{t}^{2}[l]]$ 
before consensus propagation is maximized for all $l\in\mathcal{L}$.  

To solve this SNR maximization problem, 
we consider a scalar measurement model for the outer denoiser, 
\begin{equation}
Y[l] = g[l](Z[l], W[l]),  
\end{equation}
\begin{equation}
Z_{0}[l] = 0, \quad Z_{\tau}[l] = Z[l] + B_{\tau}[l], \quad 
Z_{t}[l] = Z[l] + B_{t}[l]
\end{equation}
for $t\geq\tau>0$, where $B_{\tau}[l]$ and $B_{t}[l]$ are 
independent of $W[l]$ and zero-mean Gaussian random variables with covariance 
\begin{equation} \label{ZB}
\mathbb{E}[Z[l]B_{t}[l]] 
= \mathbb{E}[Z[l]Z_{t}[l]] - \mathbb{E}[Z^{2}[l]], 
\end{equation}
\begin{equation} \label{BB}
\mathbb{E}[B_{\tau}[l]B_{t}[l]] = \mathbb{E}[(Z_{\tau}[l]-Z[l])(Z_{t}[l]-Z[l])],
\end{equation} 
defined with (\ref{Z}), (\ref{Z_Zt}), and (\ref{Z_covariance}) for $t>0$. 
While the outer denoiser in iteration~$t$ is defined with only $Z_{t}[l]$, 
the two random variables $Z_{\tau}[l]$ and $Z_{t}[l]$ are considered to 
evaluate the covariance $\mathbb{E}[M_{\tau}[l]M_{t}[l]]$ in 
(\ref{H_variance}), which is needed in the long-memory proof 
strategy~\cite{Takeuchi222}. 

We evaluate the covariance in (\ref{ZB}) and (\ref{BB}). Using the 
definitions~(\ref{Z_Zt}), (\ref{Z_covariance}), (\ref{error_covariance}), 
and (\ref{error_covariance0}) yields 
\begin{equation} \label{v_t_bar_true0}
- \mathbb{E}[Z[l]B_{t}[l]] 
= \frac{1}{L\delta[l]}\mathrm{cov}_{0,t}[l]
\equiv \bar{v}_{0,t}[l], 
\end{equation}
\begin{equation}\label{v_t_bar_true}
\mathbb{E}[B_{\tau}[l]B_{t}[l]] = \frac{1}{L\delta[l]}\mathrm{cov}_{\tau,t}[l]
\equiv \bar{v}_{\tau,t}[l], 
\end{equation}
with $\bar{v}_{0,0}[l]=(L\delta[l])^{-1}\mathrm{cov}_{0,0}[l]$. 
The following lemma presents the optimal solution to the SNR maximization 
problem: 
\begin{lemma} \label{lemma_outer_denoiser}
Let $\hat{Z}_{t}[l](\theta, y; \bar{v}_{t,t}[l])$ denote 
the posterior mean estimator of $Z[l]$ 
given $Z_{t}[l]=\theta$ and $Y[l]=y$, 
\begin{equation} \label{PME_Z}
\hat{Z}_{t}[l](\theta, y; \bar{v}_{t,t}[l]) 
= \frac{\int zP_{Y[l]|Z[l]}(y|z)e^{-\frac{(z-\theta)^{2}}{2\bar{v}_{t,t}[l]}}dz}
{\int P_{Y[l]|Z[l]}(y|z)e^{-\frac{(z-\theta)^{2}}{2\bar{v}_{t,t}[l]}}dz}, 
\end{equation}
where $P_{Y[l]|Z[l]}(y|z)$ represents the conditional distribution\footnote{
The conditional probability density function should be used 
if $Y[l]$ is a continuous random variable. 
} of $Y[l]$ given $Z[l]$, induced from the randomness of $W[l]$ through 
$Y[l]=g[l](Z[l],W[l])$.    
If $\bar{v}_{0,t}[l]=\bar{v}_{t,t}[l]$ holds, then 
for the individual SNR $\bar{\zeta}_{\Out,t}^{2}[l]/\mathbb{E}[M_{t}^{2}[l]]$ 
we have the following inequality: 
\begin{equation}
\frac{\bar{\zeta}_{\Out,t}^{2}[l]}{\mathbb{E}[M_{t}^{2}[l]]}
\leq \mathbb{E}\left[
 \left\{
  \frac{Z_{t}[l] - \hat{Z}_{t}[l](Z_{t}[l], Y[l]; \bar{v}_{t,t}[l])}
  {\bar{v}_{t,t}[l]}
 \right\}^{2} 
\right], 
\end{equation}
where the equality holds if and only if the outer denoiser is given by 
\begin{equation} \label{outer_denoiser}
f_{\Out}[l](\theta,y;\bar{v}_{t,t}[l]) 
= C\left(
 \frac{\theta - \hat{Z}_{t}[l](\theta, y; \bar{v}_{t,t}[l])}
 {\bar{v}_{t,t}[l]}
\right) 
\end{equation}  
for any constant $C\in\mathbb{R}$. 
\end{lemma}
\begin{IEEEproof}
See Appendix~\ref{proof_lemma_outer_denoiser}. 
\end{IEEEproof}

Use of the Bayes-optimal inner denoiser~(\ref{inner_denoiser}) justifies 
the assumption $\bar{v}_{0,t}[l]=\bar{v}_{t,t}[l]$ for the 
Bayes-optimal inner denoiser~(\ref{inner_denoiser}), as shown shortly. 
This paper uses the Bayes-optimal outer denoiser~(\ref{outer_denoiser}) 
with $C=1$ while \cite{Rangan11} and \cite{Barbier19} used $C=-1$ and 
$C=-\bar{v}_{t,t}^{1/2}[l]$, respectively. Of course, these choices of the 
arbitrary constant $C$ do not provide any impacts on the performance of 
D-GAMP. 

It is open whether any Bayes-optimal denoiser is piecewise 
Lipschitz-continuous. Thus, we postulate the following assumption instead of 
Assumption~\ref{assumption_Lipschitz}: 

\begin{assumption} \label{assumption_Bayes}
The composition $f_{\Out}[l](\theta, 
g[l](z, w); \bar{v}_{t,t}[l])$ of the measurement function $g[l]$ in 
(\ref{measurement}) and Bayes-optimal outer denoiser $f_{\Out}[l]$ in 
(\ref{outer_denoiser}) is 
piecewise Lipschitz-continuous with respect to 
$(\theta, z, w)\in\mathbb{R}^{3}$. 
The Bayes-optimal inner denoiser 
$f_{\In}[l](u; \bar{\eta}_{t}[l], \bar{\Sigma}_{t,t}[l])$ 
in (\ref{inner_denoiser}) is piecewise Lipschitz-continuous with respect to 
$u\in\mathbb{R}$. 
\end{assumption}

The following lemma is a key result to evaluate the covariance 
$\bar{\Sigma}_{\tau,t}[l]$ given by (\ref{H_variance}) in the long-memory 
proof strategy~\cite{Takeuchi222}.

\begin{lemma} \label{lemma_outer_covariance}
Suppose that Assumption~\ref{assumption_Bayes} holds and consider 
the Bayes-optimal outer denoiser~(\ref{outer_denoiser}) with $C=1$. 
For given $t\geq0$, assume $\bar{v}_{\tau,t}[l]=\bar{v}_{t,t}[l]$ 
in (\ref{v_t_bar_true}) for all $\tau\in\{0,\ldots,t\}$ and 
$\bar{v}_{\tau',\tau'}[l]>\bar{v}_{\tau,\tau}[l]$ for all $\tau'<\tau\leq t$. 
If $\bar{v}_{t}[l]$ in (\ref{v_t_bar}) is equal to $\bar{v}_{t,t}[l]$, 
then we have 
\begin{equation}
\mathbb{E}[M_{\tau}[l]M_{t}[l]] 
= \bar{\xi}_{\Out,\tau}[l]
\end{equation}
for all $\tau\in\{0,\ldots,t\}$. 
\end{lemma}
\begin{IEEEproof}
See Appendix~\ref{proof_lemma_outer_covariance}. 
\end{IEEEproof}

The state evolution recursion for D-GAMP is simplified when the Bayes-optimal 
inner denoiser~(\ref{inner_denoiser}) and outer 
denoiser~(\ref{outer_denoiser}) are used. We first present 
the simplified state evolution recursion. 
For the outer module, we have 
\begin{equation} 
\mathbb{E}[M_{t}^{2}[l]] 
= \mathbb{E}[f_{\Out}^{2}[l](Z_{t}[l], Y[l]; \bar{v}_{t,t}[l])],
\label{M_t_expectation}
\end{equation}
\begin{equation} \label{H_variance_opt}
\bar{\Sigma}_{\tau,t}[l]
= \frac{1}{L\mathbb{E}[M_{t}^{2}[l]]}
+ \sum_{l'\in\mathcal{N}[l]}\underline{\bar{\Sigma}}_{t,t,J}[l'\rightarrow l]
\end{equation}
for all $\tau\in\{0,\ldots,t\}$, with  
\begin{equation} \label{H_variance_ll'_opt}
\underline{\bar{\Sigma}}_{t,t,j}[l\rightarrow l']
= \frac{1}{L\mathbb{E}[\underline{M}_{t}^{2}[l]]}
+ \sum_{\tilde{l}'\in\mathcal{N}[l]\backslash\{l'\}}
\underline{\bar{\Sigma}}_{t,t,j-1}[\tilde{l}'\rightarrow l]. 
\end{equation}
As initial conditions, $\bar{v}_{0,0}[l]
=(L\delta[l])^{-1}\mathbb{E}[X^{2}]$ and 
$\underline{\bar{\Sigma}}_{t,t,0}[l'\rightarrow l]
=\underline{\bar{\Sigma}}_{t-T,t-T,J}[l'\rightarrow l]$ 
are used, as well as 
$\underline{\bar{\Sigma}}_{t-T,t-T,J}[l'\rightarrow l]=0$ for all $t<T$. 
Here, the expectation in (\ref{M_t_expectation}) is over 
$Y[l]=g[l](Z[l], W[l])$ and $Z_{t}[l]$ defined in (\ref{Z}), 
(\ref{Z_Zt}), and (\ref{Z_covariance}).  

For the inner module, we have 
\begin{align}
&\mathrm{mse}_{t+1}[l]
\nonumber \\
&= \mathbb{E}\left[
 \left\{
  f_{\In}[l]\left(
   \frac{\eta_{t}[l]}{L}X + \tilde{H}_{t}[l];
   \eta_{t}[l], \bar{\Sigma}_{t,t}[l]
  \right) - X
 \right\}^{2}
\right],
\end{align}
\begin{equation} \label{v_t_true} 
\bar{v}_{\tau,t+1}[l] = \frac{1}{L\delta[l]}\mathrm{mse}_{t+1}[l] 
\end{equation}
for all $\tau\in\{0,\ldots,t+1\}$,  
with $\eta_{t}[l]$ given in (\ref{eta}), 
where $\tilde{H}_{t}[l]\sim\mathcal{N}(0,\bar{\Sigma}_{t,t}[l])$  
is independent of $X$. 

The following theorem shows that the update rules in Bayes-optimal D-GAMP 
are matched to the state evolution recursion. As a result, the state 
evolution recursion is simplified. Furthermore, the state evolution 
recursion for Bayes-optimal D-GAMP is guaranteed to converge toward 
a fixed point. 

\begin{theorem} \label{theorem_convergence}
Suppose that Assumptions~\ref{assumption_x}, \ref{assumption_w}, 
\ref{assumption_A}, \ref{assumption_tree} and~\ref{assumption_Bayes} hold 
and consider the Bayes-optimal inner denoiser~(\ref{inner_denoiser}) and 
outer denoiser~(\ref{outer_denoiser}) with $C=1$. Then, 
we have the following results: 
\begin{itemize}
\item Bayes-optimal D-GAMP is consistent: $\bar{v}_{t}[l]=\bar{v}_{t,t}[l]$, 
$\bar{\eta}_{t}[l]=\eta_{t}[l]$, and $\bar{\sigma}_{t}^{2}[l]
=\bar{\Sigma}_{t,t}[l]$ hold for all $t$. 
\item The error covariance 
$N^{-1}(\hat{\boldsymbol{x}}_{\tau}[l]-\boldsymbol{x}[l])^{\mathrm{T}}
(\hat{\boldsymbol{x}}_{t}[l]-\boldsymbol{x}[l])$ for Bayes-optimal D-GAMP 
converges almost surely to $\mathrm{mse}_{t}[l]$ in the large system limit 
for all $\tau\in\{0,\ldots,t\}$, in which $\mathrm{mse}_{t}[l]$ is 
given via the simplified state evolution 
recursion~(\ref{M_t_expectation})--(\ref{v_t_true}). 

\item  The state evolution recursion~(\ref{M_t_expectation})--(\ref{v_t_true}) 
for Bayes-optimal D-GAMP converges to a fixed point as $t\to\infty$. 
\end{itemize}
\end{theorem}
\begin{IEEEproof}
See Appendix~\ref{proof_theorem_convergence}. 
\end{IEEEproof}

Theorem~\ref{theorem_convergence} provides an alternative proof 
for the convergence guarantee~\cite{Cobo23} of the state evolution 
recursion in general settings for Bayes-optimal centralized 
GAMP~\cite{Rangan11}. Since we know the optimality of Bayes-optimal 
centralized GAMP~\cite{Barbier19}, from Theorems~\ref{theorem_fixed_point} 
and \ref{theorem_convergence} we can conclude that the state evolution 
recursion for Bayes-optimal D-GAMP converges to the Bayes-optimal fixed 
point for the homogeneous measurements in Theorem~\ref{theorem_fixed_point} 
when the fixed point is unique.  

\section{Numerical Results} \label{sec5}
\subsection{Numerical Conditions}
\begin{figure}[t]
\begin{center}
\includegraphics[width=\hsize]{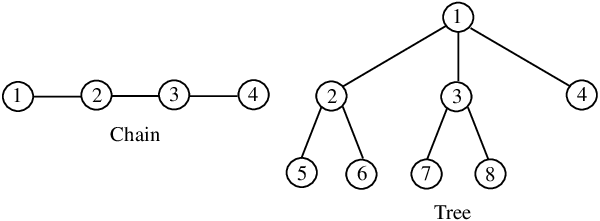}
\caption{
Tree-structured networks with no central nodes. 
}
\label{fig1} 
\end{center}
\end{figure}

In all numerical results, the i.i.d.\ Bernoulli-Gaussian signals 
with signal density~$\rho\in(0, 1]$ are considered: $x_{n}$ is independently 
sampled from the Gaussian distribution $\mathcal{N}(0, \rho^{-1})$ with 
probability~$\rho$. Otherwise, $x_{n}$ is set to zero. This signal has the 
unit power $\mathbb{E}[x_{n}^{2}]=1$. The noise vector 
$\boldsymbol{w}[l]\sim\mathcal{N}(\boldsymbol{0}, 
\sigma^{2}\boldsymbol{I}_{M[l]})$ in the measurement model~(\ref{measurement}) 
has independent zero-mean Gaussian 
elements with variance $\sigma^{2}>0$. The sensing matrix 
$\boldsymbol{A}[l]\in\mathbb{R}^{M[l]\times N}$ in node~$l$ has independent 
zero-mean Gaussian elements with variance~$(LM[l])^{-1}$. 
As examples of tree-structured networks, a one-dimensional chain and a tree 
with no central nodes in Fig.~\ref{fig1} are considered. 
These assumptions satisfy Assumptions~\ref{assumption_x}, \ref{assumption_w}, 
\ref{assumption_A}, and~\ref{assumption_tree}. 

This paper considers two measurement functions $g[l](z, w)$: One is the 
linear measurement $g[l](z, w)=z+w$ to test D-AMP. The other is clipping 
with threshold~$A>0$,  
\begin{equation} \label{clipping}
g[l](z, w) = \left\{
\begin{array}{cl}
A & \hbox{for $z+w>A$} \\
z + w & \hbox{for $|z+w|\leq A$} \\
-A & \hbox{for $z+w<-A$,}
\end{array}
\right.
\end{equation} 
which is used to evaluate D-GAMP. 

Bayes-optimal D-GAMP is used. When the linear measurement $g[l](z,w)=z+w$ 
is considered, the posterior mean estimator~(\ref{PME_Z}) in the outer 
denoiser~(\ref{outer_denoiser}) reduces to  
\begin{equation}
\hat{Z}_{t}[l](\theta, y; v_{t}[l]) 
= \theta + \frac{v_{t}[l]}{v_{t}[l] + \sigma^{2}}(y - \theta).
\end{equation}
Thus, the Bayes-optimal outer denoiser~(\ref{outer_denoiser}) with $C=1$ 
is given by 
\begin{equation}
f_{\Out}[l](\theta, y; v_{t}[l]) 
= \frac{\theta - y}{v_{t}[l] + \sigma^{2}}.
\end{equation}
In this case, D-GAMP is essentially\footnote{
D-AMP~\cite{Hayakawa18} replaced the quantity $\xi_{\Out,t}^{-1}[l]$ given in 
(\ref{xi_out}) with the well-known estimator 
$M^{-1}[l]\|\tilde{\boldsymbol{z}}_{t}[l] 
- \boldsymbol{y}[l]\|^{2}$ in (\ref{z_tilde}). However, numerical simulations 
showed that this estimator or its robust alternative had large errors for 
finite-sized systems with a non-negligible probability. Thus, this paper 
uses the original definition~(\ref{xi_out}) in D-AMP. 
} equivalent to D-AMP~\cite{Hayakawa18}. 

For the clipping case~(\ref{clipping}), we have the following Bayes-optimal 
outer denoiser~(\ref{outer_denoiser}) with $C=1$: 
\begin{equation}
f_{\Out}[l](\theta, y; v_{t}[l]) 
= \frac{\theta - y}{v_{t}[l] + \sigma^{2}} 
\quad \hbox{for $|y|<A$,}
\end{equation}
\begin{equation}
f_{\Out}[l](\theta, y; v_{t}[l]) 
= - \frac{p_{\mathrm{G}}(A-\theta, v_{t}[l] + \sigma^{2})}
{Q\left(
 (A - \theta)(v_{t}[l] + \sigma^{2})^{-1/2}
\right)}
\end{equation}
for $y>A$, and 
\begin{equation}
f_{\Out}[l](\theta, y; v_{t}[l]) 
= \frac{p_{\mathrm{G}}(A+\theta, v_{t}[l] + \sigma^{2})}
{Q\left(
 (A + \theta)(v_{t}[l] + \sigma^{2})^{-1/2}
\right)}
\end{equation}
for $y<-A$. In these expressions, $p_{\mathrm{G}}(\cdot;v)$ and $Q(x)$ denote the 
zero-mean Gaussian probability density function with variance $v$ and 
complementary cumulative distribution function of the standard Gaussian 
distribution, respectively. It is an exercise to confirm the piecewise 
Lipschitz-continuity of the composition 
$f_{\Out}[l](\theta, g[l](z, w); v_{t}[l])$ in 
Assumption~\ref{assumption_Bayes}, by using the well-known inequalities 
$x(1+x^{2})^{-1}p_{\mathrm{G}}(x; 1) < Q(x) < x^{-1}p_{\mathrm{G}}(x; 1)$ for all 
$x>0$.   

Damping~\cite{Murphy99,Vila15,Rangan19} is a heuristic technique to improve 
the convergence property of message-passing algorithms for finite-sized 
systems. While a tree-structured network is postulated, the overall graph 
for signal estimation is not a tree, so that damping is needed for D-GAMP 
as long as the system size is finite.  
In this paper, damping was used just after inner denoising in each 
node: The update rules~(\ref{x_hat}) and (\ref{v_t}) were replaced with
\begin{equation}
\hat{\boldsymbol{x}}_{t+1}[l] 
= \chi f_{\In}[l](\tilde{\boldsymbol{x}}_{t}[l]; \eta_{t}[l], \sigma_{t}^{2}[l])
+ (1 - \chi)\hat{\boldsymbol{x}}_{t}[l], 
\end{equation} 
\begin{equation}
v_{t+1}[l] = \chi\frac{N}{M[l]}\frac{\sigma_{t}^{2}[l]\xi_{\In,t}[l]}{\eta_{t}[l]}
+ (1 - \chi)v_{t}[l], 
\end{equation}
with damping factor $\chi\in(0, 1]$. While it is possible to design 
$t$-dependent (or $l$-dependent) damping factors via deep 
learning~\cite{Yoshida23}, for simplicity, this paper considers the constant 
damping factor $\chi$ for all $t$ and $l$, which was 
optimized via exhaustive search. 

As a baseline, this paper considers centralized AMP~\cite{Donoho09} or 
GAMP~\cite{Rangan11} using $\sum_{l\in\mathcal{L}}M[l]$ measurements. The purpose 
of D-GAMP is to achieve the same MSE performance as the corresponding 
centralized GAMP. In all numerical results, $10^{4}$ independent trials 
were simulated.  

\subsection{Chain Network}
The one-dimensional chain network in Fig.~\ref{fig1} is considered. 
The linear measurement is first assumed to compare D-GAMP with conventional 
D-AMP~\cite{Hayakawa18}. D-GAMP with $T[l]=1$ is essentially equivalent to 
D-AMP~\cite{Hayakawa18} for the linear measurement. 

\begin{figure}[t]
\begin{center}
\includegraphics[width=\hsize]{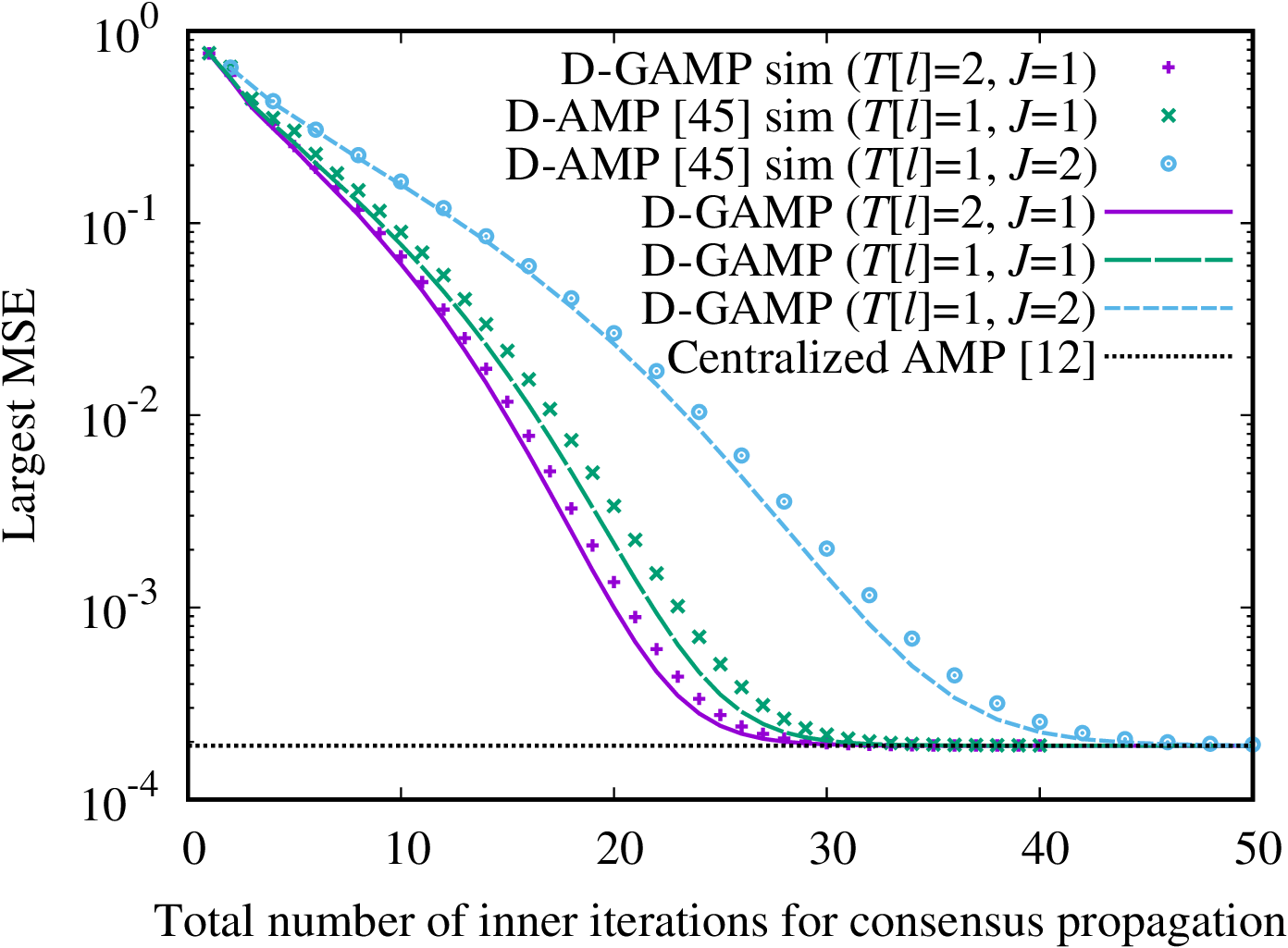}
\caption{
Largest MSE versus the total number of inner iterations for consensus 
propagation in the linear measurements. One-dimensional chain network 
with $L=4$ nodes, measurement dimension~$M[l]=480$, signal dimension~$N=6400$, 
signal density~$\rho=0.1$, SNR~$1/\sigma^{2}=30$~dB, and damping 
factor~$\chi=1$. The solid curves show state evolution results while 
numerical simulations are plotted with markers.  
}
\label{fig2} 
\end{center}
\end{figure}

Figure~\ref{fig2} shows numerical comparisons between D-GAMP and 
D-AMP~\cite{Hayakawa18} in terms of the total number of inner iterations 
for consensus propagation. As proved in Theorem~\ref{theorem_convergence}, 
the state evolution recursion for D-GAMP converges to the fixed point of the 
state evolution recursion for the corresponding centralized AMP. Furthermore, 
D-GAMP for $T[l]=2$ and $J=1$ converges more quickly than 
D-AMP~\cite{Hayakawa18} with $J=1$ or $J=2$ while $J\geq 4$ was used in 
\cite{Hayakawa18}. These observations imply that D-GAMP can reduce network 
traffic for consensus propagation compared to D-AMP~\cite{Hayakawa18}. 
 
\begin{figure}[t]
\begin{center}
\includegraphics[width=\hsize]{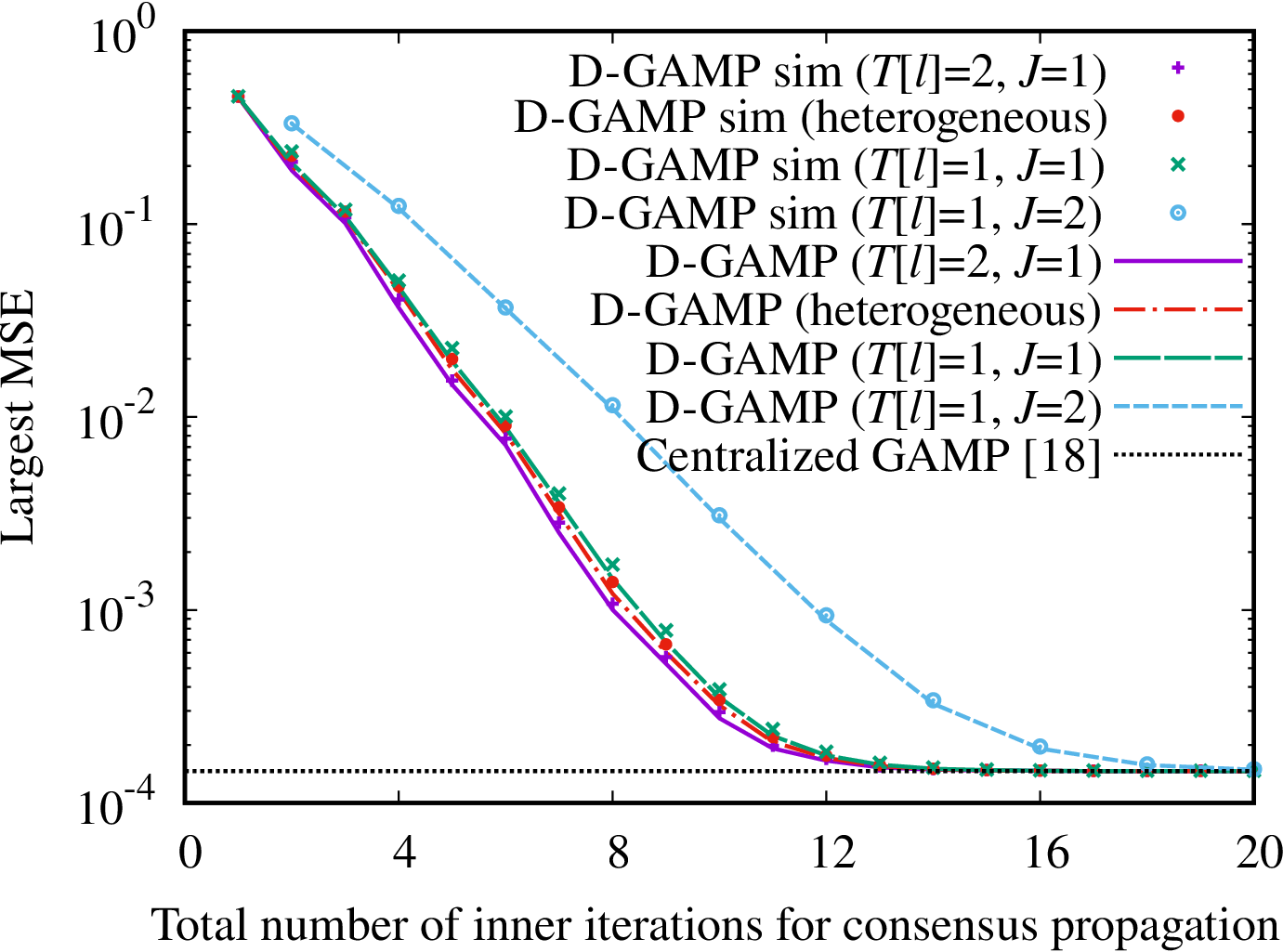}
\caption{
Largest MSE versus the total number of inner iterations for consensus 
propagation in the clipping case. One-dimensional chain network with $L=4$ 
nodes, measurement dimension~$M[l]=800$, signal dimension~$N=4000$, signal 
density~$\rho=0.1$, threshold~$A=2$, SNR~$1/\sigma^{2}=30$~dB, and 
damping factor~$\chi=1$. The solid curves show state evolution results 
while numerical simulations are plotted with markers. As a heterogeneous 
case, $T[l]=2$ for odd $l$, $T[l]=1$ for even $l$, and $J=1$ were considered.  
}
\label{fig3} 
\end{center}
\end{figure}

We next consider the clipping case in Fig.~\ref{fig3}. The basic observations 
are similar to those in Fig.~\ref{fig2}. As a heterogeneous case, D-GAMP 
with $T[l]=2$ for odd~$l$ and $T[l]=1$ for even~$l$ is shown. This case 
corresponds to a heterogeneous situation in which the odd-numbered nodes have 
twice higher processing speed than the even-numbered nodes, so that they can 
repeat two GAMP iterations while the even-numbered nodes compute a single 
GAMP iteration. The performance of D-GAMP in the heterogeneous case is 
between that in the two homogeneous cases for $J=1$: For $T[l]=1$ and 
$T[l]=2$ the odd-numbered nodes wait for the completion of one and two GAMP 
iterations in the even-numbered nodes, respectively. 

\subsection{Tree Network}
\begin{figure}[t]
\begin{center}
\includegraphics[width=\hsize]{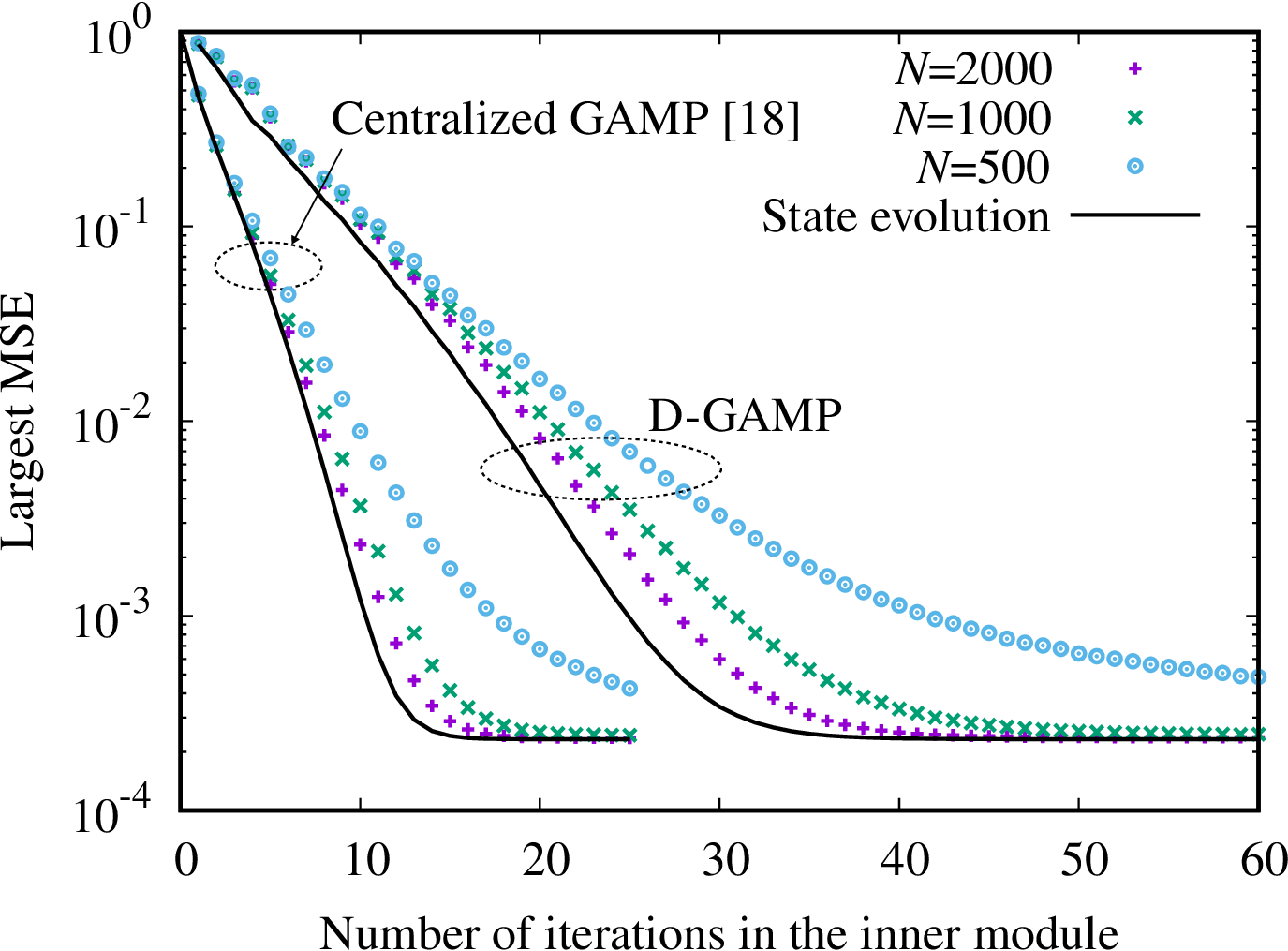}
\caption{
Largest MSE versus the number of iterations in the inner module for the 
clipping case. Tree network with $L=8$ 
nodes, compression ratio~$M[l]/N=0.05$, signal density~$\rho=0.1$, 
threshold~$A=2$, SNR~$1/\sigma^{2}=30$~dB, $T[l]=1$, and $J=1$. 
For $N=500, 1000, 2000$, D-GAMP used damping factors~$\chi=0.9, 1, 1$, 
respectively, while centralized GAMP used $\chi=0.9, 0.95, 0.95$. 
}
\label{fig4} 
\end{center}
\end{figure}

The tree network with $L=8$ nodes in Fig.~\ref{fig1} is considered. D-GAMP 
with $T[l]=1$ and $J=1$ is compared to centralized GAMP~\cite{Rangan11} in 
terms of the number of iterations~$t$. Figure~\ref{fig4} shows that D-GAMP 
converges to almost the same MSE performance as that of the corresponding 
centralized GAMP in the three cases $N=500$, $N=1000$, and $N=2000$. 
As predicted from state evolution, D-GAMP needs more iterations than the 
corresponding centralized GAMP, because of iterations for consensus 
propagation. 
Interestingly, the optimized damping factors for D-GAMP are slightly larger 
than those for the corresponding centralized GAMP. This observation is 
because consensus propagation plays a role as kind of damping 
to slow down the convergence of GAMP. Thus, consensus propagation 
does not degrade the convergence property of GAMP for finite-sized systems.  

\begin{figure}[t]
\begin{center}
\includegraphics[width=\hsize]{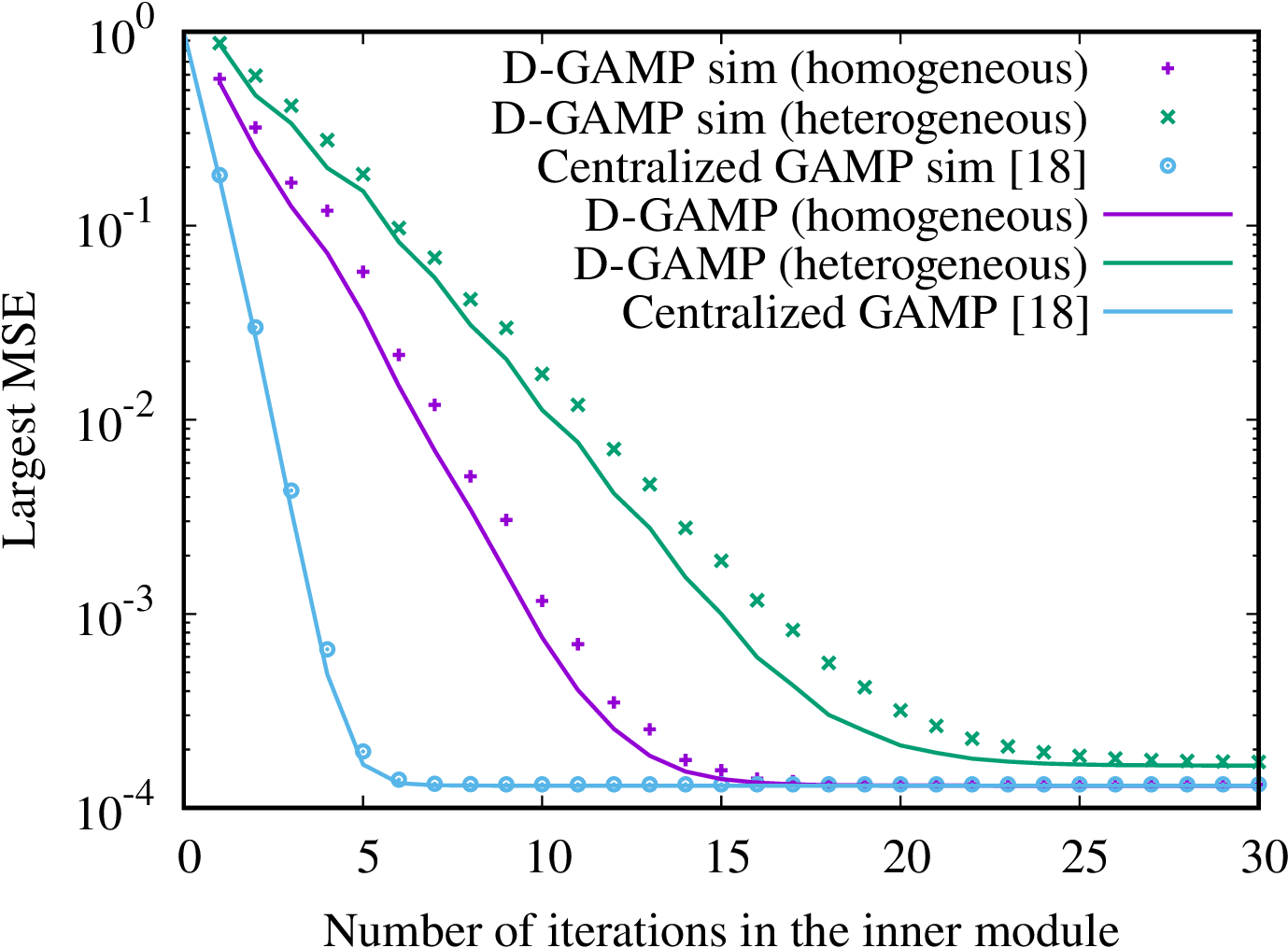}
\caption{
Largest MSE versus the number of iterations in the inner module for the 
clipping case. Tree network with $L=8$ nodes, signal 
dimension~$N=600$, signal density~$\rho=0.1$, threshold~$A=2$, 
SNR~$1/\sigma^{2}=30$~dB, $T[l]=1$, and $J=1$. $M[l]=90$ and $\chi=0.95$ were 
considered in a homogeneous case while $M[l]=150$ for odd~$l$, $M[l]=30$ 
for even~$l$, and $\chi=0.95$ were considered in a heterogeneous case. 
Centralized GAMP used damping factor $\chi=1$. 
}
\label{fig5} 
\end{center}
\end{figure}

We next consider heterogeneous measurements in Fig.~\ref{fig5}. The 
odd-numbered nodes have $M[l]=150$ measurements while the even-numbered nodes 
have $M[l]=30$ measurements. D-GAMP for the heterogeneous case cannot 
approach the same MSE performance as that of the corresponding centralized 
GAMP while D-GAMP for the homogeneous case $M[l]=90$ can achieve the same 
performance. This is because the Onsager-correction in (\ref{x}) and 
(\ref{z_tilde}) depends on the node index~$l$. To achieve the performance of 
centralized GAMP, an additional protocol is 
needed to realize the convergence of the Onsager-correction in D-GAMP 
toward that in the corresponding centralized GAMP. 

\section{Conclusions} \label{sec6}
This paper has proposed D-GAMP for signal reconstruction from distributed 
generalized linear measurements. D-GAMP is applicable to all tree-structured 
networks that do not necessarily have central nodes. State evolution has 
been used to analyze the asymptotic dynamics of D-GAMP for zero-mean i.i.d.\ 
Gaussian sensing matrices. The 
state evolution recursion for Bayes-optimal D-GAMP has been proved to converge 
toward the Bayes-optimal fixed point---achieved by the corresponding 
centralized GAMP---for homogeneous measurements with an identical dimension 
in all nodes when the fixed point is unique. 

D-GAMP has two limitations: One limitation is that zero-mean i.i.d.\ sensing 
matrices are required. As long as GAMP is used, this assumption cannot be 
weaken. To solve this issue, GAMP needs to be replaced with another 
sophisticated message-passing algorithm. 

The other limitation is in the assumption of tree-structured networks. To 
weaken this assumption, we need to replace consensus propagation with another 
sophisticated protocol for average consensus. As proved in 
Theorem~\ref{theorem_naive}, the conventional protocol~(\ref{x_tilde_naive}) 
for average consensus cannot be used for this purpose. 

\appendices

\section{Proof of Theorem~\ref{theorem_SE}}
\label{appen_proof_theorem_SE}
\subsection{Preliminaries}
We present a few technical lemmas required in proving 
Theorem~\ref{theorem_SE} via state evolution. 
We first formulate a general error model that describes the dynamics of 
estimation errors for D-GAMP. 

\begin{definition}[General Error Model]
For some $\bar{\eta}_{t}[l]\in\mathbb{R}$ and $C_{t,i}[l,l']\in\mathbb{R}$, 
the general error model is defined with five random vectors 
$\{\boldsymbol{b}_{t}[l], \boldsymbol{m}_{t}[l], \boldsymbol{h}_{t}[l], 
\tilde{\boldsymbol{h}}_{t}[l], \boldsymbol{q}_{t+1}[l]\}$,   
\begin{align} 
\boldsymbol{b}_{t}[l] 
&= \frac{\xi_{\In,t-1}[l]}{L\delta[l]}
\left\{
 \frac{\boldsymbol{m}_{t-1}[l]}{\xi_{\Out,t-1}[l]} 
 + \frac{C_{t,0}[l,l]-1}{\underline{\xi}_{\Out,t-1}[l]}
 \underline{\boldsymbol{m}}_{t-1}[l]
\right\}
\nonumber \\ 
&+ \boldsymbol{A}[l]\boldsymbol{q}_{t}[l]
\label{b}
\end{align}
\begin{equation} \label{m}
\boldsymbol{m}_{t}[l] 
= f_{\Out}[l]\left(
 \boldsymbol{b}_{t}[l] 
 + \frac{\bar{\zeta}_{t}[l]}{\bar{\xi}_{\Out,t}[l]}\boldsymbol{z}[l], 
 \boldsymbol{y}[l]; v_{t}[l]
\right), 
\end{equation}
\begin{equation}
\boldsymbol{h}_{t}[l] 
= \frac{\xi_{\Out,t}[l]}{L}\boldsymbol{q}_{t}[l]
- \boldsymbol{A}^{\mathrm{T}}[l]\boldsymbol{m}_{t}[l], 
\label{h}
\end{equation}
\begin{align}
\tilde{\boldsymbol{h}}_{t}[l] 
&= \frac{\boldsymbol{h}_{t}[l]}{\xi_{\Out,t}[l]} 
+ \frac{C_{t,0}[l,l] - 1}{\underline{\xi}_{\Out,t}[l]}
\underline{\boldsymbol{h}}_{t}[l]
\nonumber \\
&+ \sum_{i=0}^{\lfloor t/T \rfloor}\sum_{l'\neq l}C_{t,i}[l,l']
\frac{\underline{\boldsymbol{h}}_{t-iT}[l']}
{\underline{\xi}_{\Out,t-iT}[l']},  
\label{h_tilde}
\end{align}
\begin{equation} 
\boldsymbol{q}_{t+1}[l] 
= f_{\In}[l]\left(
 \frac{\bar{\eta}_{t}[l]}{L}\boldsymbol{x} 
 + \tilde{\boldsymbol{h}}_{t}[l]; \eta_{t}[l], \sigma_{t}^{2}[l]
\right) 
- \frac{\bar{\zeta}_{t+1}[l]}{\bar{\xi}_{\Out,t+1}[l]}\boldsymbol{x},
\label{q}
\end{equation}
where $\xi_{\Out,t}[l]$, $\xi_{\In,t}[l]$, $\bar{\xi}_{\Out,t}[l]$, and 
$\bar{\zeta}_{t}[l]$ are given in (\ref{xi_out}), (\ref{xi_in}), 
(\ref{xi_out_bar}), and (\ref{zeta_bar}), respectively. 
As initial conditions, we use $\boldsymbol{q}_{0}[l]
=-\bar{\xi}_{\Out,0}^{-1}[l]\bar{\zeta}_{0}[l]\boldsymbol{x}$,  
$\boldsymbol{m}_{-1}[l]=\underline{\boldsymbol{m}}_{-1}[l]=\boldsymbol{0}$. 
In particular, we have $\boldsymbol{b}_{0}[l]=-\bar{\xi}_{\Out,0}^{-1}[l]
\bar{\zeta}_{0}[l]\boldsymbol{z}[l]$. 
\end{definition}

The vector $\tilde{\boldsymbol{h}}_{t}[l]$ represents the dynamics 
of protocols for average consensus. The other vectors describe the dynamics 
of GAMP iterations. The update rule of $\tilde{\boldsymbol{h}}_{t}[l]$ 
in (\ref{h_tilde}) depends on messages in previous iterations while 
graph-based AMP~\cite{Gerbelot23} does not postulate such long-memory 
message-passing. Owing to this flexibility in the general error model, 
under Assumption~\ref{assumption_tree}, we can treat the error model of 
D-GAMP as an instance of the general error model.  

\begin{lemma} \label{lemma_error_model}
Suppose that Assumption~\ref{assumption_tree} holds, 
let $C_{t,0}[l,l]=1$, and define 
\begin{align}
&\boldsymbol{b}_{t}[l]=\tilde{\boldsymbol{z}}_{t}[l] 
- \frac{\bar{\zeta}_{t}[l]}{\bar{\xi}_{\Out,t}[l]}\boldsymbol{z}[l],
\quad 
\boldsymbol{m}_{t}[l]= \hat{\boldsymbol{z}}_{t}[l],
\nonumber \\
&\boldsymbol{h}_{t}[l]= \xi_{\Out,t}[l]\left(
 \boldsymbol{x}_{t}[l] 
 - \frac{\bar{\zeta}_{t}[l]}{L\bar{\xi}_{\Out,t}[l]}\boldsymbol{x}
\right),
\nonumber \\
&\tilde{\boldsymbol{h}}_{t}[l]
=\tilde{\boldsymbol{x}}_{t}[l] 
- \frac{\bar{\eta}_{t}[l]}{L}\boldsymbol{x}, 
\quad 
\boldsymbol{q}_{t}[l]=\hat{\boldsymbol{x}}_{t}[l] 
- \frac{\bar{\zeta}_{t}[l]}{\bar{\xi}_{\Out,t}[l]}\boldsymbol{x},  
\end{align} 
with $\xi_{\Out,t}[l]$, $\bar{\xi}_{\Out,t}[l]$, $\bar{\zeta}_{t}[l]$, 
$\bar{\eta}_{t}[l]$, and $\bar{\underline{\eta}}_{t,j}[l\rightarrow l']$ 
given in (\ref{xi_out}), (\ref{xi_out_bar}), (\ref{zeta_bar}), 
(\ref{eta_bar}), and (\ref{eta_ll'}), respectively. 
Then, there are some $\{C_{t,i}[l,l']\}$ such that these vectors satisfy 
the dynamics~(\ref{b})--(\ref{q}) in the general error model. 
\end{lemma}
\begin{IEEEproof}
The expression~(\ref{b}) with $C_{t,0}[l,l]=1$ is obtained by using 
$\boldsymbol{z}[l]=\boldsymbol{A}[l]\boldsymbol{x}$ and 
the definition of $\tilde{\boldsymbol{z}}_{t}[l]$ in (\ref{z_tilde}). 
The expression~(\ref{m}) follows from the definition
of $\hat{\boldsymbol{z}}_{t}[l]$ in (\ref{z_hat}). 
Using the definition of $\boldsymbol{x}_{t}[l]$ in (\ref{x}) yields (\ref{h}). 
The expression~(\ref{q}) follows from the definition 
of $\hat{\boldsymbol{x}}_{t+1}[l]$ in (\ref{x_hat}). 

We confirm (\ref{h_tilde}). Let 
\begin{equation}
\underline{\boldsymbol{h}}_{t,j}[l\rightarrow l']
= \underline{\boldsymbol{x}}_{t,j}[l\rightarrow l']
- \frac{\bar{\underline{\eta}}_{t,j}[l\rightarrow l']}{L}\boldsymbol{x}. 
\end{equation}
Using the definitions of $\overline{\underline{\boldsymbol{x}}_{t}[l]}$, 
$\tilde{\boldsymbol{x}}_{t}[l]$, and $\bar{\eta}_{t}[l]$ in (\ref{consensus}), 
(\ref{x_tilde}), and (\ref{eta_bar}) yields 
\begin{equation} \label{h_tilde_tmp}
\tilde{\boldsymbol{h}}_{t}[l] 
= \frac{\boldsymbol{h}_{t}[l]}{\xi_{\Out,t}[l]}  
+ \sum_{l'\in\mathcal{N}[l]}\underline{\boldsymbol{h}}_{t,J}[l'\rightarrow l]. 
\end{equation}
Furthermore, we use the definitions of 
$\underline{\boldsymbol{x}}_{t,j}[l\rightarrow l']$ and 
$\bar{\underline{\eta}}_{t,j}[l\rightarrow l']$ in (\ref{x_ll'}) and 
(\ref{eta_ll'}) to obtain 
\begin{equation} \label{h_ll'_tmp}
\underline{\boldsymbol{h}}_{t,j}[l\rightarrow l']
= \frac{\underline{\boldsymbol{h}}_{t}[l]}{\bar{\underline{\xi}}_{\Out,t}[l]}
+ \sum_{\tilde{l}'\in\mathcal{N}[l]\backslash\{l'\}}
\underline{\boldsymbol{h}}_{t,j-1}[\tilde{l}'\rightarrow l], 
\end{equation}
with $\underline{\boldsymbol{h}}_{t,0}[l'\rightarrow l]
=\underline{\boldsymbol{h}}_{t-T,J}[l'\rightarrow l]$. As an initial 
condition, $\underline{\boldsymbol{h}}_{t-T,J}[l']=\boldsymbol{0}$ are used 
for all $t<T$. The tree assumption in Assumption~\ref{assumption_tree} implies 
that $\underline{\boldsymbol{h}}_{t,J}[l'\rightarrow l]$ does not contain 
the messages $\underline{\boldsymbol{h}}_{\tau}[l]/
\bar{\underline{\xi}}_{\Out,\tau}[l]$ for any $\tau\leq t$ computed in node~$l$. 
Comparing the expression of $\tilde{\boldsymbol{h}}_{t}[l]$ with 
(\ref{h_tilde}), we have $C_{t,0}[l,l]=1$ and find that there are some 
$\{C_{t,i}[l,l']\}$ such that $\tilde{\boldsymbol{h}}_{t}[l]$ reduce to 
(\ref{h_tilde}). Thus, Lemma~\ref{lemma_error_model} holds.  
\end{IEEEproof}

The vector $\boldsymbol{b}_{t}[l]$ corresponds to the estimation errors of 
$\bar{\xi}_{\Out,t}^{-1}[l]\bar{\zeta}_{t}[l]\boldsymbol{z}[l]$ before 
outer denoising, while $\tilde{\boldsymbol{h}}_{t}[l]$ represents 
the estimation errors of $L^{-1}\bar{\eta}_{t}[l]\boldsymbol{x}$ before 
inner denoising. 

In Bolthausen's conditioning technique~\cite{Bolthausen14}, the dynamics 
of the vectors in the current iteration is evaluated via the 
conditional distribution of $\{\boldsymbol{A}[l]\}$  given 
those in all previous iterations, the signal vector $\boldsymbol{x}$, 
and the noise vector $\{\boldsymbol{w}[l]\}$. For notational convenience, 
we use the following notation in representing conditioned vectors: 

\begin{definition} \label{definition_stack}
For variables $\{a_{\tau}[l]\in\mathbb{R}\}$ associated with node~$l$, the 
column vector $\boldsymbol{a}_{t}[l]$ contains the variables for the iterations 
$\mathcal{T}_{t}[l]$ in (\ref{T_set}) where node~$l$ updates messages,  
\begin{equation}
\boldsymbol{a}_{t}[l] 
= (\boldsymbol{a}_{t}^{0}[l],\ldots,
\boldsymbol{a}_{t}^{\lfloor t/T\rfloor}[l])^{\mathrm{T}},  
\end{equation}
with $\boldsymbol{a}_{t}^{i}[l]=(a_{iT}[l],\ldots,a_{\min\{t, iT+T[l]\}-1}[l])$. 
For column vectors $\{\boldsymbol{v}_{\tau}[l]\}$, the matrix 
$\boldsymbol{V}_{t}[l]=\{\boldsymbol{v}_{\tau}[l]: \tau\in\mathcal{T}_{t}[l]\}$ 
has $|\mathcal{T}_{t}[l]|$ column vectors aligned in the same manner. 
\end{definition}

We define the conditional distribution of $\{\boldsymbol{A}[l]\}$. 
Let $\boldsymbol{B}_{t}[l]\in\mathbb{R}^{M[l]\times|\mathcal{T}_{t}[l]|}$ denote 
the matrix defined from 
$\{\boldsymbol{b}_{\tau}[l]: \tau\in\mathcal{T}_{t}[l]\}$ in 
Definition~\ref{definition_stack}. Similarly, we define 
$\boldsymbol{M}_{t}[l]$, $\boldsymbol{H}_{t}[l]$, 
$\tilde{\boldsymbol{H}}_{t}[l]$, and $\boldsymbol{Q}_{t}[l]$. 
Furthermore, let 
\begin{equation}
\mathfrak{E}_{1,0} 
= \left\{
 \{\boldsymbol{b}_{0}[l]: l\in\mathcal{L}\}, 
 \{\boldsymbol{m}_{0}[l]: l\in\mathcal{L}\}
\right\}, 
\end{equation}
\begin{align}
\mathfrak{E}_{t',t} 
&= \left\{
 \{\boldsymbol{B}_{t'}[l]: l\in\mathcal{L}\}, 
 \{\boldsymbol{M}_{t'}[l]: l\in\mathcal{L}\}, 
 \{\boldsymbol{H}_{t}[l]: l\in\mathcal{L}\}, 
\right. \nonumber \\
&\left.
 \{\tilde{\boldsymbol{H}}_{t}[l]: l\in\mathcal{L}\}, 
 \{\boldsymbol{Q}_{t+1}[l]: l\in\mathcal{L}\}
\right\}
\end{align}
for $t'>0$ and $t>0$, where the columns in the five matrices satisfy 
(\ref{b})--(\ref{q}) in the general error model. The set $\mathfrak{E}_{t,t}$ 
contains the messages that are computed just before updating 
$\boldsymbol{b}_{t}[l]$ in (\ref{b}) while $\mathfrak{E}_{t+1,t}$ includes 
them just before updating $\boldsymbol{h}_{t}[l]$ in (\ref{h}). 
The signal and noise vectors 
$\Theta=\{\boldsymbol{x}, \{\boldsymbol{w}[l]: l\in\mathcal{L}\}\}$ are 
always fixed. Thus, the conditional distribution of 
$\{\boldsymbol{A}[l]\}$ given $\mathfrak{E}_{t,t}$ and $\Theta$ is considered 
in evaluating the distribution of $\boldsymbol{b}_{t}[l]$ while the conditional 
distribution of $\{\boldsymbol{A}[l]\}$ given $\mathfrak{E}_{t+1,t}$ and 
$\Theta$ is considered in evaluating the distribution of 
$\boldsymbol{h}_{t}[l]$.

The conditional distributions of $\boldsymbol{A}[l]$ are evaluated via 
the following existing lemma:
\begin{lemma}[\cite{Bayati11}] \label{lemma_conditioning0}
Suppose that Assumption~\ref{assumption_A} holds. For some integers 
$t[l]\leq M[l]$ and $t'[l]\leq N$, let 
$\boldsymbol{X}[l]\in\mathbb{R}^{M[l]\times t[l]}$, 
$\boldsymbol{U}[l]\in\mathbb{R}^{N\times t[l]}$, 
$\boldsymbol{Y}[l]\in\mathbb{R}^{N\times t'[l]}$, and 
$\boldsymbol{V}[l]\in\mathbb{R}^{M[l]\times t'[l]}$ satisfy 
the following constraints:
\begin{equation}
\boldsymbol{X}[l] = \boldsymbol{A}[l]\boldsymbol{U}[l], \quad 
\boldsymbol{Y}[l] = \boldsymbol{A}[l]^{\mathrm{T}}\boldsymbol{V}[l]. 
\end{equation}
\begin{itemize}
\item If $\boldsymbol{U}[l]$ has full rank, 
then the conditional distribution of $\boldsymbol{A}[l]$ given  
$\boldsymbol{X}[l]$ and $\boldsymbol{U}[l]$ is represented as 
\begin{equation} \label{conditional_A0}
\boldsymbol{A}[l]\sim \boldsymbol{X}[l]\boldsymbol{U}^{\dagger}[l] 
+ \tilde{\boldsymbol{A}}[l]\boldsymbol{P}_{\boldsymbol{U}[l]}^{\perp},
\end{equation}
where $\tilde{\boldsymbol{A}}[l]$ is independent of 
$\{\boldsymbol{X}[l], \boldsymbol{U}[l]\}$ and has independent zero-mean 
Gaussian elements with variance $(LM[l])^{-1}$. 

\item If Both $\boldsymbol{U}[l]$ and $\boldsymbol{V}[l]$ have full rank, 
then the conditional distribution of $\boldsymbol{A}[l]$ given  
$\mathfrak{E}[l]=\{\boldsymbol{X}[l], \boldsymbol{U}[l], \boldsymbol{Y}[l], 
\boldsymbol{V}[l]\}$ is represented as 
\begin{equation}
\boldsymbol{A}[l] 
\sim \boldsymbol{A}_{\mathrm{bias}}[l] 
+ \boldsymbol{P}_{\boldsymbol{V}[l]}^{\perp}\tilde{\boldsymbol{A}}[l]
\boldsymbol{P}_{\boldsymbol{U}[l]}^{\perp}, 
\end{equation}
with 
\begin{align}
\boldsymbol{A}_{\mathrm{bias}}[l] 
&= \boldsymbol{X}[l]\boldsymbol{U}^{\dagger}[l]
+ (\boldsymbol{V}^{\dagger}[l])^{\mathrm{T}}
\boldsymbol{Y}^{\mathrm{T}}[l]
\boldsymbol{P}_{\boldsymbol{U}[l]}^{\perp} 
\nonumber \\
&=  (\boldsymbol{V}^{\dagger}[l])^{\mathrm{T}}
\boldsymbol{Y}^{\mathrm{T}}[l]
+ \boldsymbol{P}_{\boldsymbol{V}[l]}^{\perp}\boldsymbol{X}[l]
\boldsymbol{U}^{\dagger}[l],  
\end{align}
where $\tilde{\boldsymbol{A}}[l]$ is independent of 
$\mathfrak{E}[l]$ and has independent zero-mean 
Gaussian elements with variance $(LM[l])^{-1}$. 
\end{itemize}
\end{lemma}

Lemma~\ref{lemma_conditioning0} and Assumption~\ref{assumption_A} imply 
that $\{\boldsymbol{A}[l]\}_{l\in\mathcal{L}}$ are conditionally independent 
given $\{\mathfrak{E}[l]\}_{l\in\mathcal{L}}$. 

The following lemma is used to design the Onsager correction in D-GAMP. 
\begin{lemma}[Stein's Lemma~\cite{Stein72,Takeuchi21}] \label{lemma_Stein}
Suppose that $\{Z_{\tau}\}_{\tau=1}^{t}$ are zero-mean Gaussian random variables. 
Then, for any piecewise Lipschitz-continuous function 
$f:\mathbb{R}^{t}\to\mathbb{R}$ we have 
\begin{equation} \label{Stein}
\mathbb{E}[Z_{t'}f(Z_{1},\ldots,Z_{t})] = 
\sum_{\tau=1}^{t}\mathbb{E}[Z_{t'}Z_{\tau}]\mathbb{E}[\partial_{\tau}f(Z_{1},
\ldots,Z_{t})],
\end{equation}
where $\partial_{\tau}$ denotes the partial derivative with respect to 
the $\tau$th variable. 
\end{lemma} 
\begin{IEEEproof}
Confirm that it is possible to replace the Lipschitz-continuity in the 
proof of \cite[Lemma~2]{Takeuchi21} with the piecewise Lipschitz-continuity, 
without any changes in the proof.  
\end{IEEEproof}

Note that any piecewise Lipschitz-continuous function $f$ is almost everywhere 
differentiable. Furthermore, the definition of the piecewise 
Lipschitz-continuity implies the bounds $|f(Z)|\leq A|Z| + B$ and 
$|f'(Z)|\leq C$ for some constants $A$, $B$, and $C$. Thus, both sides 
in (\ref{Stein}) exist for any piecewise Lipschitz-continuous function $f$.   

\subsection{State Evolution} 
We define five kinds of random variables $\{B_{t}[l], M_{t}[l]$, 
$H_{t}[l], \tilde{H}_{t}[l], Q_{t+1}[l]\}$ 
to represent the asymptotic dynamics of the general error model. Consider 
the initial condition $Q_{0}[l]=-\bar{\xi}_{\Out,0}^{-1}[l]\bar{\zeta}_{0}[l]X$, 
with $X$ defined in Assumption~\ref{assumption_x}. The random variables 
$\{B_{t}[l], M_{t}[l], H_{t}[l], \tilde{H}_{t}[l], Q_{t+1}[l]\}$ are 
defined recursively.

Let $\{B_{t}[l]\}$ denote the sequence 
of zero-mean Gaussian random variables with covariance 
\begin{equation} \label{BB_QQ}
\mathbb{E}[B_{\tau'}[l']B_{\tau}[l]]
=\frac{\delta_{l,l'}}{L\delta[l]}\mathbb{E}[Q_{\tau'}[l]Q_{\tau}[l]].
\end{equation}
The random variable $M_{t}[l]$ represents the asymptotic output of 
the outer module in iteration~$t$, given by  
\begin{equation} \label{M_t}
M_{t}[l] = f_{\Out}[l](Z_{t}[l], Y[l]; \bar{v}_{t}[l]), 
\end{equation}
with $Z[l] = -\bar{\xi}_{\Out,0}[l]\bar{\zeta}_{0}^{-1}[l]B_{0}[l]$ and 
\begin{equation} \label{Z_t}
Z_{t}[l] 
= B_{t}[l] + \frac{\bar{\zeta}_{t}[l]}{\bar{\xi}_{\Out,t}[l]}Z[l],
\end{equation}
where $\bar{\xi}_{\Out,t}[l]$, $\bar{\zeta}_{t}[l]$, and $\bar{v}_{t}[l]$ 
are given by (\ref{xi_out_bar}), (\ref{zeta_bar}), and (\ref{v_t_bar}), 
respectively. 

Similarly, let $\{H_{t}[l]\in\mathbb{R}\}$ denote zero-mean Gaussian 
random variables with covariance 
\begin{equation}
\mathbb{E}[H_{\tau'}[l']H_{\tau}[l]]
= \frac{\delta_{l,l'}}{L}\mathbb{E}[M_{\tau'}[l]M_{\tau}[l]].
\end{equation}
We define the random variable $\tilde{H}_{t}[l]$ recursively as 
\begin{align} 
\tilde{H}_{t}[l] &= \frac{H_{t}[l]}{\bar{\xi}_{\Out,t}[l]} 
+ \frac{C_{t,0}[l,l] - 1}{\underline{\bar{\xi}}_{\Out,t}[l]}
\underline{H}_{t}[l]
\nonumber \\
&+ \sum_{i=0}^{\lfloor t/T \rfloor}\sum_{l'\neq l}C_{t,i}[l,l']
\frac{\underline{H}_{t-iT}[l']}{\underline{\bar{\xi}}_{\Out,t-iT}[l']}. 
\label{H_tilde}
\end{align}
Then, the random variable $Q_{t+1}[l]$ describes 
the asymptotic output of the inner module, given by 
\begin{equation} \label{Q_t} 
Q_{t+1}[l] = f_{\In}[l]\left(
 \frac{\bar{\eta}_{t}[l]}{L}X + \tilde{H}_{t}[l]; 
 \eta_{t}[l], \bar{\sigma}_{t}^{2}[l] 
\right) 
- \frac{\bar{\zeta}_{t+1}[l]}{\bar{\xi}_{\Out,t+1}[l]}X,
\end{equation}
with $\bar{\xi}_{\Out,t}[l]$, $\bar{\zeta}_{t}[l]$, 
$\bar{\eta}_{t}[l]$ and $\bar{\sigma}_{t}^{2}[l]$ defined in 
(\ref{xi_out_bar}), (\ref{zeta_bar}), (\ref{eta_bar}) and (\ref{sigma_t_bar}). 
State evolution recursion is obtained via these random variables. 

We are ready for presenting state evolution results for the general 
error model. Note that Assumption~\ref{assumption_tree} is not required 
in state evolution analysis. Thus, the following theorem is applicable 
to general ad hoc networks. 

\begin{theorem} \label{theorem_SE_tech}
Postulate Assumptions~\ref{assumption_x}, \ref{assumption_w}, 
\ref{assumption_A}, and \ref{assumption_Lipschitz}. 
Then, the outer module satisfies the following properties 
for all $\tau\in\{0,1,\ldots\}$ in the large system limit:  
\begin{enumerate}[label=(O\alph*)]
\item \label{Oa}
Let $\boldsymbol{\beta}_{\tau}[l]
=\boldsymbol{Q}_{\tau}^{\dagger}[l]\boldsymbol{q}_{\tau}[l]$ and 
$\boldsymbol{q}_{\tau}^{\perp}[l]
=\boldsymbol{P}_{\boldsymbol{Q}_{\tau}[l]}^{\perp}\boldsymbol{q}_{\tau}[l]$. 
Then, for all $l\in\mathcal{L}$ and $\tau>0$ we have 
\begin{equation}
\boldsymbol{b}_{\tau}[l]
\sim \boldsymbol{B}_{\tau}[l]\boldsymbol{\beta}_{\tau}[l]
+ \boldsymbol{M}_{\tau}[l]\boldsymbol{o}(1)
+ \tilde{\boldsymbol{A}}[l]\boldsymbol{q}_{\tau}^{\perp}[l]
\end{equation}
conditioned on $\mathfrak{E}_{\tau,\tau}$ and $\Theta$, 
where $\{\tilde{\boldsymbol{A}}[l]\}$ 
are independent random matrices and independent of 
$\{\mathfrak{E}_{\tau,\tau}, \Theta\}$. 
Each $\tilde{\boldsymbol{A}}[l]$ has independent zero-mean Gaussian elements 
with variance $(LM[l])^{-1}$. 

\item \label{Ob}
For all $l', l\in\mathcal{L}$, and $\tau'\in\mathcal{T}_{\tau+1}[l]$, 
\begin{equation}
\frac{1}{M[l]}\boldsymbol{b}_{\tau'}^{\mathrm{T}}[l']
\boldsymbol{b}_{\tau}[l] 
- \frac{\delta_{l,l'}}{NL\delta[l]}\boldsymbol{q}_{\tau'}^{\mathrm{T}}[l]
\boldsymbol{q}_{\tau}[l] \ato 0.  
\end{equation}

\item \label{Oc}
For $\mathcal{B}_{\tau+1}[l]=\{B_{\tau'}[l]\in\mathbb{R}: 
\tau'\in\mathcal{T}_{\tau+1}[l]\}$,  
suppose that $\{\mathcal{B}_{\tau+1}[l]\}_{l\in\mathcal{L}}$
are independent with respect to $l$ and independent of 
$\{W[l]\}_{l\in\mathcal{L}}$. For each $l$, 
$\mathcal{B}_{\tau+1}[l]$ are zero-mean Gaussian random 
variables with covariance 
\begin{equation}
\mathbb{E}[B_{\tau'}[l]B_{\tau}[l]]
= \frac{1}{L\delta[l]}\mathbb{E}[Q_{\tau'}[l]Q_{\tau}[l]]
\end{equation}
for all $\tau'\in\mathcal{T}_{\tau+1}[l]$. Then, we have 
\begin{align}
&(\{\boldsymbol{b}_{\tau'}[l]: \tau'\in\mathcal{T}_{\tau+1}[l]\}_{l\in\mathcal{L}}, 
\{\boldsymbol{w}[l]\}_{l\in\mathcal{L}})
\nonumber \\
&\plto (\{\mathcal{B}_{\tau+1}[l]\}_{l\in\mathcal{L}}, 
\{W[l]\}_{l\in\mathcal{L}}),
\end{align}
\begin{equation}
\xi_{\Out,\tau}[l]\ato \bar{\xi}_{\Out,\tau}[l].
\end{equation}

\item \label{Od} 
For all $l\in\mathcal{L}$ and $\tau'\in\mathcal{T}_{\tau+1}[l]$, we have 
\begin{equation} \label{bm}
\frac{1}{M[l]}\boldsymbol{b}_{\tau'}^{\mathrm{T}}[l]
\boldsymbol{m}_{\tau}[l] 
\ato \frac{\bar{\xi}_{\Out,\tau}[l]}{L\delta[l]}
\mathbb{E}[Q_{\tau'}[l]Q_{\tau}[l]].
\end{equation}

\item \label{Oe}
For $\epsilon>0$ used in Assumptions~\ref{assumption_x} and 
\ref{assumption_w}, the vector $\boldsymbol{m}_{\tau}[l]$ has bounded 
$(2+\epsilon)$th moments in the large system limit. Furthermore, 
the minimum eigenvalue of 
$M^{-1}[l]\boldsymbol{M}_{\tau+1}^{\mathrm{T}}[l]\boldsymbol{M}_{\tau+1}[l]$ 
is strictly positive in the large system limit. 
\end{enumerate}

On the other hand, the inner module satisfies the following properties 
for all $\tau\in\{0,1,\ldots\}$ in the large system limit:  
\begin{enumerate}[label=(I\alph*)]
\item \label{Ia}
Let $\boldsymbol{\alpha}_{\tau}[l]
=\boldsymbol{M}_{\tau}^{\dagger}[l]\boldsymbol{m}_{\tau}[l]$ and 
$\boldsymbol{m}_{\tau}^{\perp}[l]
=\boldsymbol{P}_{\boldsymbol{M}_{\tau}[l]}^{\perp}\boldsymbol{m}_{\tau}[l]$. 
Then, for all $l\in\mathcal{L}$ we have 
\begin{equation} 
\boldsymbol{h}_{0}[l]
\sim \tilde{\boldsymbol{A}}^{\mathrm{T}}[l]\boldsymbol{m}_{0}[l] 
+ o(1)\boldsymbol{q}_{0}[l]
\label{h0}
\end{equation}
conditioned on $\mathfrak{E}_{1,0}$ and $\Theta$. For $\tau>0$, 
\begin{equation}
\boldsymbol{h}_{\tau}[l]
\sim \boldsymbol{H}_{\tau}[l]\boldsymbol{\alpha}_{\tau}[l]
+ \boldsymbol{Q}_{\tau+1}[l]\boldsymbol{o}(1)
+ \tilde{\boldsymbol{A}}^{\mathrm{T}}[l]\boldsymbol{m}_{\tau}^{\perp}[l] 
\label{h_t}
\end{equation}
conditioned on $\mathfrak{E}_{\tau+1,\tau}$ and $\Theta$. 
Here, $\{\tilde{\boldsymbol{A}}[l]\}$ are independent random matrices 
and independent of $\{\mathfrak{E}_{\tau+1,\tau}, \Theta\}$. 
Each $\tilde{\boldsymbol{A}}[l]$ has independent zero-mean Gaussian elements 
with variance $(LM[l])^{-1}$. 

\item \label{Ib}
For all $l', l\in\mathcal{L}$, and $\tau'\in\mathcal{T}_{\tau+1}[l]$, 
\begin{equation}
\frac{1}{N}\boldsymbol{h}_{\tau'}^{\mathrm{T}}[l']
\boldsymbol{h}_{\tau}[l] 
- \frac{\delta_{l,l'}}{LM[l]}\boldsymbol{m}_{\tau'}^{\mathrm{T}}[l]
\boldsymbol{m}_{\tau}[l] \ato 0.
\end{equation}

\item \label{Ic}
For $\mathcal{H}_{\tau+1}[l]=\{H_{\tau'}[l]\in\mathbb{R}: 
\tau'\in\mathcal{T}_{\tau+1}[l]\}$, suppose that 
$\{\mathcal{H}_{\tau+1}[l]\}_{l\in\mathcal{L}}$ are independent with respect to 
$l$ and independent of $X$. 
For each $l$, $\mathcal{H}_{\tau+1}[l]$ are zero-mean Gaussian random variables 
with covariance 
\begin{equation}
\mathbb{E}[H_{\tau'}[l]H_{\tau}[l]] 
= \frac{1}{L}\mathbb{E}[M_{\tau'}[l]M_{\tau}[l]].  
\end{equation}
Then, for all $l\in\mathcal{L}$ we have 
\begin{equation}
(\{\boldsymbol{h}_{\tau'}[l]: \tau'\in\mathcal{T}_{\tau+1}[l]\}_{l\in\mathcal{L}}, 
\boldsymbol{x})
\plto
(\{\mathcal{H}_{\tau+1}[l]\}_{l\in\mathcal{L}}, X),
\end{equation}
\begin{equation}
\xi_{\In,\tau}[l]\ato \bar{\xi}_{\In,\tau}[l]. 
\end{equation}

\item \label{Id}
For all $l\in\mathcal{L}$ and $\tau'\in\mathcal{T}_{\tau+1}[l]$, we have 
\begin{align}
&\frac{1}{N}\boldsymbol{h}_{\tau'}^{\mathrm{T}}[l]\boldsymbol{q}_{\tau+1}[l]
\ato \frac{\bar{\xi}_{\In,\tau}[l]}{L}
\nonumber \\
&\cdot\mathbb{E}\left[M_{\tau'}[l]
 \left(
  \frac{M_{\tau}[l]}{\bar{\xi}_{\Out,\tau}[l]} 
  + \frac{C_{t,0}[l,l]-1}
  {\underline{\bar{\xi}}_{\Out,\tau}[l]}\underline{M}_{\tau}[l]
 \right)
\right].
\end{align}

\item \label{Ie}
For $\epsilon>0$ used in Assumption~\ref{assumption_x}, 
the vector $\boldsymbol{q}_{\tau+1}[l]$ has bounded 
$(2+\epsilon)$th moments in the large system limit. Furthermore, 
the minimum eigenvalue of 
$N^{-1}\boldsymbol{Q}_{\tau+2}^{\mathrm{T}}[l]\boldsymbol{Q}_{\tau+2}[l]$ 
is strictly positive in the large system limit. 
\end{enumerate}
\end{theorem}
\begin{IEEEproof}
The proof by induction consists of four steps. A first step is the proof of  
Properties~\ref{Oa}--\ref{Oe} for $\tau=0$ presented in 
Appendix~\ref{proof_outer0}, while a second step is the proof of 
Properties~\ref{Ia}--\ref{Ie} for $\tau=0$ 
presented in Appendix~\ref{proof_inner0}. 

For some $t\in\mathbb{N}$, suppose that Properties~\ref{Oa}--\ref{Oe} 
and Properties~\ref{Ia}--\ref{Ie} are correct for all $\tau<t$. We need to 
prove Properties~\ref{Oa}--\ref{Oe} for $\tau=t$ as a third step. See 
Appendix~\ref{proof_outer} for the details. 

The last step is the proof of Properties~\ref{Ia}--\ref{Ie} for $\tau=t$ 
under the induction hypotheses~\ref{Oa}--\ref{Oe} and 
\ref{Ia}--\ref{Ie} for all $\tau<t$, as well as Properties~\ref{Oa}--\ref{Oe} 
for $\tau=t$ proved in the third step. See Appendix~\ref{proof_inner} 
for the details. From these four steps 
we arrive at Theorem~\ref{theorem_SE_tech}. 
\end{IEEEproof}

Theorem~\ref{theorem_SE_tech} implies that the general error 
model~(\ref{b})--(\ref{q}) satisfies the asymptotic Gaussianity in each node, 
i.e.\ Properties~\ref{Oc} and \ref{Ic}. Since no additional assumptions 
on the network are postulated, Theorem~\ref{theorem_SE_tech} is available 
as a framework to design distributed protocols for average consensus in 
general networks, instead of consensus propagation for tree-structured 
networks. 

We use Theorem~\ref{theorem_SE_tech} to prove Theorem~\ref{theorem_SE}. 

\begin{IEEEproof}[Proof of Theorem~\ref{theorem_SE}]
From Lemma~\ref{lemma_error_model} and Theorem~\ref{theorem_SE_tech}, 
it is sufficient to prove that (\ref{BB_QQ})--(\ref{Q_t}) to represent 
state evolution recursion for the general error model reduces to 
that in Theorem~\ref{theorem_SE} for D-GAMP. 
The expression~(\ref{H_covariance}) follows from Property~\ref{Ic} 
in Theorem~\ref{theorem_SE_tech}. Using (\ref{h_tilde_tmp}) and 
(\ref{h_ll'_tmp}) in Lemma~\ref{lemma_error_model}, we find the equivalence 
between $\tilde{H}_{t}[l]$ in (\ref{H_tilde_tree}) and $\tilde{H}_{t}[l]$ 
in (\ref{H_tilde}). 

We derive the covariance for $\{Z[l]\}$ and $\{Z_{t}[l]\}$. 
Using the definition of $Z_{t}[l]$ in (\ref{Z_t}) and   
$Z[l]=-\bar{\xi}_{\Out,0}[l]\bar{\zeta}_{0}^{-1}[l]B_{0}[l]$ yields 
\begin{align}
&\mathbb{E}[Z_{\tau}[l']Z_{t}[l]]
= \delta_{l,l'}\mathbb{E}\left[
 \left(
  B_{\tau}[l] - \frac{\bar{\zeta}_{\tau}[l]\bar{\xi}_{\Out,0}[l]}
  {\bar{\xi}_{\Out,\tau}[l]\bar{\zeta}_{0}[l]}B_{0}[l]
 \right)
\right. \nonumber \\
&\left.
 \cdot\left(
  B_{t}[l] - \frac{\bar{\zeta}_{t}[l]\bar{\xi}_{\Out,0}[l]}
  {\bar{\xi}_{\Out,t}[l]\bar{\zeta}_{0}[l]}B_{0}[l]
 \right)
\right]
\nonumber \\
&= \frac{\delta_{l,l'}}{L\delta[l]}\mathbb{E}\left[
 \left(
  Q_{\tau}[l] + \frac{\bar{\zeta}_{\tau}[l]}{\bar{\xi}_{\Out,\tau}[l]}X
 \right)
 \left(
  Q_{t}[l] + \frac{\bar{\zeta}_{t}[l]}{\bar{\xi}_{\Out,t}[l]}X
 \right)
\right]
\nonumber \\
&= \frac{\delta_{l,l'}}{L\delta[l]}\mathbb{E}\left[
 f_{\In,\tau-1}[l]f_{\In,t-1}[l]
\right],
\end{align}
with $f_{\In,\tau}[l]
=f_{\In}[l](L^{-1}\bar{\eta}_{\tau}[l]X + \tilde{H}_{\tau}[l]; \eta_{\tau}[l], 
\bar{\sigma}_{\tau}^{2}[l])$. Here, the second equality follows from 
Property~\ref{Oc} in Theorem~\ref{theorem_SE_tech} and 
$Q_{0}[l]=-\bar{\xi}_{\Out,0}^{-1}[l]\bar{\zeta}_{0}[l]X$. The 
last equality is obtained from the definition of $Q_{t}[l]$ in (\ref{Q_t}). 
 
Similarly, we obtain the other covariance, 
\begin{equation}
\mathbb{E}[Z[l']Z[l]]
= \frac{\delta_{l,l'}}{L\delta[l]}\mathbb{E}[X^{2}], 
\end{equation} 
\begin{align}
&\mathbb{E}[Z[l']Z_{t+1}[l]]
\nonumber \\
&= \frac{\delta_{l,l'}}{L\delta[l]}\mathbb{E}\left[
 Xf_{\In}[l]\left(
  \frac{\bar{\eta}_{t}[l]}{L}X + \tilde{H}_{t}[l]; \eta_{t}[l], 
  \bar{\sigma}_{t}^{2}[l]
 \right)
\right].
\end{align}
Thus, we arrive at Theorem~\ref{theorem_SE}. 
\end{IEEEproof}

\subsection{Outer Module for $\tau=0$} \label{proof_outer0}
\begin{IEEEproof}[Proof of Property~\ref{Ob}]
From Assumption~\ref{assumption_A} and the definition 
$\boldsymbol{b}_{0}[l]=\boldsymbol{A}[l]\boldsymbol{q}_{0}[l]$ in (\ref{b}), 
the vectors $\{\boldsymbol{b}_{0}[l]: l\in\mathcal{L}\}$ 
conditioned on $\{\boldsymbol{q}_{0}[l]: l\in\mathcal{L}\}$ are 
independent. Furthermore, $\boldsymbol{b}_{0}[l]$ conditioned on 
$\{\boldsymbol{q}_{0}[l]: l\in\mathcal{L}\}$  
has independent zero-mean Gaussian elements with variance 
$(LM[l])^{-1}\|\boldsymbol{q}_{0}[l]\|^{2}$. 
Assumption~\ref{assumption_x} implies that the variance 
$(LM[l])^{-1}\|\boldsymbol{q}_{0}[l]\|^{2}=(LM[l])^{-1}
(\bar{\xi}_{\Out,0}^{-1}[l]\bar{\zeta}_{0}[l])^{2}\|\boldsymbol{x}\|^{2}$ 
converges almost surely to 
$(L\delta[l])^{-1}(\bar{\xi}_{\Out,0}^{-1}[l]\bar{\zeta}_{0}[l])^{2}
\mathbb{E}[X^{2}]$ in the large system limit. Thus, the strong law of 
large numbers implies Property~\ref{Ob} for $\tau=0$. 
\end{IEEEproof}

\begin{IEEEproof}[Proof of Property~\ref{Oc}]
We first prove the former convergence in Property~\ref{Oc} 
with \cite[Lemma~1]{Takeuchi201}. In proving \cite[Lemma~1]{Takeuchi201}, 
pseudo-Lipschitz functions were considered. We only present the main idea 
for generalizing them to piecewise pseudo-Lipschitz functions. 

As an example, consider the expectation $\mathbb{E}[f(z)]$ of a piecewise 
pseudo-Lipschitz function $f(z)$ for an absolutely continuous random variable 
$z\in\mathbb{R}$. We separate the domain of $f$ into 
two sets: the set of all discontinuous points $\mathcal{D}$ and 
the remainder $\mathbb{R}\backslash\mathcal{D}$. By definition, 
$z$ is in $\mathbb{R}\backslash\mathcal{D}$ with probability~$1$. 
We evaluate the expectation as 
$\mathbb{E}[f(z)]=\mathbb{E}[f(z) | z\in\mathcal{D}]P(z\in\mathcal{D}) 
+ \mathbb{E}[f(z) | z\not\in\mathcal{D}]P(z\notin\mathcal{D})$. 
The former term does not contribute to the expectation, because of 
$P(z\in\mathcal{D})=0$. The latter term can be bounded in the same manner 
as in \cite[Lemma~1]{Takeuchi201} since conditioning $z\notin\mathcal{D}$ 
does not affect the distribution of absolutely continuous $z$. 
According to this argument, we can generalize \cite[Lemma~1]{Takeuchi201} 
to a lemma for piecewise pseudo-Lipschitz functions. Thus, we regard 
\cite[Lemma~1]{Takeuchi201} as a result for piecewise pseudo-Lipschitz 
functions. 

From Assumption~\ref{assumption_A} and the definition 
$\boldsymbol{b}_{0}[l]=\boldsymbol{A}[l]\boldsymbol{q}_{0}[l]$ in (\ref{b}), 
using \cite[Lemma~1]{Takeuchi201} and Property~\ref{Ob} for 
$\tau=0$ yields $\{\boldsymbol{b}_{0}[l]\}\plto \{B_{0}[l]\}$. 
Since the noise vectors $\{\boldsymbol{w}[l]\}$ are independent of 
$\{\boldsymbol{b}_{0}[l]\}$, we use Assumption~\ref{assumption_w} 
to arrive at the former convergence in Property~\ref{Oc} for $\tau=0$. 

We next prove the latter convergence in Property~\ref{Oc}. 
In proving \cite[Lemma 5]{Bayati11}, two properties of Lipschitz-continuous 
functions were used: almost everywhere differentiability and the boundedness 
of derivatives, which are satisfied for any piecewise Lipschitz-continuous 
function. Thus, \cite[Lemma 5]{Bayati11} is available for all piecewise 
Lipschitz-continuous functions. 
Since $f_{\Out}[l](0,g[l](z,w);v_{0}[l])$ is a piecewise Lipschitz-continuous 
function of $(z, w)$ with $v_{0}[l]=\bar{v}_{0}[l]$, we use 
\cite[Lemma 5]{Bayati11} and 
the former convergence in Property~\ref{Oc} for $\tau=0$ to find that 
$\xi_{\Out,0}[l]$ in (\ref{xi_out}) converges almost surely to 
$\bar{\xi}_{\Out,0}[l]$ in (\ref{xi_out_bar}). 
\end{IEEEproof}

\begin{IEEEproof}[Proof of Property~\ref{Od}]
Under Assumption~\ref{assumption_Lipschitz}, we can use 
Property~\ref{Oc} for $\boldsymbol{m}_{0}[l]
=f_{\Out}[l](\boldsymbol{0}, g[l](\boldsymbol{z}[l], \boldsymbol{w}[l]); 
v_{0}[l])$ with $v_{0}[l]=\bar{v}_{0}[l]$ and 
$\boldsymbol{z}[l]=-\bar{\xi}_{\Out,0}[l]
\bar{\zeta}_{0}^{-1}[l]\boldsymbol{b}_{0}[l]$ to obtain 
\begin{align} 
&\frac{1}{M[l]}\boldsymbol{b}_{0}^{\mathrm{T}}[l]
\boldsymbol{m}_{0}[l]
\ato \mathbb{E}\left[
 B_{0}[l]f_{\Out}[l]\left(
  0, g[l]\left(
   Z[l], W[l]
  \right); \bar{v}_{0}[l]
 \right)
\right]
\nonumber \\
&= - \frac{\bar{\xi}_{\Out,0}[l]}{\bar{\zeta}_{0}[l]}
\mathbb{E}\left[
 B_{0}[l]B_{0}[l]
\right](-\bar{\zeta}_{0}[l])
= \frac{\bar{\xi}_{\Out,0}[l]}
{L\delta[l]}\mathbb{E}[(Q_{0}[l])^{2}]
\label{bm0_tmp}
\end{align}
for $Z[l]=-\bar{\xi}_{\Out,0}[l]\bar{\zeta}_{0}^{-1}[l]B_{0}[l]$, where 
the first and second equalities in (\ref{bm0_tmp}) follow from 
Lemma~\ref{lemma_Stein} the definition of $\bar{\zeta}_{0}[l]$ in 
(\ref{zeta_bar}) and from Property~\ref{Oc} for $\tau=0$, 
respectively. Thus, Property~\ref{Od} holds for $\tau=0$. 
\end{IEEEproof}

\begin{IEEEproof}[Proof of Property~\ref{Oe}]
See \cite[Proof of Property~(A4) for $\tau=0$ in Theorem~4]{Takeuchi201}. 
According to the argument in the proof of the former property in 
Property~\ref{Oc} for $\tau=0$, we can generalize pseudo-Lipschitz functions 
in \cite{Takeuchi201} to piecewise pseudo-Lipschitz functions. 
\end{IEEEproof}

\subsection{Inner Module for $\tau=0$} \label{proof_inner0}
\begin{IEEEproof}[Proof of Property~\ref{Ia}]
We evaluate the distribution of $\boldsymbol{h}_{0}[l]$ in (\ref{h})  
conditioned on $\mathfrak{E}_{1,0}$ and $\Theta$. 
We use Lemma~\ref{lemma_conditioning0} under the constraints 
$\boldsymbol{b}_{0}[l]=\boldsymbol{A}[l]\boldsymbol{q}_{0}[l]$ 
for all $l\in\mathcal{L}$ to obtain 
\begin{equation}
\boldsymbol{A}[l]
\sim \frac{\boldsymbol{b}_{0}[l]\boldsymbol{q}_{0}^{\mathrm{T}}[l]}
{\|\boldsymbol{q}_{0}[l]\|^{2}}
- \tilde{\boldsymbol{A}}[l]
\boldsymbol{P}_{\boldsymbol{q}_{0}[l]}^{\perp}
\end{equation}
conditioned on $\mathfrak{E}_{1,0}$ and $\Theta$, in which 
$\{\tilde{\boldsymbol{A}}[l]\}$ are independent matrices and 
independent of $\{\mathfrak{E}_{1,0}$, $\Theta\}$. 
Each $\tilde{\boldsymbol{A}}[l]$ has 
independent zero-mean Gaussian elements with variance $(LM[l])^{-1}$. 
Applying this expression 
to $\boldsymbol{h}_{0}[l]$ in (\ref{h}), we have 
\begin{equation}
\boldsymbol{h}_{0}[l]
\sim 
\frac{\xi_{\Out,0}[l]}{L}\boldsymbol{q}_{0}[l]
- \frac{\boldsymbol{b}_{0}^{\mathrm{T}}[l]\boldsymbol{m}_{0}[l]}
{\|\boldsymbol{q}_{0}[l]\|^{2}}\boldsymbol{q}_{0}[l]
+ \boldsymbol{P}_{\boldsymbol{q}_{0}[l]}^{\perp}
\tilde{\boldsymbol{A}}^{\mathrm{T}}[l]\boldsymbol{m}_{0}[l]
\end{equation}
conditioned on $\mathfrak{E}_{1,0}$ and $\Theta$. 
From the definition $\boldsymbol{P}_{\boldsymbol{q}_{0}[l]}^{\perp}
=\boldsymbol{I} - \|\boldsymbol{q}_{0}[l]\|^{-2}
\boldsymbol{q}_{0}[l]\boldsymbol{q}_{0}^{\mathrm{T}}[l]$, we use 
\cite[Lemma 3(c)]{Bayati11} to have 
\begin{equation}
\boldsymbol{P}_{\boldsymbol{q}_{0}[l]}^{\perp}
\tilde{\boldsymbol{A}}^{\mathrm{T}}[l]\boldsymbol{m}_{0}[l]
\aeq \tilde{\boldsymbol{A}}^{\mathrm{T}}[l]\boldsymbol{m}_{0}[l]
+ o(1)\boldsymbol{q}_{0}[l]. 
\end{equation}

To prove Property~\ref{Ia} for $\tau=0$, it is sufficient to prove 
that $\boldsymbol{b}_{0}^{\mathrm{T}}[l]\boldsymbol{m}_{0}[l]
/\|\boldsymbol{q}_{0}[l]\|^{2}$ converges almost surely to 
$L^{-1}\xi_{\Out,0}[l]$ in the large system limit. Using Property~\ref{Od} 
for $\tau=0$ yields 
$N^{-1}\boldsymbol{b}_{0}^{\mathrm{T}}[l]\boldsymbol{m}_{0}[l]
\ato L^{-1}\bar{\xi}_{\Out,0}[l]\mathbb{E}[Q_{0}^{2}[l]]$. 
Furthermore, Assumption~\ref{assumption_x} implies 
$N^{-1}\|\boldsymbol{q}_{0}[l]\|^{2}\ato\mathbb{E}[(Q_{0}[l])^{2}]$. 
From these observations we find that 
$\boldsymbol{b}_{0}^{\mathrm{T}}[l]\boldsymbol{m}_{0}[l]
/\|\boldsymbol{q}_{0}[l]\|^{2}$ converges almost surely to 
$L^{-1}\bar{\xi}_{\Out,0}[l]$ in the large system limit, which is almost surely 
equal to $L^{-1}\xi_{\Out,0}[l]$, because of Property~\ref{Oc} for $\tau=0$. 
Thus, Property~\ref{Ia} holds for $\tau=0$. 
\end{IEEEproof}

\begin{IEEEproof}[Proof of Property~\ref{Ib}]
Property~\ref{Ia} for $\tau=0$ implies 
the conditional independence of $\{\boldsymbol{h}_{0}[l]: 
l\in\mathcal{L}\}$ given $\mathfrak{E}_{1,0}$ and $\Theta$ 
in the large system limit. Thus, we use Property~\ref{Ia} for $\tau=0$ 
and \cite[Lemma~3]{Takeuchi21} to obtain 
\begin{equation}
\frac{1}{N}\boldsymbol{h}_{0}^{\mathrm{T}}[l']
\boldsymbol{h}_{0}[l]
\aeq \frac{\delta_{l,l'}}{LM[l]}\|\boldsymbol{m}_{0}[l]\|^{2} + o(1).
\end{equation}
Thus, Property~\ref{Ib} holds for $\tau=0$. 
\end{IEEEproof}

\begin{IEEEproof}[Proof of Property~\ref{Ic}]
We find the almost sure convergence $\sigma_{0}^{2}[l]\ato
\bar{\sigma}_{0}^{2}[l]$ for (\ref{sigma_t}) and (\ref{sigma_t_bar}), 
because of Property~\ref{Oc} for $\tau=0$.  
We only prove the former convergence because the latter convergence 
$\xi_{\In,0}[l]\ato\bar{\xi}_{\In,0}[l]$ follows from the former convergence 
in Property~\ref{Ic} for $\tau=0$, $\sigma_{0}^{2}[l]\ato
\bar{\sigma}_{0}^{2}[l]$, and \cite[Lemma~5]{Bayati11}. 
Using Properties~\ref{Ia} and \ref{Ib} for $\tau=0$, 
Assumption~\ref{assumption_x}, and \cite[Lemma 1]{Takeuchi201}, we 
obtain the convergence $(\{\boldsymbol{h}_{0}[l]\}_{l\in\mathcal{L}}, 
\boldsymbol{x})\plto(\{H_{0}[l]\}_{l\in\mathcal{L}}, X)$, 
in which $\{H_{0}[l]\}$ are independent of $X$ and zero-mean Gaussian random 
variables with covariance 
$\mathbb{E}[H_{0}[l]H_{0}[l']]=L^{-1}\delta_{l,l'}\mathbb{E}[M_{0}^{2}[l]]$. 
Thus, Property~\ref{Ic} holds for $\tau=0$.
\end{IEEEproof}

\begin{IEEEproof}[Proof of Property~\ref{Id}]
Using the definition of $\boldsymbol{q}_{1}[l]$ in (\ref{q}) yields 
\begin{align}
&\frac{1}{N}\boldsymbol{h}_{0}^{\mathrm{T}}[l]\boldsymbol{q}_{1}[l]
= \frac{1}{N}\boldsymbol{h}_{0}^{\mathrm{T}}[l]
f_{\In}[l]\left(
 \frac{\bar{\eta}_{0}[l]}{L}\boldsymbol{x} 
 + \tilde{\boldsymbol{h}}_{0}[l]; \eta_{0}[l], \sigma_{0}^{2}[l]
\right) 
\nonumber \\
&- \frac{\bar{\zeta}_{1}[l]}{\bar{\xi}_{\Out,1}[l]}
\frac{1}{N}\boldsymbol{h}_{0}^{\mathrm{T}}[l]\boldsymbol{x}
\nonumber \\
&\ato \mathbb{E}\left[
 H_{0}[l]f_{\In}[l]\left(
  \frac{\bar{\eta}_{0}[l]}{L}X + \tilde{H}_{0}[l]; 
  \eta_{t}[l], \bar{\sigma}_{0}^{2}[l]
 \right)
\right] 
\nonumber \\
&- \frac{\bar{\zeta}_{1}[l]}{\bar{\xi}_{\Out,1}[l]}\mathbb{E}[H_{0}[l]X], 
\end{align}
where the last follows from the definition of $\tilde{\boldsymbol{h}}_{0}[l]$ 
in (\ref{h_tilde}) and Property~\ref{Ic} for $\tau=0$. 
Since we have $\mathbb{E}[H_{0}[l]X]=\mathbb{E}[H_{0}[l]]
\mathbb{E}[X]=0$ in the second term, we use the definition of 
$\tilde{H}_{0}[l]$ in (\ref{H_tilde}) and 
Lemma~\ref{lemma_Stein} to evaluate the first term as
\begin{align}
&\mathbb{E}\left[
 H_{0}[l]f_{\In}[l](L^{-1}\bar{\eta}_{0}[l]X + \tilde{H}_{0}[l]; \eta_{0}[l], 
 \bar{\sigma}_{0}^{2}[l])
\right]
\nonumber \\
&= \frac{\bar{\xi}_{\In,0}[l]}{\bar{\xi}_{\Out,0}[l]}C_{0,0}[l,l]
\mathbb{E}[H_{0}^{2}[l]] 
\nonumber \\
&+ \bar{\xi}_{\In,0}[l]\sum_{l'\neq l}
\frac{C_{0,0}[l,l']}{\underline{\bar{\xi}}_{\Out,0}[l']}
\mathbb{E}[H_{0}[l]H_{0}[l']]
\nonumber \\
&= C_{0,0}[l,l]\frac{\bar{\xi}_{\In,0}[l]}{L\bar{\xi}_{\Out,0}[l]}
\mathbb{E}[M_{0}^{2}[l]],
\end{align}
where the last equality follows from Property~\ref{Ib} for $\tau=0$.  
Thus, we arrive at Property~\ref{Id} for $\tau=0$. 
\end{IEEEproof}

\begin{IEEEproof}[Proof of Property~\ref{Ie}]
See \cite[Proof of Property~(B4) for $\tau=0$ in Theorem~4]{Takeuchi201}. 
\end{IEEEproof}

\subsection{Outer Module for $\tau>0$} \label{proof_outer}
For some $t\in\mathbb{N}$, suppose that Properties~\ref{Oa}--\ref{Oe} 
and Properties~\ref{Ia}--\ref{Ie} are correct for all $\tau<t$. 
We prove Properties~\ref{Oa}--\ref{Oe} for $\tau=t$ under these 
induction hypotheses. 

\begin{IEEEproof}[Proof of Property~\ref{Oa}]
Let 
\begin{equation}
\boldsymbol{\Lambda}_{\In,t}[l]
= \mathrm{diag}\left\{
 \frac{\xi_{\In,\tau}[l]}{L\delta[l]\xi_{\Out,\tau}[l]}: \tau\in\mathcal{T}_{t}[l]
\right\},
\end{equation} 
\begin{equation}
\underline{\boldsymbol{\Lambda}}_{\In,t}[l]
= (C_{t,0}[l,l] - 1)\mathrm{diag}\left\{
 \frac{\xi_{\In,\tau}[l]}
 {L\delta[l]\underline{\xi}_{\Out,\tau}[l]}: \tau\in\mathcal{T}_{t}[l]
\right\},
\end{equation} 
\begin{equation}
\boldsymbol{\Lambda}_{\Out,t}[l]
=\mathrm{diag}\left\{
 \frac{\xi_{\Out,\tau}[l]}{L}: \tau\in\mathcal{T}_{t}[l]
\right\}.
\end{equation} 
From the definitions of $\boldsymbol{b}_{t}[l]$ and 
$\boldsymbol{h}_{t}[l]$ in (\ref{b}) and (\ref{h}), 
$\boldsymbol{A}[l]$ conditioned on $\mathfrak{E}_{t,t}$ and $\Theta$ 
satisfies the following constraints just before 
updating $\boldsymbol{b}_{t}[l]$ for all $l\in\mathcal{L}$:
\begin{align}
&\boldsymbol{B}_{t}[l] - \left[
 \boldsymbol{0}, \boldsymbol{M}_{t-1}[l]\boldsymbol{\Lambda}_{\In,t-1}[l] 
 + \underline{\boldsymbol{M}}_{t-1}[l]
 \underline{\boldsymbol{\Lambda}}_{\In,t-1}[l] 
\right]
\nonumber \\
&= \boldsymbol{A}[l]\boldsymbol{Q}_{t}[l], 
\end{align}
\begin{equation}
\boldsymbol{Q}_{t}[l]\boldsymbol{\Lambda}_{\Out,t}[l] 
- \boldsymbol{H}_{t}[l]
= \boldsymbol{A}^{\mathrm{T}}[l]\boldsymbol{M}_{t}[l]
\end{equation}
for all $l\in\mathcal{L}$. 

We let $\boldsymbol{U}[l]=\boldsymbol{Q}_{t}[l]$ and 
$\boldsymbol{V}[l]=\boldsymbol{M}_{t}[l]$ in Lemma~\ref{lemma_conditioning0}. 
Since the induction hypotheses~\ref{Oe} and \ref{Ie} for $\tau=t-1$ imply 
that $\boldsymbol{M}_{t}[l]$ and $\boldsymbol{Q}_{t}[l]$ have full rank, 
we can use Lemma~\ref{lemma_conditioning0} to obtain  
\begin{align}
\boldsymbol{A}[l] 
&\sim
- \left[
 \boldsymbol{0},\boldsymbol{M}_{t-1}[l] 
 \boldsymbol{\Lambda}_{\In,t-1}[l]
 - \underline{\boldsymbol{M}}_{t-1}[l]
 \underline{\boldsymbol{\Lambda}}_{\In,t-1}[l]
\right]\boldsymbol{Q}_{t}^{\dagger}[l]
\nonumber \\
&+ \boldsymbol{B}_{t}[l]\boldsymbol{Q}_{t}^{\dagger}[l]
- (\boldsymbol{M}_{t}^{\dagger}[l])^{\mathrm{T}}
\boldsymbol{H}_{t}^{\mathrm{T}}[l]\boldsymbol{P}_{\boldsymbol{Q}_{t}[l]}^{\perp}
\nonumber \\
&+ \boldsymbol{P}_{\boldsymbol{M}_{t}[l]}^{\perp}\tilde{\boldsymbol{A}}[l]
\boldsymbol{P}_{\boldsymbol{Q}_{t}[l]}^{\perp}
\end{align}
conditioned on $\mathfrak{E}_{t,t}$ and $\Theta$, where we have used 
$\boldsymbol{Q}_{t}^{\mathrm{T}}[l]
\boldsymbol{P}_{\boldsymbol{Q}_{t}[l]}^{\perp}=\boldsymbol{O}$. 
Here, $\{\tilde{\boldsymbol{A}}[l]\}$ are independent matrices and 
independent of $\{\mathfrak{E}_{t,t}$, $\Theta\}$. 
Each $\tilde{\boldsymbol{A}}[l]$ has 
independent zero-mean Gaussian elements with variance $(LM[l])^{-1}$. 

Let $\boldsymbol{\beta}_{t}[l]
=\boldsymbol{Q}_{t}^{\dagger}[l]\boldsymbol{q}_{t}[l]$ and 
$\boldsymbol{q}_{t}^{\perp}[l]
=\boldsymbol{P}_{\boldsymbol{Q}_{t}[l]}^{\perp}\boldsymbol{q}_{t}[l]$. 
Furthermore, we write the $\tau$th element of $\boldsymbol{\beta}_{t}[l]$ as 
$\beta_{t,\tau}[l]$. Substituting the expression of $\boldsymbol{A}[l]$ 
into the definition of $\boldsymbol{b}_{t}[l]$ in (\ref{b}) yields 
\begin{align}
&\boldsymbol{b}_{t}[l] 
\sim \boldsymbol{B}_{t}[l]\boldsymbol{\beta}_{t}[l]
-\sum_{\tau\in\mathcal{T}_{t}[l]\backslash\{0\}}
\frac{\beta_{t,\tau}[l]\xi_{\In,\tau-1}[l]}
{L\delta[l]\xi_{\Out,\tau-1}[l]}
\boldsymbol{m}_{\tau-1}[l]
\nonumber \\ 
&- (C_{t,0}[l,l]-1)\sum_{\tau\in\mathcal{T}_{t}[l]\backslash\{0\}}
\beta_{t,\tau}[l]\frac{\xi_{\In,\tau-1}[l]}
{L\delta[l]\underline{\xi}_{\Out,\tau-1}[l]}
\underline{\boldsymbol{m}}_{\tau-1}[l]
\nonumber \\
&- (\boldsymbol{M}_{t}^{\dagger}[l])^{\mathrm{T}}
\boldsymbol{H}_{t}^{\mathrm{T}}[l]\boldsymbol{q}_{t}^{\perp}[l]
+ \boldsymbol{P}_{\boldsymbol{M}_{t}[l]}^{\perp}\tilde{\boldsymbol{A}}[l]
\boldsymbol{q}_{t}^{\perp}[l]
\nonumber \\
&+ \frac{\xi_{\In,t-1}[l]}{L\delta[l]}
\left\{
 \frac{\boldsymbol{m}_{t-1}[l]}{\xi_{\Out,t-1}[l]} 
 + \frac{C_{t,0}[l,l]-1}{\underline{\xi}_{\Out,t-1}[l]}
 \underline{\boldsymbol{m}}_{t-1}[l]
\right\}
\label{b_tmp}
\end{align}
conditioned on $\mathfrak{E}_{t,t}$ and $\Theta$. 

We simplify the expression of $\boldsymbol{b}_{t}[l]$. 
Using $\boldsymbol{q}_{t}^{\perp}[l]
= \boldsymbol{q}_{t}[l]- \boldsymbol{Q}_{t}[l]\boldsymbol{\beta}_{t}[l]$  
and the induction hypothesis \ref{Id} for $\tau=t-1$ yields 
\begin{align}
&(\boldsymbol{M}_{t}^{\dagger}[l])^{\mathrm{T}}
\boldsymbol{H}_{t}^{\mathrm{T}}[l]\boldsymbol{q}_{t}^{\perp}[l]
\nonumber \\
&\aeq \frac{\bar{\xi}_{\In,t-1}[l]}{L\delta[l]}
\left\{
 \frac{\boldsymbol{m}_{t-1}[l]}{\bar{\xi}_{\Out,t-1}[l]} 
 + \frac{C_{t,0}[l,l]-1}{\underline{\bar{\xi}}_{\Out,t-1}[l]}
 \underline{\boldsymbol{m}}_{t-1}[l]
\right\}
\nonumber \\
&+ \boldsymbol{M}_{t}[l]\boldsymbol{o}(1)
- \sum_{\tau\in\mathcal{T}_{t}[l]\backslash\{0\}}
\frac{\beta_{t,\tau}[l]\bar{\xi}_{\In,\tau-1}[l]}
{L\delta[l]\bar{\xi}_{\Out,\tau-1}[l]}
\boldsymbol{m}_{\tau-1}[l] 
\nonumber \\
&- (C_{t,0}[l,l]-1)\sum_{\tau\in\mathcal{T}_{t}[l]\backslash\{0\}}
\frac{\beta_{t,\tau}[l]\bar{\xi}_{\In,\tau-1}[l]}
{L\delta[l]\underline{\bar{\xi}}_{\tau-1}[l]}
\underline{\boldsymbol{m}}_{\tau-1}[l]. 
\end{align}
Substituting this expression into the representation of $\boldsymbol{b}_{t}[l]$ 
in (\ref{b_tmp}) and using the induction hypotheses~\ref{Oc} and 
\ref{Ic} for all $\tau<t$, we arrive at 
\begin{equation}
\boldsymbol{b}_{t}[l] 
\sim \boldsymbol{B}_{t}[l]\boldsymbol{\beta}_{t}[l]
+ \boldsymbol{M}_{t}[l]\boldsymbol{o}(1)
+ \boldsymbol{P}_{\boldsymbol{M}_{t}[l]}^{\perp}\tilde{\boldsymbol{A}}[l]
\boldsymbol{q}_{t}^{\perp}[l]
\end{equation}
conditioned on $\mathfrak{E}_{t,t}$ and $\Theta$, which is equivalent 
to Property~\ref{Oa} for $\tau=t$, because 
$\boldsymbol{P}_{\boldsymbol{M}_{t}[l]}^{\perp}\tilde{\boldsymbol{A}}[l]
\boldsymbol{q}_{t}^{\perp}[l]
=\tilde{\boldsymbol{A}}[l]
\boldsymbol{q}_{t}^{\perp}[l] + \boldsymbol{M}_{t}[l]\boldsymbol{o}(1)$ 
holds due to \cite[Lemma 3(c)]{Bayati11}. 
\end{IEEEproof}

\begin{IEEEproof}[Proof of Property~\ref{Ob}]
For all $\tau'<t$, using Property~\ref{Oa} for $\tau=t$ yields 
\begin{align}
\frac{1}{M[l]}\boldsymbol{b}_{\tau'}^{\mathrm{T}}[l']\boldsymbol{b}_{t}[l]
&\aeq \frac{1}{M[l]}\boldsymbol{b}_{\tau'}^{\mathrm{T}}[l']
\boldsymbol{B}_{t}[l]\boldsymbol{\beta}_{t}[l] + o(1)
\nonumber \\
&\aeq \frac{\delta_{l,l'}}{NL\delta[l]}
\boldsymbol{q}_{\tau'}^{\mathrm{T}}[l]
\boldsymbol{Q}_{t}[l]\boldsymbol{\beta}_{t}[l]
+ o(1)
\nonumber \\
&= \frac{\delta_{l,l'}}{NL\delta[l]}
\boldsymbol{q}_{\tau'}^{\mathrm{T}}[l]\boldsymbol{q}_{t}[l]
+ o(1),
\end{align}
where the second and last equalities follow from the 
induction hypothesis~\ref{Ob} for $\tau<t$ 
and $\boldsymbol{\beta}_{t}[l]
=\boldsymbol{Q}_{t}^{\dagger}[l]\boldsymbol{q}_{t}[l]$, respectively. 

For $\tau'=t$ we use Property~\ref{Oa} for $\tau=t$ and 
\cite[Lemma~3]{Takeuchi21} to obtain 
\begin{align}
\frac{\boldsymbol{b}_{t}^{\mathrm{T}}[l']\boldsymbol{b}_{t}[l]}{M[l]}
&\aeq \frac{1}{M[l]}
\boldsymbol{\beta}_{t}^{\mathrm{T}}[l']\boldsymbol{B}_{t}^{\mathrm{T}}[l']
\boldsymbol{B}_{t}[l]\boldsymbol{\beta}_{t}[l]
\nonumber \\
&+ \frac{\delta_{l,l'}}{NL\delta[l]}\|\boldsymbol{q}_{t}^{\perp}[l]\|^{2}
+ o(1)
\nonumber \\
&\aeq \frac{\delta_{l,l'}}{NL\delta[l]}
(\|\boldsymbol{q}_{t}^{\parallel}[l]\|^{2}
+ \|\boldsymbol{q}_{t}^{\perp}[l]\|^{2})
+ o(1)
\nonumber \\
&\aeq \frac{\delta_{l,l'}}{NL\delta[l]}\|\boldsymbol{q}_{t}[l]\|^{2} + o(1), 
\end{align}
with $\boldsymbol{q}_{t}^{\parallel}[l]=\boldsymbol{P}_{\boldsymbol{Q}_{t}[l]}^{\parallel}
\boldsymbol{q}_{t}[l]$, 
where the second equality follows from the induction hypothesis~\ref{Ob} 
for $\tau<t$ and the definition $\boldsymbol{\beta}_{t}[l]
=\boldsymbol{Q}_{t}^{\dagger}[l]\boldsymbol{q}_{t}[l]$. 
Thus, Property~\ref{Ob} holds for $\tau=t$. 
\end{IEEEproof}

\begin{IEEEproof}[Proof of Property~\ref{Oc}]
The former convergence in Property~\ref{Oc} for $\tau=t$ follows from 
Property~\ref{Ob} for $\tau=t$, \cite[Lemma~1]{Takeuchi201}, and 
Assumption~\ref{assumption_w}. 
We find the convergence $v_{t}[l]\ato\bar{v}_{t}[l]$ for 
(\ref{v_t}) and (\ref{v_t_bar}) from the 
induction hypotheses~\ref{Oc} and \ref{Ic} for all $\tau<t$. 
The latter convergence $\xi_{\Out,t}[l]\ato\bar{\xi}_{\Out,t}[l]$ follows 
from the former convergence in Property~\ref{Oc} for $\tau=t$, 
$v_{t}[l]\ato\bar{v}_{t}[l]$, and \cite[Lemma~5]{Bayati11}. 
\end{IEEEproof}

\begin{IEEEproof}[Proof of Property~\ref{Od}]
Using the definition of $\boldsymbol{m}_{t}[l]$ in (\ref{m}) and 
Property~\ref{Oc} for $\tau=t$ yields 
\begin{align}
&\frac{1}{M[l]}\boldsymbol{b}_{\tau'}^{\mathrm{T}}[l]\boldsymbol{m}_{t}[l]
\nonumber \\
&\ato \mathbb{E}\left[
 B_{\tau'}[l]f_{\Out}[l]\left(
  B_{t}[l] + \frac{\bar{\zeta}_{t}[l]}{\bar{\xi}_{\Out,t}[l]}Z[l], Y[l]; 
  \bar{v}_{t}[l] 
 \right)
\right]
\nonumber \\
&= \bar{\xi}_{\Out,t}[l]\mathbb{E}[B_{\tau'}[l]B_{t}[l]] 
\nonumber \\
&+ \left(
 \frac{\bar{\zeta}_{t}[l]}{\bar{\xi}_{\Out,t}[l]}\bar{\xi}_{\Out,t}[l] 
 - \bar{\zeta}_{t}[l] 
\right)\mathbb{E}[B_{\tau'}[l]Z[l]]
\nonumber \\
&= \frac{\bar{\xi}_{\Out,t}[l]}{L\delta[l]}
\mathbb{E}[Q_{\tau'}[l]Q_{t}[l]],
\end{align}
with $\bar{\zeta}_{t}[l]$ given in (\ref{zeta_bar}), 
where the first and second equalities follow from Lemma~\ref{lemma_Stein} 
and $Y[l]=g[l](Z[l],W[l])$ and from Property~\ref{Oc} for $\tau=t$, 
respectively. Thus, Property~\ref{Od} holds for $\tau=t$. 
\end{IEEEproof}

\begin{IEEEproof}[Proof of Property~\ref{Oe}]
See \cite[Proof of Property~(A4) for $\tau=t$ in Theorem~4]{Takeuchi201}. 
\end{IEEEproof}

\subsection{Inner Module for $\tau>0$} \label{proof_inner}
Suppose that Properties~\ref{Oa}--\ref{Oe} and 
\ref{Ia}--\ref{Ie} for all $\tau<t$ are correct. To complete the proof of 
Theorem~\ref{theorem_SE_tech} by induction, we prove 
Properties~\ref{Ia}--\ref{Ie} for $\tau=t$ under these induction hypotheses, 
as well as Properties~\ref{Oa}--\ref{Oe} for $\tau=t$. 

\begin{IEEEproof}[Proof of Property~\ref{Ia}]
From the definitions of $\boldsymbol{b}_{t}[l]$ and 
$\boldsymbol{h}_{t}[l]$ in (\ref{b}) and (\ref{h}), 
$\boldsymbol{A}[l]$ conditioned on $\mathfrak{E}_{t+1,t}$ and $\Theta$ 
satisfies the following constraints just before 
updating $\boldsymbol{h}_{t}[l]$: 
\begin{equation}
\boldsymbol{B}_{t+1}[l]
- \left[
 \boldsymbol{0},\boldsymbol{M}_{t}[l]\boldsymbol{\Lambda}_{\In,t}[l] 
 + \underline{\boldsymbol{M}}_{t}[l]
 \underline{\boldsymbol{\Lambda}}_{\In,t}[l]
\right]
=\boldsymbol{A}[l]\boldsymbol{Q}_{t+1}[l], 
\end{equation}
\begin{equation}
\boldsymbol{Q}_{t}[l]\boldsymbol{\Lambda}_{\Out,t}[l] 
- \boldsymbol{H}_{t}[l] 
= \boldsymbol{A}^{\mathrm{T}}[l]\boldsymbol{M}_{t}[l]
\end{equation}
for all $l\in\mathcal{L}$. Since the induction 
hypotheses~\ref{Oe} and \ref{Ie} for $\tau=t-1$ imply 
that $\boldsymbol{M}_{t}[l]$ and $\boldsymbol{Q}_{t+1}[l]$ have full rank, 
we use Lemma~\ref{lemma_conditioning0} to obtain 
\begin{align} 
\boldsymbol{A}[l]
&\sim (\boldsymbol{M}_{t}^{\dagger}[l])^{\mathrm{T}}
(\boldsymbol{Q}_{t}[l]\boldsymbol{\Lambda}_{\Out,t}[l] - \boldsymbol{H}_{t}[l]
)^{\mathrm{T}}
\nonumber \\
&+ \boldsymbol{P}_{\boldsymbol{M}_{t}[l]}^{\perp}
\boldsymbol{B}_{t+1}[l]\boldsymbol{Q}_{t+1}^{\dagger}[l] 
- \boldsymbol{P}_{\boldsymbol{M}_{t}[l]}^{\perp}
\tilde{\boldsymbol{A}}[l]\boldsymbol{P}_{\boldsymbol{Q}_{t+1}[l]}^{\perp}
\end{align} 
conditioned on $\mathfrak{E}_{t+1,t}$ and $\Theta$. Here,
$\{\tilde{\boldsymbol{A}}[l]\}$ are independent matrices and 
independent of $\{\mathfrak{E}_{t+1,t}$, $\Theta\}$. 
Each $\tilde{\boldsymbol{A}}[l]$ has independent zero-mean Gaussian 
elements with variance $(LM[l])^{-1}$. 

Let $\boldsymbol{\alpha}_{t}[l]=\boldsymbol{M}_{t}^{\dagger}[l]
\boldsymbol{m}_{t}[l]$ and 
$\boldsymbol{m}_{t}^{\perp}[l]=\boldsymbol{P}_{\boldsymbol{M}_{t}[l]}^{\perp}
\boldsymbol{m}_{t}[l]$. 
Substituting the expression of $\boldsymbol{A}[l]$ into 
the definition of $\boldsymbol{h}_{t}[l]$ in (\ref{h}), we have 
\begin{equation}
\boldsymbol{h}_{t}[l] 
\sim \boldsymbol{H}_{t}[l]\boldsymbol{\alpha}_{t}[l] 
+ \boldsymbol{h}_{t}^{\mathrm{bias}}[l]
+ \boldsymbol{P}_{\boldsymbol{Q}_{t+1}[l]}^{\perp}
\tilde{\boldsymbol{A}}^{\mathrm{T}}[l]\boldsymbol{m}_{t}^{\perp}[l] 
\end{equation}
conditioned on $\mathfrak{E}_{t+1,t}$ and $\Theta$, with 
\begin{align}
&\boldsymbol{h}_{t}^{\mathrm{bias}}[l] 
= \frac{\xi_{\Out,t}[l]}{L}\boldsymbol{q}_{t}[l]
- \boldsymbol{Q}_{t}[l]\boldsymbol{\Lambda}_{\Out,t}[l]\boldsymbol{\alpha}_{t}[l] 
\nonumber \\
&- (\boldsymbol{Q}_{t+1}^{\dagger}[l])^{\mathrm{T}}\boldsymbol{B}_{t+1}^{\mathrm{T}}[l]
\boldsymbol{m}_{t}[l]
+ (\boldsymbol{Q}_{t+1}^{\dagger}[l])^{\mathrm{T}}\boldsymbol{B}_{t+1}^{\mathrm{T}}[l]
\boldsymbol{M}_{t}[l]\boldsymbol{\alpha}_{t}[l], 
\label{h_bias_l}
\end{align}
where we have used $\boldsymbol{m}_{t}^{\perp}[l]=\boldsymbol{m}_{t}[l]
-\boldsymbol{M}_{t}[l]\boldsymbol{\alpha}_{t}[l]$. 
In particular, we use \cite[Lemma 3(c)]{Bayati11} to obtain 
\begin{equation}
\boldsymbol{P}_{\boldsymbol{Q}_{t+1}[l]}^{\perp}
\tilde{\boldsymbol{A}}^{\mathrm{T}}[l]
\boldsymbol{m}_{t}^{\perp}[l]
\aeq \tilde{\boldsymbol{A}}^{\mathrm{T}}[l]
\boldsymbol{m}_{t}^{\perp}[l]
+ \boldsymbol{Q}_{t+1}[l]\boldsymbol{o}(1).
\end{equation}
Thus, it is sufficient to prove 
$\boldsymbol{h}_{t}^{\mathrm{bias}}[l]\aeq
\boldsymbol{Q}_{t+1}[l]\boldsymbol{o}(1)$. 

We evaluate $\boldsymbol{h}_{t}^{\mathrm{bias}}[l]$. 
Using Property~\ref{Od} for $\tau=t$ yields 
\begin{align}
&(\boldsymbol{Q}_{t+1}^{\dagger}[l])^{\mathrm{T}}\boldsymbol{B}_{t+1}^{\mathrm{T}}[l]
\boldsymbol{m}_{t}[l]
\nonumber \\
&\aeq \frac{\bar{\xi}_{\Out,t}[l]}{L}
(\boldsymbol{Q}_{t+1}^{\dagger}[l])^{\mathrm{T}}
\boldsymbol{Q}_{t+1}^{\mathrm{T}}[l]\boldsymbol{q}_{t}[l] 
+ \boldsymbol{Q}_{t+1}[l]\boldsymbol{o}(1)
\nonumber \\
&= \frac{\bar{\xi}_{\Out,t}[l]}{L}\boldsymbol{q}_{t}[l] 
+ \boldsymbol{Q}_{t+1}[l]\boldsymbol{o}(1). 
\end{align}
Similarly, we obtain 
\begin{align}
&(\boldsymbol{Q}_{t+1}^{\dagger}[l])^{\mathrm{T}}\boldsymbol{B}_{t+1}^{\mathrm{T}}[l]
\boldsymbol{M}_{t}[l]\boldsymbol{\alpha}_{t}[l] 
\nonumber \\
&= \sum_{\tau\in\mathcal{T}_{t}[l]}
\alpha_{t,\tau}[l]
(\boldsymbol{Q}_{t+1}^{\dagger}[l])^{\mathrm{T}}\boldsymbol{B}_{t+1}^{\mathrm{T}}[l]
\boldsymbol{m}_{\tau}[l]
\nonumber \\
&\aeq \sum_{\tau\in\mathcal{T}_{t}[l]}\alpha_{t,\tau}[l] 
\frac{\bar{\xi}_{\Out,\tau}[l]}{L}\boldsymbol{q}_{\tau}[l] 
+ \boldsymbol{Q}_{t+1}[l]\boldsymbol{o}(1),
\end{align}
with $\alpha_{t,\tau}[l]$ denoting the $\tau$th element of 
$\boldsymbol{\alpha}_{t}[l]$. 
Substituting these expressions into the definition of 
$\boldsymbol{h}_{t}^{\mathrm{bias}}[l]$ in (\ref{h_bias_l}) and using 
Property~\ref{Oc} for all $\tau\leq t$, 
we arrive at $\boldsymbol{h}_{t}^{\mathrm{bias}}[l]\aeq
\boldsymbol{Q}_{t+1}[l]\boldsymbol{o}(1)$. Thus, 
Property~\ref{Ia} holds for $\tau=t$. 
\end{IEEEproof}

\begin{IEEEproof}[Proof of Property~\ref{Ib}]
Property~\ref{Ia} for $\tau=t$ implies the conditional independence of 
$\{\boldsymbol{h}_{t}[l]: l\in\mathcal{L}\}$ given 
$\mathfrak{E}_{t+1,t}$ and $\Theta$. We first consider the case of 
$\tau'<t$. Using Property~\ref{Ia} 
for $\tau=t$ and the induction hypothesis~\ref{Ib} for $\tau<t$ yields 
\begin{align}
\frac{1}{N}\boldsymbol{h}_{\tau'}^{\mathrm{T}}[l']
\boldsymbol{h}_{t}[l] 
&\aeq \frac{\delta_{l,l'}}{LM[l]}\boldsymbol{m}_{\tau'}^{\mathrm{T}}[l]
\boldsymbol{M}_{t}[l]\boldsymbol{\alpha}_{t}[l] + o(1)
\nonumber \\
&= \frac{\delta_{l,l'}}{LM[l]}\boldsymbol{m}_{\tau'}^{\mathrm{T}}[l]
\boldsymbol{m}_{t}[l] + o(1)
\label{h_tau'_t_divisible}
\end{align}
for all $\tau'<t$, where the last equality follows from the definition 
$\boldsymbol{\alpha}_{t}[l]=\boldsymbol{M}_{t}^{\dagger}[l]\boldsymbol{m}_{t}[l]$. 

We next consider the case of $\tau'=t$. Using Property~\ref{Ia} for $\tau=t$ 
and the induction hypothesis~\ref{Ib} for $\tau<t$ yields 
\begin{align}
\frac{\boldsymbol{h}_{t}^{\mathrm{T}}[l']
\boldsymbol{h}_{t}[l]}{N} 
&\aeq \frac{\delta_{l,l'}}{LM[l]}
\boldsymbol{\alpha}_{t}^{\mathrm{T}}[l]
\boldsymbol{M}_{t}^{\mathrm{T}}[l]\boldsymbol{M}_{t}[l]
\boldsymbol{\alpha}_{t}[l] 
\nonumber \\
&+ \frac{\delta_{l,l'}}{LM[l]}\|\boldsymbol{m}_{t}^{\perp}[l]\|^{2}
+ o(1)
\nonumber \\
&= \frac{\delta_{l,l'}}{LM[l]}\left(
 \|\boldsymbol{P}_{\boldsymbol{M}_{t}[l]}^{\parallel}\boldsymbol{m}_{t}[l]\|^{2}
 + \|\boldsymbol{m}_{t}^{\perp}[l]\|^{2}
\right) + o(1) 
\nonumber \\
&= \frac{\delta_{l,l'}}{LM[l]}\|\boldsymbol{m}_{t}[l]\|^{2} + o(1). 
\label{h_t_t_divisible}
\end{align}
Thus, Property~\ref{Ib} holds for $\tau=t$. 
\end{IEEEproof}

\begin{IEEEproof}[Proof of Property~\ref{Ic}]
The former convergence in Property~\ref{Ic} for $\tau=t$ follows from 
Assumption~\ref{assumption_x}, Properties~\ref{Ia} and \ref{Ib} 
for $\tau=t$, and \cite[Lemma 1]{Takeuchi201}.  
We find the convergence $\sigma_{t}^{2}[l]\ato\bar{\sigma}_{t}^{2}[l]$ 
for (\ref{sigma_t}) and (\ref{sigma_t_bar}) 
from Property~\ref{Oc} for $\tau=t$. 
The latter convergence $\xi_{\In,t}[l]\ato\bar{\xi}_{\In,t}[l]$ is obtained 
from the former convergence in Property~\ref{Ic} for $\tau=t$, 
$\sigma_{t}^{2}[l]\ato\bar{\sigma}_{t}^{2}[l]$, and \cite[Lemma~5]{Bayati11}. 
\end{IEEEproof}

\begin{IEEEproof}[Proof of Property~\ref{Id}]
Using the definition of $\boldsymbol{q}_{t+1}[l]$ 
in (\ref{q}) and Property~\ref{Ic} for $\tau=t$ yields 
\begin{align}
&\frac{\boldsymbol{h}_{\tau'}^{\mathrm{T}}[l]\boldsymbol{q}_{t+1}[l]}{N}
\ato \mathbb{E}\left[
 H_{\tau'}[l]f_{\In}[l]\left(
  \frac{\bar{\eta}_{t}}{L}X + \tilde{H}_{t}[l]; \eta_{t}[l], 
  \bar{\sigma}_{t}^{2}[l]  
 \right)
\right]
\nonumber \\
&= \bar{\xi}_{\In,t}[l]\mathbb{E}[H_{\tau'}[l]\tilde{H}_{t}[l]]
+ o(1) 
\nonumber \\
&= \bar{\xi}_{\In,t}[l]\mathbb{E}\left[
 H_{\tau'}[l]\left\{
  \frac{H_{t}[l]}{\bar{\xi}_{\Out,t}[l]} 
  + \frac{C_{t,0}[l,l]-1}{\underline{\bar{\xi}}_{\Out,t}[l]}\underline{H}_{t}[l]
 \right\}
\right]
\nonumber \\
&+ \bar{\xi}_{\In,t}[l]\sum_{i=0}^{\lfloor t/T \rfloor}\sum_{l'\neq l}
\frac{C_{t,i}[l,l']}{\underline{\bar{\xi}}_{\Out,t-iT}[l']}
\mathbb{E}[H_{\tau'}[l]\underline{H}_{t-iT}[l']] 
+ o(1),
\end{align}
where the first and second equalities follow from Lemma~\ref{lemma_Stein} and 
the definition of $\tilde{H}_{t}[l]$ in (\ref{H_tilde}), respectively. We use 
Property~\ref{Ib} for $\tau\leq t$ to arrive at Property~\ref{Id} for $\tau=t$. 
\end{IEEEproof}

\begin{IEEEproof}[Proof of Property~\ref{Ie}]
See \cite[Proof of Property~(B4) for $\tau=t$ in Theorem~4]{Takeuchi201}. 
\end{IEEEproof}

\section{Proof of Theorem~\ref{theorem_variance}}
\label{proof_theorem_variance} 
The tree assumption in Assumption~\ref{assumption_tree} is used in 
the proof of Theorem~\ref{theorem_variance}. 
Focus a root node~$l_{0}\in\mathcal{L}$ and consider message-passing that 
aggregate messages from leaf nodes toward the root node~$l_{0}$. We analyze 
properties of message-passing with respect to a path that passes through 
the root node~$l_{0}$.  

Let $(l, l', \tilde{l}', \tilde{l})$ 
denote four different nodes $l$, $l'$, $\tilde{l}'$, and $\tilde{l}$ that 
are located in this order on a path. We first prove the following 
lemma for the path length longer than or equal to $3$:

\begin{lemma} \label{lemma_length3}
For four different nodes $(l, l', \tilde{l}', \tilde{l})$ on a path, 
\begin{equation} \label{length3}
\mathbb{E}\left[
 \underline{H}_{\tau,j}[l\rightarrow l']
 \underline{H}_{\tau',j}[\tilde{l}\rightarrow \tilde{l}']
\right] = 0
\end{equation}
with all $\tau'\in\{0,\ldots,\tau\}$. 
\end{lemma} 
\begin{IEEEproof}
The proof is by induction with respect to the total number of 
inner iterations for consensus propagation. 
For the first inner iteration $\tau=0$ and $j=1$, the definition of 
$\underline{H}_{0,1}[l\rightarrow l']$ in (\ref{H_ll'_tree}) implies 
$H_{0,1}[l\rightarrow l']=\underline{\bar{\xi}}_{\Out,0}^{-1}[l]H_{0}[l]$. 
Thus, we use Property~\ref{Ic} in Theorem~\ref{theorem_SE_tech} to obtain 
(\ref{length3}) for $\tau=0$ and $j=1$ due to $l\neq\tilde{l}$.  
For some integers $t\geq T$ and $i\leq J+1$, suppose that (\ref{length3}) is 
correct for $\tau=t$ and $j=i-1$. If $i\leq J$ holds, we need to prove 
(\ref{length3}) for $\tau=t$ and $j=i$. Otherwise, we need to prove 
(\ref{length3}) for $\tau=t+1$ and $j=1$.

We only consider the case of $\tau=t$ and $j=i$ since (\ref{length3}) for 
$\tau=t+1$ and $j=1$ can be proved in the same manner. 
Since we have a path that connect the nodes $l$, $l'$, $\tilde{l'}$, and 
$\tilde{l}$, the tree assumption in Assumption~\ref{assumption_tree} 
implies that $\underline{H}_{\tau,i-1}[l''\rightarrow l]$ does not contain 
the message $\underline{H}_{\tau'}[\tilde{l}]
/\underline{\bar{\xi}}_{\Out,\tau'}[\tilde{l}]$ computed in node~$\tilde{l}$ 
for any $l''\in\mathcal{N}[l]\backslash\{l'\}$ and that 
$\underline{H}_{\tau',i-1}[\tilde{l}''\rightarrow \tilde{l}]$ does not include 
the message $\underline{H}_{\tau}[l]/\underline{\bar{\xi}}_{\Out,\tau}[l]$ 
computed in node~$l$ for any 
$\tilde{l}''\in\mathcal{N}[\tilde{l}]\backslash\{l'\}$. 
Using the definition of $\underline{H}_{t,i}[l\rightarrow l']$ in 
(\ref{H_ll'_tree}) and Property~\ref{Ic} in Theorem~\ref{theorem_SE_tech} 
yields 
\begin{align}
&\mathbb{E}\left[
 \underline{H}_{t,i}[l\rightarrow l']
 \underline{H}_{\tau',i}[\tilde{l}\rightarrow \tilde{l}']
\right]
\nonumber \\
&= \sum_{l''\in\mathcal{N}[l]\backslash\{l'\}}
\sum_{\tilde{l}''\in\mathcal{N}[\tilde{l}]\backslash\{\tilde{l}'\}}
\mathbb{E}\left[
 \underline{H}_{t,i-1}[l''\rightarrow l]
 \underline{H}_{\tau',i-1}[\tilde{l}''\rightarrow \tilde{l}]
\right]. 
\end{align}
Since Assumption~\ref{assumption_tree} implies 
$\mathcal{N}[l]\cap\mathcal{N}[\tilde{l}]=\emptyset$ for the 
nodes $(l, l', \tilde{l}', \tilde{l})$, we have 
$l''\neq \tilde{l}''$ for the indices of the summation. For 
four nodes $(l, l', \tilde{l}', \tilde{l})
=(l'', l, \tilde{l}, \tilde{l}'')$ on a path, we use the 
induction hypothesis~(\ref{length3}) for $\tau=t$ and $j=i-1$ to obtain 
(\ref{length3}) for $\tau=t$ and $j=i$. 
Thus, (\ref{length3}) is correct for all $\tau$ and~$j$. 
\end{IEEEproof}

We use Lemma~\ref{lemma_length3} to prove the following lemma 
for a path of length~$2$. 

\begin{lemma} \label{lemma_length2}
For three different nodes $l$, $l'$, and $\tilde{l}$ that are located 
in this order on a path of length~$2$, we have 
\begin{equation} \label{length2}
\mathbb{E}\left[
 \underline{H}_{t,j}[l\rightarrow l']
 \underline{H}_{\tau,j}[\tilde{l}\rightarrow l']
\right] = 0
\end{equation}
for all $\tau\in\{0,\ldots,t\}$. 
\end{lemma}
\begin{IEEEproof}
Since Assumption~\ref{assumption_tree} implies that 
$\underline{H}_{t,j-1}[\tilde{l}'\rightarrow l]$ does not contain 
$\underline{H}_{t}[\tilde{l}]/\underline{\bar{\xi}}_{\Out,t}[\tilde{l}]$ 
for any $\tilde{l}'\in\mathcal{N}[l]\backslash\{l'\}$ and that 
$\underline{H}_{t,j-1}[\tilde{l}'\rightarrow \tilde{l}]$ does not include 
$\underline{H}_{t}[l]/\underline{\bar{\xi}}_{\Out,t}[l]$ 
for $\tilde{l}'\in\mathcal{N}[\tilde{l}]\backslash\{l'\}$, 
we use the definition of $\underline{H}_{t,j}[l\rightarrow l']$ in 
(\ref{H_ll'_tree}) and Property~\ref{Ib} in Theorem~\ref{theorem_SE_tech} 
to obtain   
\begin{align}
&\mathbb{E}\left[
 \underline{H}_{t,j}[l\rightarrow l']
 \underline{H}_{\tau,j}[\tilde{l}\rightarrow l']
\right]
\nonumber \\
&= \sum_{l''\in\mathcal{N}[l]\backslash\{l'\}}
\sum_{\tilde{l}''\in\mathcal{N}[\tilde{l}]\backslash\{l'\}}
\mathbb{E}\left[
 \underline{H}_{t,j-1}[l''\rightarrow l]
 \underline{H}_{\tau,j-1}[\tilde{l}''\rightarrow \tilde{l}]
\right]. 
\end{align}
Assumption~\ref{assumption_tree} implies 
$\mathcal{N}[l]\cap\mathcal{N}[\tilde{l}]=\{l'\}$, so that we have 
$l''\neq \tilde{l}''$ for the indices of the summation. For four nodes 
$(l, l', \tilde{l}', \tilde{l})=(l'', l, \tilde{l}, \tilde{l}'')$ on a path,  
we use Lemma~\ref{lemma_length3} to arrive at (\ref{length2}).  
\end{IEEEproof}

We prove Theorem~\ref{theorem_variance}. Since 
Assumption~\ref{assumption_tree} implies that 
$\underline{H}_{t,J}[l'\rightarrow l]$ does not contain the message 
$H_{t}[l]/\bar{\xi}_{\Out,t}[l]$ computed in node~$l$, 
we use the definition of $\tilde{H}_{t}[l]$ in (\ref{H_tilde_tree}) and 
Lemma~\ref{lemma_length2} yields 
\begin{align}
\mathbb{E}[\tilde{H}_{\tau}[l]\tilde{H}_{t}[l]]
&= \frac{\mathbb{E}[M_{\tau}[l]M_{t}[l]]}
{L\bar{\xi}_{\Out,\tau}[l]\bar{\xi}_{\Out,t}[l]}
\nonumber \\
&+ \sum_{l'\in\mathcal{N}[l]}
\mathbb{E}\left[
 \underline{H}_{\tau,J}[l'\rightarrow l]
 \underline{H}_{t,J}[l'\rightarrow l]
\right]. 
\end{align}
Similarly, for $\underline{H}_{t,j}[l\rightarrow l']$ in (\ref{H_ll'_tree}) 
we obtain 
\begin{align}
\mathbb{E}[\underline{H}_{\tau,j}[l\rightarrow l']
\underline{H}_{t,j}[l\rightarrow l'] ]
= \frac{\mathbb{E}[\underline{M}_{\tau}[l]\underline{M}_{t}[l]]}
{L\underline{\bar{\xi}}_{\Out,\tau}[l]\underline{\bar{\xi}}_{\Out,t}[l]}
\nonumber \\
+ \sum_{\tilde{l}'\in\mathcal{N}[l]\backslash\{l'\}}
\mathbb{E}[\underline{H}_{\tau,j-1}[\tilde{l}'\rightarrow l]
\underline{H}_{t,j-1}[\tilde{l}'\rightarrow l]]. 
\end{align}
Thus, Theorem~\ref{theorem_variance} holds. 

\section{Proof of Theorem~\ref{theorem_fixed_point}}
\label{proof_theorem_fixed_point}
To represent a fixed point of the state evolution 
recursion with respect to the variance variables, we replace the iteration 
index~$t$ with an asterisk for all variables. We first focus on the inner 
module.  Since a unique fixed point of (unnormalized) consensus 
propagation~\cite{Moallemi06} is summation consensus 
under Assumption~\ref{assumption_tree}, from (\ref{eta_bar}), 
(\ref{sigma_t_bar}), and Theorem~\ref{theorem_variance} we have  
\begin{equation} \label{eta_FP}
\frac{1}{L}\bar{\eta}_{*}[l]
= \frac{1}{L}\sum_{l'\in\mathcal{L}}\frac{\bar{\zeta}_{*}[l']}
{\bar{\xi}_{\Out,*}[l']}
\equiv \bar{\eta}_{*},  
\end{equation}
\begin{equation} \label{sigma_FP}
\bar{\sigma}_{*}^{2}[l] 
=  \frac{1}{L}\sum_{l'\in\mathcal{L}}\frac{1}{\bar{\xi}_{\Out,*}[l']}
\equiv \bar{\sigma}_{*}^{2}, 
\end{equation}
\begin{equation} \label{H_variance_FP}
\bar{\Sigma}_{*}[l]
= \sum_{l'\in\mathcal{L}}\frac{\mathbb{E}[M_{*}^{2}[l']]}{L\bar{\xi}_{\Out,*}^{2}[l']}
\equiv \bar{\Sigma}_{*}
\end{equation}
for all $l\in\mathcal{L}$. 
Thus, all variables fed back from the inner 
module are identical for all $l$, i.e.\ 
for (\ref{xi_in_bar}), (\ref{v_t_bar}), (\ref{Z_Zt}), and 
(\ref{Z_covariance}) 
\begin{equation}
\bar{\xi}_{\In,*}[l] 
= \mathbb{E}[\partial_{1}f_{\In,*}(\bar{\eta}_{*}X 
+ \tilde{H}_{*}; L, \bar{\sigma}_{*}^{2})] 
\equiv \bar{\xi}_{\In,*}, 
\end{equation}
\begin{equation} 
\bar{v}_{*}[l] = \frac{\bar{\sigma}_{*}^{2}\bar{\xi}_{\In,*}}{L\delta}
\equiv \bar{v}_{*}, 
\end{equation}
\begin{equation}
\mathbb{E}[Z[l]Z_{*}[l]] 
= \frac{1}{L\delta}\mathbb{E}\left[
 Xf_{\In}\left(
  \bar{\eta}_{*}X + \tilde{H}_{*}; L, \bar{\sigma}_{*}^{2}
 \right)
\right]
\equiv \frac{\mathbb{E}[ZZ_{*}]}{L\delta},
\end{equation}
\begin{equation}
\mathbb{E}[Z_{*}^{2}[l]] 
= \frac{1}{L\delta}\mathbb{E}\left[
 f_{\In}^{2}\left(
  \bar{\eta}_{*}X + \tilde{H}_{*}; L, \bar{\sigma}_{*}^{2}
 \right)
\right]
\equiv \frac{\mathbb{E}[Z_{*}^{2}]}{L\delta},  
\end{equation}
with $Z\sim\mathcal{N}(0, (L\delta)^{-1}\mathbb{E}[X^{2}])$ independent 
of $W$ and $\tilde{H}_{*}\sim\mathcal{N}(0,\bar{\Sigma}_{*})$ 
independent of $X$. In the derivation of these expressions, we have 
used $\eta_{*}[l]=L$ obtained from the definition of $\eta_{t}[l]$ in 
(\ref{eta}). 

We next focus on the outer module. Let $Z$ and $Z_{*}$ denote zero-mean 
Gaussian random variable with covariance $\mathbb{E}[Z^{2}]$, 
$\mathbb{E}[ZZ_{*}]$, and $\mathbb{E}[Z_{*}^{2}]$, independent of $W$. 
Using (\ref{xi_out_bar}) and (\ref{zeta_bar}) yields 
\begin{equation} 
\bar{\xi}_{\Out,*}[l] 
= \mathbb{E}\left[
 \partial_{1}f_{\Out}(Z_{*}, Y; \bar{v}_{*})
\right]
\equiv \bar{\xi}_{\Out,*},
\end{equation}
\begin{equation} 
\bar{\zeta}_{*}[l] = - \mathbb{E}\left[
 \left.
  \frac{\partial}{\partial z}f_{\Out}(Z_{*},g(z,W);\bar{v}_{*})
 \right|_{z=Z}
\right]
\equiv \bar{\zeta}_{*},
\end{equation}
with $Y = g(Z, W)$. Substituting these expressions into 
(\ref{eta_FP}), (\ref{sigma_FP}), and (\ref{H_variance_FP}), we have 
$\bar{\eta}_{*}=\bar{\zeta}_{*}/\bar{\xi}_{\Out,*}$, 
$\bar{\sigma}_{*}^{2}=\bar{\xi}_{\Out,*}^{-1}$, and 
\begin{equation}
\bar{\Sigma}_{*}
= \frac{1}{\bar{\xi}_{\Out,*}^{2}}\mathbb{E}\left[
 f_{\Out}^{2}(Z_{*}, Y; \bar{v}_{*})
\right]. 
\end{equation}
From these observations, we find that the fixed point of the state evolution 
recursion for D-GAMP is equivalent to that for centralized 
GAMP~\cite{Rangan11}. 

\section{Proof of Theorem~\ref{theorem_naive}}
\label{proof_theorem_naive} 
We first derive the error model of D-GAMP with the distributed 
protocol~(\ref{x_tilde_naive}). Replace $\bar{\eta}_{t}[l]$ in 
Lemma~\ref{lemma_error_model} with  
\begin{equation} \label{eta_naive}
\bar{\eta}_{t}[l] = \frac{\bar{\zeta}_{t}[l]}{\bar{\xi}_{\Out,t}[l]} 
+ \gamma\sum_{l'\in\mathcal{N}[l]}
\left(
 \frac{\bar{\zeta}_{t}[l']}{\bar{\xi}_{\Out,t}[l']} 
 -  \frac{\bar{\zeta}_{t}[l]}{\bar{\xi}_{\Out,t}[l]} 
\right). 
\end{equation}
Applying $\tilde{\boldsymbol{x}}_{t}[l]$ in 
(\ref{x_tilde_naive}) to the definition of $\tilde{\boldsymbol{h}}_{t}[l]$ 
in Lemma~\ref{lemma_error_model} for $T=1$ yields 
\begin{equation} \label{h_tilde_naive} 
\tilde{\boldsymbol{h}}_{t}[l] 
= \frac{\boldsymbol{h}_{t}[l]}{\xi_{\Out,t}[l]} + \gamma\sum_{l'\in\mathcal{N}[l]}
\left(
 \frac{\boldsymbol{h}_{t}[l']}{\xi_{\Out,t}[l']} 
 - \frac{\boldsymbol{h}_{t}[l]}{\xi_{\Out,t}[l]} 
\right),
\end{equation}
with $\boldsymbol{h}_{t}[l]$ defined in Lemma~\ref{lemma_error_model}. 
Thus, the error model of D-GAMP with the distributed 
protocol~(\ref{x_tilde_naive}) is equal to the 
error model~(\ref{b})--(\ref{q}) with $C_{t,0}[l,l]=1$, in which 
$\tilde{\boldsymbol{h}}_{t}[l]$ in (\ref{h_tilde}) is replaced with 
(\ref{h_tilde_naive}). 

Define $\Sigma_{t,t}[l]=N^{-1}\|\tilde{\boldsymbol{h}}_{t}[l]\|^{2}$. 
It is sufficient to confirm that the effective SNR 
$L^{-2}\bar{\eta}_{t}^{2}[l]/\Sigma_{t,t}[l]$ for the inner denoiser is 
different from that of centralized GAMP. 
Suppose that $\bar{\xi}_{\Out,t}[l]$ and $\bar{\zeta}_{t}[l]$ converge 
to $\bar{\xi}_{\Out,*}[l]$ and $\bar{\zeta}_{*}[l]$ as $t\to\infty$, 
respectively. 
When $\gamma>0$ is set appropriately~\cite{Xiao04}, the definition of 
$\bar{\eta}_{t}[l]$ in (\ref{eta_naive}) implies that 
the following consensus is achieved:  
\begin{equation}
\lim_{t\to\infty}\frac{\bar{\eta}_{t}[l]}{L} 
= \frac{1}{L}\sum_{l'\in\mathcal{L}}\frac{\bar{\zeta}_{*}[l']}
{\bar{\xi}_{\Out,*}[l']}
= \bar{\eta}_{*} 
\end{equation}
for all $l\in\mathcal{L}$, with $\bar{\eta}_{*}$ given in (\ref{eta_FP}). 

We next evaluate $\Sigma_{t,t}[l]=N^{-1}\|\tilde{\boldsymbol{h}}_{t}[l]\|^{2}$. 
Applying Properties~\ref{Oc} and \ref{Ib} 
in Theorem~\ref{theorem_SE_tech} to $\tilde{\boldsymbol{h}}_{t}[l]$ 
in (\ref{h_tilde_naive}) yields 
\begin{equation} \label{H_variance_naive}
\Sigma_{t,t}[l] 
\ato (1 - \gamma|\mathcal{N}[l])^{2}
\frac{\mathbb{E}[M_{t}^{2}[l]]}{L\bar{\xi}_{\Out,t}^{2}[l]} 
+ \gamma^{2}\sum_{l'\in\mathcal{N}[l]}\frac{\mathbb{E}[M_{t}^{2}[l']]}
{L\bar{\xi}_{\Out,t}^{2}[l']},  
\end{equation}
which cannot converge to $\bar{\Sigma}_{*}[l]$ in (\ref{H_variance_FP}), 
because (\ref{H_variance_naive}) is different from the distributed 
protocol for average consensus~\cite{Xiao04}. These observations imply that 
the fixed point of the effective SNR 
$L^{-2}\bar{\eta}_{t}^{2}[l]/\Sigma_{t,t}[l]$ is different from that for 
centralized GAMP. Thus, Theorem~\ref{theorem_naive} holds.

\section{Properties for the Bayes-optimal Denoisers}
\subsection{Proof of Lemma~\ref{lemma_outer_denoiser}}
\label{proof_lemma_outer_denoiser}

We first confirm the following proposition: 
\begin{proposition} \label{proposition1}
\begin{itemize}
\item $\bar{v}_{0,t}[l]=\bar{v}_{t,t}[l]$ implies 
$\mathbb{E}[Z[l]Z_{t}[l]]=\mathbb{E}[Z_{t}^{2}[l]]$. 
\item $\bar{v}_{0,\tau}[l]=\bar{v}_{\tau,\tau}[l]$, 
$\bar{v}_{0,t}[l]=\bar{v}_{t,t}[l]$, and $\bar{v}_{\tau,t}[l]=\bar{v}_{t,t}[l]$ 
imply $\mathbb{E}[Z_{\tau}[l]Z_{t}[l]]=\mathbb{E}[Z_{\tau}^{2}[l]]$. 
\end{itemize}
\end{proposition}
\begin{IEEEproof}
We first prove the former property. 
Applying the assumption $\bar{v}_{0,t}[l]=\bar{v}_{t,t}[l]$ to the 
definitions of (\ref{v_t_bar_true0}) and (\ref{v_t_bar_true}) yields 
$\mathbb{E}[(Z[l]+B_{t}[l])B_{t}[l]]=0$, which implies 
$\mathbb{E}[Z_{t}[l]B_{t}[l]]=0$. Thus, we have $\mathbb{E}[Z_{t}^{2}[l]]
=\mathbb{E}[Z_{t}[l]Z[l]]$.  

We next prove the latter property. Using the 
definitions of (\ref{v_t_bar_true0}) and (\ref{v_t_bar_true}) yields 
\begin{align}
&\mathbb{E}[Z_{\tau}^{2}[l]] - \mathbb{E}[Z_{\tau}[l]Z_{t}[l]]
= \mathbb{E}[(Z[l] + B_{\tau}[l])(B_{\tau}[l] - B_{t}[l])]
\nonumber \\
&= - \bar{v}_{0,\tau}[l] + \bar{v}_{0,t}[l] + \bar{v}_{\tau,\tau}[l] 
- \bar{v}_{\tau,t}[l]
= 0,
\end{align}
where the last equality follows from the assumptions. 
\end{IEEEproof}

We prove Lemma~\ref{lemma_outer_denoiser}. 
From Proposition~\ref{proposition1} 
under the assumption $\bar{v}_{0,t}[l]=\bar{v}_{t,t}[l]$, 
it is straightforward to find the representation 
$Z[l]\sim Z_{t}[l] + N_{t}[l]$ for all $t\geq0$, with 
$N_{t}[l]\sim\mathcal{N}(0,\bar{v}_{t,t}[l])$ independent of $Z_{t}[l]$. 
This representation justifies the expression of the posterior mean 
estimator $\hat{Z}_{t}[l]$ in (\ref{PME_Z}). 

Let $f_{\Out,t}[l]=f_{\Out}[l](Z_{t}[l], g[l](Z[l], W[l]); \bar{v}_{t,t}[l])$ 
with $\bar{v}_{t}[l]=\bar{v}_{t,t}[l]$. 
We use the representation $Z[l]\sim Z_{t}[l] + N_{t}[l]$,  
Lemma~\ref{lemma_Stein}, and the definition of $\bar{\zeta}_{t}[l]$ in 
(\ref{zeta_bar}) to obtain 
\begin{equation}
\mathbb{E}[N_{t}[l]f_{\Out,t}[l]] = - \bar{v}_{t,t}[l]\bar{\zeta}_{t}[l]. 
\end{equation}
Thus, we have the following identity: 
\begin{align}
\bar{\zeta}_{t}[l]
&= - \mathbb{E}\left[
 \frac{Z[l] - Z_{t}[l]}{\bar{v}_{t,t}[l]}f_{\Out,t}[l]
\right] 
\nonumber \\
&= \mathbb{E}\left[
 \frac{Z_{t}[l] - \hat{Z}_{t}[l]}{\bar{v}_{t,t}[l]}f_{\Out,t}[l]
\right], 
\end{align}
with the posterior mean estimator 
$\hat{Z}_{t}[l]=\mathbb{E}[Z[l] | Z_{t}[l], Y[l]]$. 
Using the Cauchy–Schwarz inequality for $\zeta_{t}^{2}[l]$, we arrive at 
\begin{equation}
\frac{\bar{\zeta}_{t}^{2}[l]}{\mathbb{E}[M_{t}^{2}[l]]}
\leq \mathbb{E}\left[
 \left(
  \frac{Z_{t}[l] - \hat{Z}_{t}[l]}{\bar{v}_{t,t}[l]}
 \right)^{2} 
\right],
\end{equation}
where the equality holds if and only if there is some constant 
$C\in\mathbb{R}$ such that 
$f_{\Out,t}[l]=C(Z_{t}[l] - \hat{Z}_{t}[l])/\bar{v}_{t,t}[l]$ is satisfied. 
Thus, Lemma~\ref{lemma_outer_denoiser} holds. 

\subsection{Proof of Lemma~\ref{lemma_outer_covariance}}
\label{proof_lemma_outer_covariance}
We first prove basic properties of the Bayes-optimal denoisers.  

\begin{lemma}[\cite{Rangan11}] \label{lemma_posterior} 
Consider estimation of $X$ based on the Gaussian measurement 
$Y=aX+Z$ with $Z\sim\mathcal{N}(0,\sigma^{2})$ independent of $X$. 
Let the posterior mean estimator $f(y)=\mathbb{E}[X|Y=y]$. Then, we 
have 
\begin{equation}
\mathbb{E}[Xf(Y)] = \mathbb{E}[f^{2}(Y)], 
\end{equation}
\begin{equation}
f'(y) = \frac{a\mathbb{E}[\{X - f(Y)\}^{2} | Y=y]}{\sigma^{2}}. 
\end{equation}
\end{lemma}
\begin{IEEEproof}
The former identity is trivial: 
$\mathbb{E}[Xf(Y)] = \mathbb{E}[\mathbb{E}[Xf(Y) | Y]] 
= \mathbb{E}[f^{2}(Y)]$. The latter identity is obtained from direct 
computation, 
\begin{align}
f'(y) &= \frac{d}{dy}\frac{\int xe^{-\frac{(y-ax)^{2}}{2\sigma^{2}}}dP(x)}
{\int e^{-\frac{(y-ax)^{2}}{2\sigma^{2}}}dP(x)}
\nonumber \\
&= - \frac{\langle x \rangle y-a\langle x^{2} \rangle}{\sigma^{2}}
+ \langle x\rangle
\frac{y-a\langle x \rangle}{\sigma^{2}}
= a\frac{\langle x^{2}\rangle - \langle x \rangle^{2}}{\sigma^{2}}, 
\end{align}
with 
\begin{equation}
\langle x^{i} \rangle 
= \frac{\int x^{i}e^{-\frac{(y-ax)^{2}}{2\sigma^{2}}}dP(x)}
{\int e^{-\frac{(y-ax)^{2}}{2\sigma^{2}}}dP(x)}. 
\end{equation}
Thus, the latter identity holds. 
\end{IEEEproof}

We follow \cite{Rangan11} to represent the Bayes-optimal outer 
denoiser~(\ref{outer_denoiser}) with $C=1$ as the (negative) score function 
\begin{equation} \label{outer_denoiser_score}
f_{\Out}[l](\theta,y;\bar{v}_{t,t}[l]) 
= - \frac{\partial}{\partial \theta}\log P[l](y|\theta;\bar{v}_{t,t}[l]),
\end{equation}
with 
\begin{equation}
P[l](y|\theta; v) = \int P_{Y[l]|Z[l]}(y|z)\frac{1}{\sqrt{2\pi v}}
e^{-\frac{(z - \theta)^{2}}{2v}}dz,  
\end{equation}
where $P_{Y[l]|Z[l]}(y|z)$ denotes the conditional distribution of $Y[l]$ 
given $Z[l]$. Note that the distribution $P[l](y|\theta; v)$ is normalized 
with respect to $y$. It is straightforward to confirm the equivalence 
between the two representations. 

We reproduce an existing lemma~\cite{Rangan11}. 

\begin{lemma}[\cite{Rangan11}]  \label{lemma_identity} 
Suppose that Assumption~\ref{assumption_Bayes} holds and 
consider the Bayes-optimal outer denoiser~(\ref{outer_denoiser_score}) 
with $C=1$. 
Furthermore, suppose that $\bar{v}_{t}[l]$ in (\ref{v_t_bar}) is equal to 
$\bar{v}_{t,t}[l]$ in (\ref{v_t_bar_true}). If $\bar{v}_{0,t}[l]$ is 
equal to $\bar{v}_{t,t}[l]$, then we have 
\begin{equation}
\bar{\zeta}_{t}[l] = \bar{\xi}_{\Out,t}[l].
\end{equation}
\end{lemma}
\begin{IEEEproof}
We first consider the case of $t>0$. 
Utilizing the well-known identity for the score function 
\begin{align}
&\mathbb{E}\left[
 \left. 
  \frac{\partial}{\partial \theta}
  \log P[l](Y[l] | \theta; \bar{v}_{t,t}[l])
 \right| Z_{t}[l] = \theta 
\right] = 0
\end{align} 
and the representation of the Bayes-optimal outer denoiser in 
(\ref{outer_denoiser_score}) with $\bar{v}_{t}[l]=\bar{v}_{t,t}[l]$ yields 
\begin{equation}
\mathbb{E}[Z_{t}[l]f_{\Out}[l](Z_{t}[l], Y[l]; \bar{v}_{t,t}[l])]
= 0. 
\end{equation}
We use Lemma~\ref{lemma_Stein} and 
the definitions of $\bar{\xi}_{\Out,t}[l]$ and $\bar{\zeta}_{t}[l]$ 
in (\ref{xi_out_bar}) and (\ref{zeta_bar}) to obtain 
\begin{align}
0 &= \mathbb{E}[Z_{t}[l]f_{\Out}[l](Z_{t}[l], g[l](Z[l], W[l]); 
\bar{v}_{t,t}[l])]
\nonumber \\
&= \mathbb{E}[Z_{t}^{2}[l]]\bar{\xi}_{\Out,t}[l] 
- \mathbb{E}[Z_{t}[l]Z[l]]\bar{\zeta}_{t}[l]
\nonumber \\
&= \mathbb{E}[Z_{t}^{2}[l]]
(\bar{\xi}_{\Out,t}[l] - \bar{\zeta}_{t}[l]),
\end{align}
where the last equality follows from Proposition~\ref{proposition1} 
under the assumption $\bar{v}_{0,t}[l]=\bar{v}_{t,t}[l]$. 
Since $\mathbb{E}[Z_{t}^{2}]>0$ holds for $t>0$, we arrive at 
Lemma~\ref{lemma_identity} for $t>0$. 

We next consider the case of $t=0$, in which we have $Z_{0}[l]=0$.  
To extract information about the partial derivative of 
$f_{\Out}[l](\theta, Y[l];\bar{v}_{0,0}[l])$, we inject an 
independent weak Gaussian noise $Z_{0}^{(n)}[l]
\sim\mathcal{N}(0,n^{-1})$ to $\theta$ for sufficiently large $n\in\mathbb{N}$, 
\begin{equation}
\bar{\xi}_{\Out,0}^{(n)}[l] = 
\mathbb{E}\left[
 \partial_{1}f_{\Out}[l](Z_{0}^{(n)}[l], Y[l]; \bar{v}_{0}[l])
\right],
\end{equation}  
\begin{equation}
\bar{\zeta}_{0}^{(n)}[l] = - \mathbb{E}\left[
 \left.
  \frac{\partial}{\partial z}f_{\Out}[l](Z_{0}^{(n)}[l],g[l](z,W[l]);
  \bar{v}_{0}[l])
 \right|_{z=Z[l]}
\right],
\end{equation}
associated with $\bar{\xi}_{\Out,0}[l]$ and $\bar{\zeta}_{0}[l]$ 
in (\ref{xi_out_bar}) and (\ref{zeta_bar}), respectively. 
Repeating the proof of Lemma~\ref{lemma_identity} for $t>0$, we obtain 
$\bar{\xi}_{\Out,0}^{(n)}[l]=\bar{\zeta}_{0}^{(n)}[l]$ for any $n\in\mathbb{N}$. 

To prove $\bar{\xi}_{\Out,0}[l]=\bar{\zeta}_{0}[l]$, we show 
$\lim_{n\to\infty}\bar{\xi}_{\Out,0}^{(n)}[l]=\bar{\xi}_{\Out,0}[l]$ and 
$\lim_{n\to\infty}\bar{\zeta}_{0}^{(n)}[l]=\bar{\zeta}_{0}[l]$. 
Assumption~\ref{assumption_Bayes} implies that 
$f_{\Out,0}[l]$ is differentiable almost everywhere and has bounded 
partial derivatives. Thus, we use the dominated convergence theorem to 
arrive at these limits. 
\end{IEEEproof}

We prove Lemma~\ref{lemma_outer_covariance}. 
From the assumption $\bar{v}_{\tau',\tau'}[l]>\bar{v}_{\tau,\tau}[l]$ and 
Proposition~\ref{proposition1} under the assumption $\bar{v}_{\tau,t}[l]
=\bar{v}_{t,t}[l]$ for all $\tau\in\{0,\ldots,t\}$, we have 
the following cascaded representation of $Z_{\tau}[l]$ and $Z_{t}[l]$:  
\begin{equation} \label{Z_representation}
Z[l] = Z_{t}[l] + N_{t}[l], 
\quad Z_{t}[l] = Z_{\tau}[l] + \Delta N_{\tau,t}[l]  
\end{equation}
for $\tau\geq 0$, where $N_{t}[l]\sim\mathcal{N}(0, \bar{v}_{t,t}[l])$ and 
$\Delta N_{\tau,t}[l]\sim\mathcal{N}(0, \bar{v}_{\tau,\tau}[l] - \bar{v}_{t,t}[l])$ 
are independent of all random variables. It is straightforward to confirm 
$\mathbb{E}[Z_{\tau}[l]Z_{t}[l]]=\mathbb{E}[Z_{\tau}^{2}[l]]$ 
and $\mathbb{E}[(Z_{\tau}[l] - Z[l])^{2}]=\bar{v}_{\tau,\tau}[l]$. Since 
$Z_{\tau}[l]$ depends on $Z[l]$ only through $Z_{t}[l]$ for all $\tau>0$, 
as well as $Z_{0}[l]=0$, we have 
\begin{equation} \label{PME_Z_LM}
\mathbb{E}\left[
 Z | Z_{t}[l], Z_{\tau}[l], Y[l]
\right] 
= \mathbb{E}\left[
 Z | Z_{t}[l], Y[l] 
\right].
\end{equation} 

For $M_{t}[l] = f_{\Out}[l](Z_{t}[l], Y[l]; \bar{v}_{t,t}[l])$, 
we use the definition of the Bayes-optimal outer denoiser in 
(\ref{outer_denoiser}) with $C=1$ to evaluate 
$\mathbb{E}[M_{\tau}[l]M_{t}[l]]$ as 
\begin{align}
&\bar{v}_{t,t}[l]\mathbb{E}[M_{\tau}[l]M_{t}[l]] 
= \mathbb{E}\left[
 M_{\tau}[l](Z_{t}[l] - \hat{Z}_{t}[l])
\right]
\nonumber \\
&= \mathbb{E}\left[
 M_{\tau}[l]\left(
  Z_{t}[l] 
  - \mathbb{E}[Z[l] | Z_{\tau}[l], Z_{t}[l], Y[l]]
 \right)
\right]
\nonumber \\
&= \mathbb{E}\left[
 \mathbb{E}\left[
  M_{\tau}[l](Z_{t}[l] - Z[l]) 
  \left|  
   Z_{\tau}[l],
  \right. Z_{t}[l], Y[l]
 \right]
\right]
\nonumber \\
&= \mathbb{E}\left[
 M_{\tau}[l](Z_{t}[l] - Z[l]) 
\right], 
\end{align}
where the second and third equalities follow from the 
identity~(\ref{PME_Z_LM}) and from the fact that $M_{\tau}[l]$ is a 
deterministic function of $Z_{\tau}[l]$ and $Y[l]$. 
Using Lemma~\ref{lemma_Stein} yields 
\begin{align}
&\bar{v}_{t,t}[l]\mathbb{E}[M_{\tau}[l]M_{t}[l]] 
\nonumber \\
&= \mathbb{E}[(Z_{t}[l] - Z[l])Z_{\tau}[l]]\bar{\xi}_{\Out,\tau}[l] 
- \mathbb{E}[(Z_{t}[l] - Z[l])Z[l]]\bar{\zeta}_{\tau}[l]
\nonumber \\
&= \bar{\xi}_{\Out,\tau}[l]\mathbb{E}[(Z_{t}[l] - Z[l])
(Z_{\tau}[l] - Z[l])]
\nonumber \\
&= \bar{\xi}_{\Out,\tau}[l]\bar{v}_{\tau,t}[l]
\end{align}
with $\bar{\xi}_{\Out,\tau}[l]$ and $\bar{\zeta}_{\tau}[l]$ given in 
(\ref{xi_out_bar}) and (\ref{zeta_bar}), where the second and last 
equalities follow from Lemma~\ref{lemma_identity} and the definition of 
$\bar{v}_{\tau,t}[l]$ in (\ref{v_t_bar_true}), respectively. 
Using the assumption $\bar{v}_{\tau,t}[l]=\bar{v}_{t,t}[l]$ for all 
$\tau\in\{0,\ldots,t\}$, we arrive at Lemma~\ref{lemma_outer_covariance}.

\section{Proof of Theorem~\ref{theorem_convergence}}
\label{proof_theorem_convergence}
\subsection{Long-Memory Proof Strategy}  
In the long-memory proof strategy~\cite{Takeuchi222}, the covariance 
matrix of estimation errors for D-GAMP is utilized to prove the convergence 
of the state evolution recursion with respect to the variance of the 
estimation errors. When the covariance matrix has a special structure, the 
positive definiteness of the covariance matrix implies the convergence 
of its diagonal elements from basic properties in linear algebra. 
Furthermore, the Bayes-optimality of the denoisers produces the special 
structure in the covariance matrix naturally. As a result, the convergence 
of the state evolution recursion can be proved, without utilizing concrete 
properties in the measurement model. 

As shown in Theorem~\ref{theorem_SE_tech}, rigorous state evolution with 
respect to the MSE has already required evaluation of the error covariance 
matrix and a guarantee for its positive definiteness. To prove the 
convergence in the long-memory strategy, thus, the only additional tasks  
are evaluation of the error covariance matrices for the Bayes-optimal 
denoisers, as presented in Lemmas~\ref{lemma_inner_covariance} and 
\ref{lemma_outer_covariance}. In this sense, we are ready for proving 
Theorem~\ref{theorem_convergence}. 

The following lemma is a technical result to guarantee the monotonicity for 
the diagonal elements of a covariance matrix: 

\begin{lemma}[\cite{Takeuchi222}] \label{lemma_monotonicity}
Suppose that a symmetric matrix 
$\boldsymbol{M}_{t}\in\mathbb{R}^{(t+1)\times (t+1)}$ 
is strictly positive definite. Let $m_{\tau,t}$ denote the $(\tau, t)$ 
element of $\boldsymbol{M}_{t}$. Then, 
\begin{itemize}
\item $m_{\tau',\tau}=m_{\tau,\tau}$ for all $\tau\in\{0,\ldots,t\}$ 
and $\tau'\in\{0,\ldots,\tau\}$ implies 
$m_{\tau',\tau'}>m_{\tau,\tau}$ for all $\tau'<\tau\leq t$. 
\item $m_{\tau',\tau}=m_{\tau',\tau'}$ for all $\tau\in\{0,\ldots,t\}$ 
and $\tau'\in\{0,\ldots,\tau\}$ implies 
$m_{\tau',\tau'}<m_{\tau,\tau}$ for all $\tau'<\tau\leq t$. 
\end{itemize}
\end{lemma}
\begin{IEEEproof}
We only prove the latter property since the former property was proved 
in \cite[Lemma~3]{Takeuchi222}. The proof is by induction. For $t=1$, 
we use $\det\boldsymbol{M}_{1}=m_{0,0}m_{1,1}-m_{0,0}^{2}$ and the positive 
definiteness of $\boldsymbol{M}_{1}$ to obtain $m_{1,1}>m_{0,0}>0$. 
Suppose that the latter property is correct for some $t$. We need to 
prove the latter property for $\boldsymbol{M}_{t+1}$. 

The positive definiteness of $\boldsymbol{M}_{t+1}$ implies that of 
$\boldsymbol{M}_{1}$. Thus, we have $0<m_{0,0}<m_{1,1}$. 
Subtracting the first row in $\boldsymbol{M}_{t+1}$ from the other rows, 
we find $\det\boldsymbol{M}_{t+1}=m_{0,0}\det\tilde{\boldsymbol{M}}_{t}$, 
with $\tilde{m}_{\tau',\tau}\equiv[\tilde{\boldsymbol{M}}_{t}]_{\tau',\tau}
=m_{\tau'+1,\tau+1}-m_{0,0}$ for $\tau',\tau\in\{0,\ldots,t\}$. The positive 
definiteness of $\boldsymbol{M}_{t+1}$ implies the positive 
definiteness of $\tilde{\boldsymbol{M}}_{t}$. Since 
$\tilde{m}_{\tau',\tau}=\tilde{m}_{\tau',\tau'}$ holds for $\tau'\leq \tau$, 
we use the induction hypothesis for $\tilde{\boldsymbol{M}}_{t}$ to obtain 
$m_{\tau'+1,\tau'+1}<m_{\tau+1,\tau+1}$ for all $0<\tau'<\tau<t$. Combining these 
results, we arrive at the latter property for $\boldsymbol{M}_{t+1}$.   
\end{IEEEproof}

As presented in Lemmas~\ref{lemma_inner_covariance} and 
\ref{lemma_outer_covariance}, the structure $m_{\tau',\tau}=m_{\tau,\tau}$ or 
$m_{\tau',\tau}=m_{\tau',\tau'}$ in Lemma~\ref{lemma_monotonicity} appears 
when the Bayes-optimal denoisers are used. To prove 
Theorem~\ref{theorem_convergence}, we need the following lemma:  

\begin{lemma} \label{lemma_consistency}
Suppose that Assumption~\ref{assumption_Bayes} holds and 
consider the Bayes-optimal inner denoiser~(\ref{inner_denoiser}) and 
outer denoiser~(\ref{outer_denoiser}) with $C=1$. Then, for all $t=0,1,\ldots$ 
we have the following properties for the outer module: 
\begin{itemize}
\item $\bar{\Sigma}_{\tau',t}[l]$ satisfies (\ref{H_variance_opt}). 
In particular, $\bar{\Sigma}_{\tau',t}[l]=\bar{\Sigma}_{t,t}[l]$ holds 
for all $\tau'\in\{0,\ldots,t\}$. 

\item $\bar{\eta}_{t}[l]=\eta_{t}[l]$ and 
$\bar{\sigma}_{t}^{2}[l]=\bar{\Sigma}_{t,t}[l]$ hold. 

\item $\bar{\Sigma}_{\tau',\tau'}>\bar{\Sigma}_{\tau,\tau}[l]$ holds 
for all $\tau'<\tau\leq t$
\end{itemize}
On the other hand, for the inner module we have 
\begin{itemize}
\item $\mathrm{cov}_{\tau',t+1}[l]=\mathrm{cov}_{t+1,t+1}[l]$ holds 
for all $\tau'\in\{0,\ldots,t+1\}$.

\item $\bar{v}_{t+1}[l]=\bar{v}_{t+1,t+1}[l]$ and 
$\bar{\zeta}_{t+1}[l]=\bar{\xi}_{\Out,t+1}[l]$ hold. 

\item $\mathrm{cov}_{\tau',\tau'}[l]>\mathrm{cov}_{\tau,\tau}[l]$ 
holds for all $\tau'<\tau\leq t+1$. 
\end{itemize}
\end{lemma}
\begin{IEEEproof}
The proof of Lemma~\ref{lemma_consistency} is by induction with respect 
to $t$. The proof for $t=0$ is omitted because it is the same as 
for the general case. For some $\tau$, suppose that 
Lemma~\ref{lemma_consistency} is correct for  
all $t<\tau$. We need to prove Lemma~\ref{lemma_consistency} for $t=\tau$. 
See Appendices~\ref{proof_lemma_consistency_outer} and 
\ref{proof_lemma_consistency_inner} for the proofs of 
the properties in the outer and inner modules, respectively.  
\end{IEEEproof}

We prove Theorem~\ref{theorem_convergence}. The first properties in 
Theorem~\ref{theorem_convergence} are part of Lemma~\ref{lemma_consistency}. 
The second property follows from Theorems~\ref{theorem_SE}, 
\ref{theorem_variance}, and Lemma~\ref{lemma_consistency}. The last property 
is obtained from the monotonicity in Lemma~\ref{lemma_consistency}: 
$\{\mathrm{cov}_{t,t}[l]\}$ are a monotonically decreasing sequence 
with respect to $t$. Since the MSE $\mathrm{cov}_{t,t}[l]$ is non-negative, 
we conclude that $\mathrm{cov}_{t,t}[l]$ converges to a non-negative constant 
as $t\to\infty$. 

\subsection{Proof for the Outer Module}
\label{proof_lemma_consistency_outer}
For some $\tau$, suppose that Lemma~\ref{lemma_consistency} is correct for  
all $t<\tau$. We confirm the conditions in Lemma~\ref{lemma_outer_covariance} 
for all $t\leq \tau$. 
Applying the induction hypotheses $\mathrm{cov}_{\tau',t}[l]
=\mathrm{cov}_{t,t}[l]$ for all $t\leq \tau$ and 
$\tau'\in\{0,\ldots,t\}$ 
and $\mathrm{cov}_{\tau',\tau'}[l]>\mathrm{cov}_{t,t}[l]$ for 
all $\tau'<t\leq \tau$ to the relationship between $\bar{v}_{\tau',t}[l]$ 
and $\mathrm{cov}_{\tau',t}[l]$ in (\ref{v_t_bar_true}), respectively, 
we obtain $\bar{v}_{\tau',t}[l]=\bar{v}_{t,t}[l]$ for all 
$t\leq \tau$ and $\tau'\in\{0,\ldots,t\}$ and 
$\bar{v}_{\tau',\tau'}[l]>\bar{v}_{t,t}[l]$ for all $\tau'<t\leq \tau$. 
From these properties and the induction hypothesis 
$\bar{v}_{t}[l]=\bar{v}_{t,t}[l]$ for all $t\leq \tau$, 
as well as Assumption~\ref{assumption_Bayes}, 
we can use Lemma~\ref{lemma_outer_covariance} for all $t\leq \tau$. 

We first derive the representation of $\bar{\Sigma}_{\tau',\tau}[l]$ in 
(\ref{H_variance_opt}) for $\tau'\in\{0,\ldots,\tau\}$. 
Lemma~\ref{lemma_outer_covariance} for $t=\tau$ 
implies that (\ref{H_variance}) and (\ref{H_variance_ll'}) reduce to 
\begin{equation} 
\bar{\Sigma}_{\tau',\tau}[l]
= \frac{1}{L\bar{\xi}_{\Out,\tau}[l]}
+ \sum_{l'\in\mathcal{N}[l]}\underline{\bar{\Sigma}}_{\tau',\tau,J}[l'\rightarrow l],
\end{equation}
with 
\begin{equation} 
\underline{\bar{\Sigma}}_{\tau',\tau,j}[l\rightarrow l']
= \frac{1}{L\underline{\bar{\xi}}_{\Out,\tau}[l]}
+ \sum_{\tilde{l}'\in\mathcal{N}[l]\backslash\{l'\}}
\underline{\bar{\Sigma}}_{\tau',\tau,j-1}[\tilde{l}'\rightarrow l], 
\end{equation}
which are equivalent to (\ref{H_variance_opt}) and 
(\ref{H_variance_ll'_opt}), because of the identity 
$\bar{\xi}_{\Out,\tau}[l]=\mathbb{E}[M_{\tau}^{2}[l]]$ 
obtained from Lemma~\ref{lemma_outer_covariance}. 
The identity $\bar{\Sigma}_{\tau',\tau}[l]=\bar{\Sigma}_{\tau,\tau}[l]$ is trivial 
for $\tau'\in\{0,\ldots,\tau\}$ from the definitions of 
$\bar{\Sigma}_{\tau',\tau}[l]$ 
and $\underline{\bar{\Sigma}}_{\tau',\tau}[l\rightarrow l']$. 

We next prove $\bar{\eta}_{\tau}[l]=\eta_{\tau}[l]$ and 
$\bar{\sigma}_{\tau}^{2}[l]=\bar{\Sigma}_{\tau,\tau}[l]$. 
We use the induction hypothesis $\bar{\zeta}_{\tau}[l]=\bar{\xi}_{\Out,\tau}[l]$ 
for the definitions of $\bar{\eta}_{\tau}[l]$ and 
$\underline{\bar{\eta}}_{\tau,j}[l\rightarrow l']$ in (\ref{eta_bar}) and 
(\ref{eta_ll'}) 
to obtain $\bar{\eta}_{\tau}[l]=\eta_{\tau}[l]$ given in (\ref{eta}). 
The identity $\bar{\sigma}_{\tau}^{2}[l]=\bar{\Sigma}_{\tau,\tau}[l]$ follows from 
the definitions of $\bar{\sigma}_{\tau}^{2}[l]$,  
$\underline{\bar{\sigma}}_{\tau,j}^{2}[l\rightarrow l']$, 
$\bar{\Sigma}_{\tau,\tau}[l]$, 
and $\underline{\bar{\Sigma}}_{\tau,\tau,j}[l\rightarrow l']$ 
in (\ref{sigma_t_bar}), (\ref{sigma_ll'_bar}), (\ref{H_variance_opt}), and 
(\ref{H_variance_ll'_opt}), as well as Lemma~\ref{lemma_outer_covariance}. 

Finally, we prove $\bar{\Sigma}_{\tau',\tau'}>\bar{\Sigma}_{t,t}[l]$ for all 
$\tau'<t\leq \tau$. From the definition of $\bar{\Sigma}_{\tau',\tau}[l]$ 
in (\ref{H_variance_opt}), 
it is sufficient to prove $\mathbb{E}[M_{\tau'}^{2}[l]]
<\mathbb{E}[M_{t}^{2}[l]]$ for all $\tau'<t\leq \tau$. 
From Properties~\ref{Oc} and~\ref{Oe} in Theorem~\ref{theorem_SE_tech}, we 
find the positive definiteness of the covariance matrix that has 
$\mathbb{E}[M_{\tau'}[l]M_{t}[l]]$ as the $(\tau', t)$ element for all 
$\tau',t\in\{0,\ldots,\tau\}$. Using 
Lemma~\ref{lemma_monotonicity} for this covariance matrix with 
$\mathbb{E}[M_{\tau'}[l]M_{t}[l]]=\mathbb{E}[M_{\tau'}^{2}[l]]$ for all 
$\tau'\in\{0,\ldots,t\}$, obtained from Lemma~\ref{lemma_outer_covariance} 
for all $t\leq \tau$, 
we arrive at $\mathbb{E}[M_{\tau'}^{2}[l]]<\mathbb{E}[M_{t}^{2}[l]]$ for all 
$\tau'<t\leq \tau$, which implies 
$\bar{\Sigma}_{\tau',\tau'}>\bar{\Sigma}_{t,t}[l]$ for all 
$\tau'<t\leq \tau$. 

\subsection{Proof for the Inner Module}
\label{proof_lemma_consistency_inner}
We prove the identity $\mathrm{cov}_{\tau',\tau+1}[l]
=\mathrm{cov}_{\tau+1,\tau+1}[l]$ 
for all $\tau'\in\{0,\ldots,\tau+1\}$. We first consider the case of 
$\tau'=0$. Let $f_{\In,\tau}[l]= f_{\In}[l](L^{-1}\bar{\eta}_{\tau}[l]X 
+ \tilde{H}_{\tau}[l]; \eta_{\tau}[l], \bar{\Sigma}_{\tau,\tau}[l])$. 
Using the definition of $\mathrm{cov}_{0,\tau+1}[l]$ in 
(\ref{error_covariance0}) and $\bar{\sigma}_{\tau}^{2}[l]
=\bar{\Sigma}_{\tau,\tau}[l]$ yields 
\begin{align}
\mathrm{cov}_{0,\tau+1}[l]
&= \mathbb{E}\left[
 (X - f_{\In,\tau}[l] + f_{\In,\tau}[l])(X - f_{\In,\tau}[l])
\right]
\nonumber \\
&= \mathbb{E}[(X - f_{\In,\tau}[l])^{2}], 
\end{align}
where the last equality follows from the well-known property 
$\mathbb{E}[f_{\In,\tau}[l](X - f_{\In,\tau}[l])]=0$ for the Bayes-optimal inner 
denoiser~(\ref{inner_denoiser}). Thus, we use the definition of 
$\mathrm{cov}_{\tau+1,\tau+1}[l]$ in (\ref{error_covariance}) to have 
$\mathrm{cov}_{0,\tau+1}[l]=\mathrm{cov}_{\tau+1,\tau+1}[l]$. 

We next consider the case of $\tau'>0$. From 
$\bar{\Sigma}_{\tau',\tau}[l]=\bar{\Sigma}_{\tau,\tau}[l]$ 
for all $\tau'\in\{0,\ldots,\tau\}$, 
$\bar{\Sigma}_{\tau',\tau'}>\bar{\Sigma}_{t,t}$ for all $\tau'<t\leq \tau$, 
and $\bar{\sigma}_{\tau}^{2}[l]=\bar{\Sigma}_{\tau,\tau}[l]$, we use 
Lemma~\ref{lemma_inner_covariance} to obtain 
$\mathrm{cov}_{\tau'+1,\tau+1}[l]=\mathrm{cov}_{\tau+1,\tau+1}[l]$ for all 
$\tau'\in\{0,\ldots,\tau\}$. Combining these results, we arrive at 
$\mathrm{cov}_{\tau',\tau+1}[l]=\mathrm{cov}_{\tau+1,\tau+1}[l]$ for all 
$\tau'\in\{0,\ldots,\tau+1\}$. 

Let us prove $\bar{v}_{\tau+1}[l]=\bar{v}_{\tau+1,\tau+1}[l]$. From 
$\bar{\eta}_{\tau}[l]=\eta_{\tau}[l]$ and 
$\bar{\sigma}_{\tau}^{2}[l]=\bar{\Sigma}_{\tau,\tau}[l]$, we use 
Lemma~\ref{lemma_posterior} for the Bayes-optimal inner 
denoiser~(\ref{inner_denoiser}) and the definition of $\bar{\xi}_{\In,\tau}[l]$ 
in (\ref{xi_in_bar}) to obtain 
\begin{equation}
\bar{\xi}_{\In,\tau}[l]
=\frac{\eta_{\tau}[l]\mathrm{cov}_{\tau+1,\tau+1}[l]}{L\bar{\Sigma}_{\tau,\tau}[l]}, 
\end{equation}  
with $\mathrm{cov}_{\tau+1,\tau+1}[l]$ given in (\ref{error_covariance}). 
Applying this identity and $\bar{\sigma}_{\tau}^{2}[l]
=\bar{\Sigma}_{\tau,\tau}[l]$ to $\bar{v}_{\tau+1}[l]$ in (\ref{v_t_bar}), we have  
\begin{equation}
\bar{v}_{\tau+1}[l] = \frac{1}{L\delta[l]}\mathrm{cov}_{\tau+1,\tau+1}[l],  
\end{equation}
which is equal to $\bar{v}_{\tau+1,\tau+1}[l]$ in (\ref{v_t_bar_true}). 

We prove the identity $\bar{\zeta}_{\tau+1}[l]=\bar{\xi}_{\Out,\tau+1}[l]$. 
Using $\mathrm{cov}_{0,\tau+1}[l]=\mathrm{cov}_{\tau+1,\tau+1}[l]$ and the 
definition~(\ref{v_t_bar_true0}) yields $\bar{v}_{0,\tau+1}[l]
=\bar{v}_{\tau+1,\tau+1}[l]$. 
From this identity,  $\bar{v}_{\tau+1}[l]=\bar{v}_{\tau+1,\tau+1}[l]$, and 
Assumption~\ref{assumption_Bayes}, we use Lemma~\ref{lemma_identity} 
to obtain the identity $\bar{\zeta}_{\tau+1}[l]=\bar{\xi}_{\Out,\tau+1}[l]$. 

Finally, we prove $\mathrm{cov}_{\tau',\tau'}[l]>\mathrm{cov}_{t,t}[l]$ 
for all $\tau'<t\leq \tau+1$. Using $\bar{\sigma}_{\tau}[l]
=\bar{\Sigma}_{\tau,\tau}[l]$ 
and $\bar{\zeta}_{\tau+1}[l]=\bar{\xi}_{\Out,\tau+1}[l]$, we represent 
the random variable $Q_{\tau'}$ in (\ref{Q_t}) as $Q_{0}[l] = -X$ and 
\begin{equation} \label{Q_t_representation}
Q_{\tau'+1}[l] = f_{\In}[l]\left(
 \frac{\bar{\eta}_{\tau'}[l]}{L}X + \tilde{H}_{\tau'}[l]; \eta_{\tau'}[l], 
 \bar{\Sigma}_{\tau',\tau'}[l] 
\right) - X
\end{equation}
for $\tau'\geq0$, which imply 
$\mathbb{E}[Q_{\tau'}[l]Q_{t}[l]]=\mathrm{cov}_{\tau',t}[l]$ 
from the definitions of $\mathrm{cov}_{\tau',t}[l]$ in 
(\ref{error_covariance}), (\ref{error_covariance0}), and 
(\ref{error_covariance00}). Properties~\ref{Ic} and \ref{Ie} in 
Theorem~\ref{theorem_SE_tech} imply the positive definiteness of the 
covariance matrix that has $\mathrm{cov}_{\tau',t}[l]$ as the 
$(\tau',t)$ element for all $\tau', t\in\{0,\ldots,\tau+1\}$. 
From this positive definiteness and 
$\mathrm{cov}_{\tau',\tau+1}[l]=\mathrm{cov}_{\tau+1,\tau+1}[l]$ for all 
$\tau'\in\{0,\ldots,\tau+1\}$, as well as the induction hypothesis 
$\mathrm{cov}_{\tau',t}[l]=\mathrm{cov}_{t,t}[l]$ for all 
$t\in\{0,\ldots,\tau\}$ and $\tau'\in\{0,\ldots,t\}$, we use 
Lemma~\ref{lemma_monotonicity} to obtain 
$\mathrm{cov}_{\tau',\tau'}[l]>\mathrm{cov}_{t,t}[l]$ 
for all $\tau'<t\leq \tau+1$. 
Thus, Lemma~\ref{lemma_consistency} holds for $t=\tau$. 

\section*{Acknowledgment}
The author thanks the anonymous reviewers for their suggestions that have 
improved the quality of the manuscript greatly. 

\balance 

\bibliographystyle{IEEEtran}
\bibliography{IEEEabrv,kt-it2023}

\begin{thebibliography}{10}
\providecommand{\url}[1]{#1}
\csname url@samestyle\endcsname
\providecommand{\newblock}{\relax}
\providecommand{\bibinfo}[2]{#2}
\providecommand{\BIBentrySTDinterwordspacing}{\spaceskip=0pt\relax}
\providecommand{\BIBentryALTinterwordstretchfactor}{4}
\providecommand{\BIBentryALTinterwordspacing}{\spaceskip=\fontdimen2\font plus
\BIBentryALTinterwordstretchfactor\fontdimen3\font minus
  \fontdimen4\font\relax}
\providecommand{\BIBforeignlanguage}[2]{{%
\expandafter\ifx\csname l@#1\endcsname\relax
\typeout{** WARNING: IEEEtran.bst: No hyphenation pattern has been}%
\typeout{** loaded for the language `#1'. Using the pattern for}%
\typeout{** the default language instead.}%
\else
\language=\csname l@#1\endcsname
\fi
#2}}
\providecommand{\BIBdecl}{\relax}
\BIBdecl

\bibitem{Donoho09}
D.~L. Donoho, A.~Maleki, and A.~Montanari, ``Message-passing algorithms for
  compressed sensing,'' \emph{Proc. Nat. Acad. Sci.}, vol. 106, no.~45, pp.
  18\,914--18\,919, Nov. 2009.

\bibitem{Donoho06}
D.~L. Donoho, ``Compressed sensing,'' \emph{{IEEE} Trans. Inf. Theory},
  vol.~52, no.~4, pp. 1289--1306, Apr. 2006.

\bibitem{Candes061}
E.~J. Cand\`es, J.~Romberg, and T.~Tao, ``Robust uncertainty principles: Exact
  signal reconstruction from highly incomplete frequency information,''
  \emph{{IEEE} Trans. Inf. Theory}, vol.~52, no.~2, pp. 489--509, Feb. 2006.

\bibitem{Kabashima03}
Y.~Kabashima, ``A {CDMA} multiuser detection algorithm on the basis of belief
  propagation,'' \emph{J. Phys. A: Math. Gen.}, vol.~36, no.~43, pp.
  11\,111--11\,121, Oct. 2003.

\bibitem{Som12}
S.~Som and P.~Schniter, ``Compressive imaging using approximate message passing
  and a {Markov}-tree prior,'' \emph{{IEEE} J. Sel. Topics Signal Process.},
  vol.~60, no.~7, pp. 3439--3448, Jul. 2012.

\bibitem{Tan15}
J.~Tan, Y.~Ma, and D.~Baron, ``Compressive imaging via approximate message
  passing with image denoising,'' \emph{{IEEE} Trans. Signal Process.},
  vol.~63, no.~8, pp. 2085--2092, Apr. 2015.

\bibitem{Anitori13}
L.~Anitori, A.~Maleki, M.~Otten, R.~G. Baraniuk, and P.~Hoogeboom, ``Design and
  analysis of compressed sensing radar detectors,'' \emph{{IEEE} Trans. Signal
  Process.}, vol.~61, no.~4, pp. 813--827, Feb. 2013.

\bibitem{Rush17}
C.~Rush, A.~Greig, and R.~Venkataramanan, ``Capacity-achieving sparse
  superposition codes via approximate message passing decoding,'' \emph{{IEEE}
  Trans. Inf. Theory}, vol.~63, no.~3, pp. 1476--1500, Mar. 2017.

\bibitem{Barbier17}
J.~Barbier and F.~Krzakala, ``Approximate message-passing decoder and capacity
  achieving sparse superposition codes,'' \emph{{IEEE} Trans. Inf. Theory},
  vol.~63, no.~8, pp. 4894--4927, Aug. 2017.

\bibitem{Lesieur17}
T.~Lesieur, F.~Krzakala, and L.~Zdeborov\'a, ``Constrained low-rank matrix
  estimation: phase transitions, approximate message passing and
  applications,'' \emph{J. Stat. Mech.: Theory Exp.}, vol. 2017, no.~7, p.
  073403, Jul. 2017.

\bibitem{Montanari21}
A.~Montanari and R.~Venkataramanan, ``Estimation of low-rank matrices via
  approximate message passing,'' \emph{Ann. Stat.}, vol.~49, no.~1, pp.
  321--345, Feb. 2021.

\bibitem{Bayati11}
M.~Bayati and A.~Montanari, ``The dynamics of message passing on dense graphs,
  with applications to compressed sensing,'' \emph{{IEEE} Trans. Inf. Theory},
  vol.~57, no.~2, pp. 764--785, Feb. 2011.

\bibitem{Bayati15}
M.~Bayati, M.~Lelarge, and A.~Montanari, ``Universality in polytope phase
  transitions and message passing algorithms,'' \emph{Ann. Appl. Probab.},
  vol.~25, no.~2, pp. 753--822, Apr. 2015.

\bibitem{Takeuchi19}
K.~Takeuchi, ``A unified framework of state evolution for message-passing
  algorithms,'' in \emph{Proc. 2019 IEEE Int. Symp. Inf. Theory}, Paris,
  France, Jul. 2019, pp. 151--155.

\bibitem{Bolthausen14}
E.~Bolthausen, ``An iterative construction of solutions of the {TAP} equations
  for the {Sherrington}-{Kirkpatrick} model,'' \emph{Commun. Math. Phys.}, vol.
  325, no.~1, pp. 333--366, Jan. 2014.

\bibitem{Reeves19}
G.~Reeves and H.~D. Pfister, ``The replica-symmetric prediction for random
  linear estimation with {Gaussian} matrices is exact,'' \emph{{IEEE} Trans.
  Inf. Theory}, vol.~65, no.~4, pp. 2252--2283, Apr. 2019.

\bibitem{Barbier20}
J.~Barbier, N.~Macris, M.~Dia, and F.~Krzakala, ``Mutual information and
  optimality of approximate message-passing in random linear estimation,''
  \emph{{IEEE} Trans. Inf. Theory}, vol.~66, no.~7, pp. 4270--4303, Jul. 2020.

\bibitem{Rangan11}
S.~Rangan, ``Generalized approximate message passing for estimation with random
  linear mixing,'' in \emph{Proc. 2011 IEEE Int. Symp. Inf. Theory}, Saint
  Petersburg, Russia, Aug. 2011, pp. 2168--2172.

\bibitem{Kamilov121}
U.~S. Kamilov, A.~Bourquard, A.~Amini, and M.~Unser, ``One-bit measurements
  with adaptive thresholds,'' \emph{{IEEE} Signal Process. Lett.}, vol.~19,
  no.~10, pp. 607--610, Oct. 2012.

\bibitem{Kamilov122}
U.~S. Kamilov, V.~K. Goyal, and S.~Rangan, ``Message-passing de-quantization
  with applications to compressed sensing,'' \emph{{IEEE} Trans. Signal
  Process.}, vol.~60, no.~12, pp. 6270--6281, Dec. 2012.

\bibitem{Schniter14}
P.~Schniter and S.~Rangan, ``Compressive phase retrieval via generalized
  approximate message passing,'' \emph{{IEEE} Trans. Signal Process.}, vol.~63,
  no.~4, pp. 1043--1055, Feb. 2015.

\bibitem{Ma19}
J.~Ma, J.~Xu, and A.~Maleki, ``Optimization-based {AMP} for phase retrieval:
  The impact of initialization and $\ell_{2}$ regularization,'' \emph{{IEEE}
  Trans. Inf. Theory}, vol.~65, no.~6, pp. 3600--3629, Jun. 2019.

\bibitem{Bao16}
H.~Bao, J.~Fang, Z.~Chen, H.~Li, and S.~Li, ``An efficient {Bayesian} {PAPR}
  reduction method for {OFDM}-based massive {MIMO} systems,'' \emph{{IEEE}
  Trans. Wireless Commun.}, vol.~15, no.~6, pp. 4183--4195, Jun. 2016.

\bibitem{Chen16}
J.-C. Chen, C.-J. Wang, K.-K. Wong, and C.-K. Wen, ``Low-complexity precoding
  design for massive multiuser {MIMO} systems using approximate message
  passing,'' \emph{{IEEE} Trans. Veh. Technol.}, vol.~65, no.~7, pp.
  5707--5714, Jul. 2016.

\bibitem{Javanmard13}
A.~Javanmard and A.~Montanari, ``State evolution for general approximate
  message passing algorithms, with applications to spatial coupling,''
  \emph{Inf. Inference: A Journal of the IMA}, vol.~2, no.~2, pp. 115--144,
  Dec. 2013.

\bibitem{Barbier19}
J.~Barbier, F.~Krzakala, N.~Macris, L.~Miolane, and L.~Zdeborov\'a, ``Optimal
  errors and phase transitions in high-dimensional generalized linear models,''
  \emph{Proc. Nat. Acad. Sci.}, vol. 116, no.~12, pp. 5451--5460, Mar. 2019.

\bibitem{Marzetta10}
T.~L. Marzetta, ``Noncooperative cellular wireless with unlimited numbers of
  base station antennas,'' \emph{{IEEE} Trans. Wireless Commun.}, vol.~9,
  no.~11, pp. 3590--3600, Nov. 2010.

\bibitem{Ngo17}
H.~Q. Ngo, A.~Ashikhmin, H.~Yang, E.~G. Larsson, and T.~L. Marzetta,
  ``Cell-free massive {MIMO} versus small cells,'' \emph{{IEEE} Trans. Wireless
  Commun.}, vol.~16, no.~3, pp. 1834--1850, Mar. 2017.

\bibitem{Daubechies04}
I.~Daubechies, M.~Defrise, and C.~D. Mol, ``An iterative thresholding algorithm
  for linear inverse problems with a sparsity constraint,'' \emph{Commun. Pure
  Appl. Math.}, vol.~57, no.~11, pp. 1413--1457, Aug. 2004.

\bibitem{Beck09}
A.~Beck and M.~Teboulle, ``A fast iterative shrinkage-thresholding algorithm
  for linear inverse problems,'' \emph{SIAM J. Imaging Sci.}, vol.~2, no.~1,
  pp. 183--202, Mar. 2009.

\bibitem{Blumensath09}
T.~Blumensath and M.~E. Davies, ``Iterative hard thresholding for compressed
  sensing,'' \emph{Appl. Comput. Harmon. Anal.}, vol.~27, no.~3, pp. 265--274,
  Nov. 2009.

\bibitem{Patterson13}
S.~Patterson, Y.~C. Eldar, and I.~Keidar, ``Distributed sparse signal recovery
  for sensor networks,'' in \emph{Proc. 2013 IEEE Int. Conf. Acoust. Speech
  Signal Process.}, Vancouver, BC, Canada, May 2013, pp. 4494--4498.

\bibitem{Saber04}
R.~Olfati-Saber and R.~M. Murray, ``Consensus problems in networks of agents
  with switching topology and time-delays,'' \emph{{IEEE} Trans. Autom.
  Control}, vol.~49, no.~9, pp. 1520--1533, Sep. 2004.

\bibitem{Xiao04}
L.~Xiao and S.~Boyd, ``Fast linear iterations for distributed averaging,''
  \emph{Syst. Contr. Lett.}, vol.~53, no.~1, pp. 65--78, Sep. 2004.

\bibitem{Mateos10}
G.~Mateos, J.~A. Bazerque, and G.~B. Giannakis, ``Distributed sparse linear
  regression,'' \emph{{IEEE} Trans. Signal Process.}, vol.~58, no.~10, pp.
  5262--5276, Oct. 2010.

\bibitem{Bazerque10}
J.~A. Bazerque and G.~B. Giannakis, ``Distributed spectrum sensing for
  cognitive radio networks by exploiting sparsity,'' \emph{{IEEE} Trans. Signal
  Process.}, vol.~58, no.~3, pp. 1847--1862, Mar. 2010.

\bibitem{Mota12}
J.~F.~C. Mota, J.~M.~F. Xavier, P.~M.~Q. Aguiar, and M.~P\"uschel,
  ``Distributed basis pursuit,'' \emph{{IEEE} Trans. Signal Process.}, vol.~60,
  no.~4, pp. 1942--1956, Apr. 2012.

\bibitem{Mota13}
------, ``{D-ADMM}: A communication-efficient distributed algorithm for
  separable optimization,'' \emph{{IEEE} Trans. Signal Process.}, vol.~61,
  no.~10, pp. 2718--2723, May 2013.

\bibitem{Shi14}
W.~Shi, Q.~Ling, K.~Yuan, G.~Wu, and W.~Yin, ``On the linear convergence of the
  {ADMM} in decentralized consensus optimization,'' \emph{{IEEE} Trans. Signal
  Process.}, vol.~62, no.~7, pp. 1750--1761, Apr. 2014.

\bibitem{Han14}
P.~Han, R.~Niu, M.~Ren, and Y.~C. Eldar, ``Distributed approximate message
  passing for sparse signal recovery,'' in \emph{Proc. 2014 IEEE Global Conf.
  Signal Inf. Process.}, Atlanta, GA, USA, Dec. 2014, pp. 497--501.

\bibitem{Han16}
P.~Han, J.~Zhu, R.~Niu, and D.~Baron, ``Multi-processor approximate message
  passing using lossy compression,'' in \emph{Proc. 2016 IEEE Int. Conf.
  Acoust. Speech Signal Process.}, Shanghai, China, Mar. 2016, pp. 6240--6244.

\bibitem{Ma17}
Y.~Ma, Y.~M. Lu, and D.~Baron, ``Multiprocessor approximate message passing
  with column-wise partitioning,'' in \emph{Proc. 2017 IEEE Int. Conf. Acoust.
  Speech Signal Process.}, New Orleans, LA, USA, Mar. 2017, pp. 4765--4769.

\bibitem{Guo22}
M.~Guo and M.~C. Gursoy, ``Joint activity detection and channel estimation in
  cell-free massive {MIMO} networks with massive connectivity,'' \emph{{IEEE}
  Trans. Commun.}, vol.~70, no.~1, pp. 317--331, Jan. 2022.

\bibitem{Bai22}
J.~Bai and E.~G. Larsson, ``Activity detection in distributed {MIMO}:
  Distributed {AMP} via likelihood ratio fusion,'' \emph{{IEEE} Wireless
  Commun. Lett.}, vol.~11, no.~10, pp. 2200--2204, Oct. 2022.

\bibitem{Hayakawa18}
R.~Hayakawa, A.~Nakai, and K.~Hayashi, ``Distributed approximate message
  passing with summation propagation,'' in \emph{Proc. 2018 IEEE Int. Conf.
  Acoust. Speech Signal Process.}, Calgary, AB, Canada, Apr. 2018, pp.
  4104--4108.

\bibitem{Moallemi06}
C.~C. Moallemi and B.~V. Roy, ``Consensus propagation,'' \emph{{IEEE} Trans.
  Inf. Theory}, vol.~52, no.~11, pp. 4753--4766, Nov. 2006.

\bibitem{Takeuchi221}
K.~Takeuchi, ``On the convergence of orthogonal/vector {AMP}: Long-memory
  message-passing strategy,'' in \emph{Proc. 2022 IEEE Int. Symp. Inf. Theory},
  Espoo, Finland, Jun.--Jul. 2022, pp. 1366--1371.

\bibitem{Takeuchi222}
------, ``On the convergence of orthogonal/vector {AMP}: Long-memory
  message-passing strategy,'' \emph{{IEEE} Trans. Inf. Theory}, vol.~68,
  no.~12, pp. 8121--8138, Dec. 2022.

\bibitem{Liu221}
L.~Liu, S.~Huang, and B.~M. Kurkoski, ``Sufficient statistic memory approximate
  message passing,'' in \emph{Proc. 2022 IEEE Int. Symp. Inf. Theory}, Espoo,
  Finland, Jun.--Jul. 2022, pp. 1378--1383.

\bibitem{Takeuchi202}
K.~Takeuchi, ``Convolutional approximate message-passing,'' \emph{{IEEE} Signal
  Process. Lett.}, vol.~27, pp. 416--420, 2020.

\bibitem{Takeuchi21}
------, ``Bayes-optimal convolutional {AMP},'' \emph{{IEEE} Trans. Inf.
  Theory}, vol.~67, no.~7, pp. 4405--4428, Jul. 2021.

\bibitem{Fan22}
Z.~Fan, ``Approximate message passing algorithms for rotationally invariant
  matrices,'' \emph{Ann. Statist.}, vol.~50, no.~1, pp. 197--224, Feb. 2022.

\bibitem{Venkataramanan22}
R.~Venkataramanan, K.~K\"ogler, and M.~Mondelli, ``Estimation in rotationally
  invariant generalized linear models via approximate message passing,'' in
  \emph{Proc. 39th Int. Conf. Mach. Learn.}, Baltimore, MD, USA, Jul. 2022.

\bibitem{Skuratovs22}
N.~Skuratovs and M.~E. Davies, ``Compressed sensing with upscaled vector
  approximate message passing,'' \emph{{IEEE} Trans. Inf. Theory}, vol.~68,
  no.~7, pp. 4818--4836, Jul. 2022.

\bibitem{Liu222}
L.~Liu, S.~Huang, and B.~M. Kurkoski, ``Memory {AMP},'' \emph{{IEEE} Trans.
  Inf. Theory}, vol.~68, no.~12, pp. 8015--8039, Dec. 2022.

\bibitem{Takeuchi24}
K.~Takeuchi, ``Decentralized generalized approximate message-passing for
  tree-structured networks,'' in \emph{Proc. 2024 IEEE Int. Conf. Acoust.
  Speech Signal Process.}, Seoul, Korea, Apr. 2024, pp. 12\,866--12\,870.

\bibitem{Gerbelot23}
C.~Gerbelot and R.~Berthier, ``Graph-based approximate message passing
  iterations,'' \emph{Inf. Inference: A Journal of the IMA}, vol.~12, no.~4,
  pp. 2562--2628, Dec. 2023.

\bibitem{Manoel17}
A.~Manoel, F.~Krzakala, M.~M\'ezard, and L.~Zdeborov\'a, ``Multi-layer
  generalized linear estimation,'' in \emph{Proc. 2017 IEEE Int. Symp. Inf.
  Theory}, Aachen, Germany, Jun. 2017, pp. 2098--2102.

\bibitem{Aubin20}
B.~Aubin, B.~Loureiro, A.~Maillard, F.~Krzakala, and L.~Zdeborov\'a, ``The
  spiked matrix model with generative priors,'' \emph{{IEEE} Trans. Inf.
  Theory}, vol.~67, no.~2, pp. 1156--1181, Feb. 2021.

\bibitem{Ma23}
Z.~Ma and S.~Nandy, ``Community detection with contextual multilayer
  networks,'' \emph{{IEEE} Trans. Inf. Theory}, vol.~69, no.~5, pp. 3203--3239,
  May 2023.

\bibitem{Cobo23}
P.~P. Cobo, K.~Hsieh, and R.~Venkataramanan, ``Bayes-optimal estimation in
  generalized linear models via spatial coupling,'' in \emph{Proc. 2023 IEEE
  Int. Symp. Inf. Theory}, Taipei, Taiwan, Jun. 2023, pp. 773--778.

\bibitem{Takeuchi201}
K.~Takeuchi, ``Rigorous dynamics of expectation-propagation-based signal
  recovery from unitarily invariant measurements,'' \emph{{IEEE} Trans. Inf.
  Theory}, vol.~66, no.~1, pp. 368--386, Jan. 2020.

\bibitem{Murphy99}
K.~P. Murphy, Y.~Weiss, and M.~I. Jordan, ``Loopy belief propagation for
  approximate inference: An empirical study,'' in \emph{Proc. 15th Conf.
  Uncertain. Artif. Intell.}, Stockholm, Sweden, Jul.--Aug. 1999, pp. 467--475.

\bibitem{Vila15}
J.~Vila, P.~Schniter, S.~Rangan, F.~Krzakala, and L.~Zdeborov\'a, ``Adaptive
  damping and mean removal for the generalized approximate message passing
  algorithm,'' in \emph{Proc. 2015 IEEE Int. Conf. Acoust. Speech Signal
  Process.}, South Brisbane, Australia, Apr. 2015, pp. 2021--2025.

\bibitem{Rangan19}
S.~Rangan, P.~Schniter, A.~Fletcher, and S.~Sarkar, ``On the convergence of
  approximate message passing with arbitrary matrices,'' \emph{{IEEE} Trans.
  Inf. Theory}, vol.~65, no.~9, pp. 5339--5351, Sep. 2019.

\bibitem{Yoshida23}
T.~Yoshida and K.~Takeuchi, ``Deep learning of damped {AMP} decoding networks
  for sparse superposition codes via annealing,'' \emph{IEICE Trans.
  Fundamentals.}, vol. E106-A, no.~3, pp. 414--421, Mar. 2023.

\bibitem{Stein72}
C.~Stein, ``A bound for the error in the normal approximation to the
  distribution of a sum of dependent random variables,'' in \emph{6th Berkeley
  Symp. Math. Statist. Prob.}, vol.~2, 1972, pp. 583--602.

\end{thebibliography}





\end{document}